\documentclass[12pt,english,reqno]{smfart}
\usepackage{mathptmx}
\theoremstyle{plain}
\newtheorem{thm}{Theorem}[section]
\newtheorem{lem}[thm]{Lemma}
\newtheorem{prop}[thm]{Proposition}
\newtheorem{defn}[thm]{Definition}
\theoremstyle{definition}
\newtheorem*{rem} {Remark}
\newtheorem*{ex} {Example}
\title{Symmetry classes of disordered fermions}
\author{P.\ Heinzner, A.\ Huckleberry}
\address{Fakult\"at f\"ur Mathematik, Ruhr-Universit\"at Bochum,
Germany}
\author{M.R.\ Zirnbauer}
\address{Institut f\"ur Theoretische Physik, Universit\"at zu
K\"oln, Germany}
\email{heinzner@cplx.ruhr-uni-bochum.de,
ahuck@cplx.ruhr-uni-bochum.de, zirn@thp.uni-koeln.de}
\date{June 10, 2004}
\renewcommand{\theequation}{\arabic{section}.\arabic{equation}}
\begin{document}
\begin{abstract}
Building upon Dyson's fundamental 1962 article known in random-matrix
theory as {\it the threefold way}, we classify disordered fermion
systems with quadratic Hamiltonians by their unitary and antiunitary
symmetries. Important physical examples are afforded by noninteracting
quasiparticles in disordered metals and superconductors, and by
relativistic fermions in random gauge field backgrounds.

The primary data of the classification are a Nambu space of
fermionic field operators which carry a representation of some
symmetry group.  Our approach is to eliminate all of the unitary
symmetries from the picture by transferring to an irreducible
block of equivariant homomorphisms.  After reduction, the block
data specifying a linear space of symmetry-compatible Hamiltonians
consist of a basic vector space $V$, a space of endomorphisms in
$\mathrm {End} (V \oplus V^\ast)$, a bilinear form on $V \oplus
V^\ast$ which is either symmetric or alternating, and one or two
antiunitary symmetries that may mix $V$ with $V^\ast$.  Every such
set of block data is shown to determine an irreducible classical
compact symmetric space. Conversely, every irreducible classical
compact symmetric space occurs in this way.

This proves the correspondence between symmetry classes and symmetric
spaces conjectured some time ago.\\

\noindent$\mathsf{Keywords}$: disordered electron systems, random
Dirac fermions, quantum chaos; representation theory, symmetric spaces
\end{abstract}
\maketitle

\section{Introduction}\label{sect: intro}

In a famous and influential paper published in 1962 (``The
threefold way: algebraic structure of symmetry groups and
ensembles in quantum mechanics'' \cite{dyson}), Freeman J. Dyson
classified matrix ensembles by a scheme that became fundamental to
several areas of theoretical physics, including the statistical
theory of complex many-body systems, mesoscopic physics,
disordered electron systems, and the area of quantum chaos. Being
set in the context of standard quantum mechanics, Dyson's
classification asserted that ``the most general matrix ensemble,
defined with a symmetry group that may be completely arbitrary,
reduces to a direct product of independent irreducible ensembles
each of which belongs to one of three known types.'' These three
ensembles, or rather their underlying matrix spaces, are nowadays
known as the Wigner-Dyson symmetry classes of orthogonal, unitary,
and symplectic symmetry.

Over the last ten years, various matrix spaces beyond Dyson's
threefold way have come to the fore in random-matrix physics and
mathematics. On the physics side, such spaces arise in problems of
disordered or chaotic fermions; among these are the Euclidean Dirac
operator in a stochastic gauge field background \cite{vz}, and
quasiparticle excitations in disordered superconductors or metals in
proximity to a superconductor \cite{az}. In the mathematical research
area of number theory, the study of statistical correlations in the
values of the Riemann zeta function, and more generally of families of
$L$-functions, has prompted some of the same extensions \cite{ks}.

A brief account of why new structures emerge on the physics side is as
follows. When Dirac first wrote down his famous equation in 1928, he
assumed that he was writing an equation for the \emph{wavefunction} of
the electron. Later, because of the instability caused by
negative-energy solutions, the Dirac equation was reinterpreted (via
second quantization) as an equation for the \emph{fermionic field
operators} of a quantum field theory. A similar change of viewpoint is
carried out in reverse in the Hartree-Fock-Bogoliubov mean-field
description of quasiparticle excitations in superconductors. There,
one starts from the equations of motion for linear superpositions of
the electron creation and annihilation operators, and reinterprets
them as a unitary quantum dynamics for what might be called the
quasiparticle `wavefunction'.

In both cases -- the Dirac equation and the quasiparticle dynamics of
a superconductor -- there enters a structure not present in the
standard quantum mechanics underlying Dyson's classification: the
fermionic field operators are subject to a set of conditions known as
the \emph{canonical anticommutation relations}, and these are
preserved by the quantum dynamics. Therefore, whenever second
quantization is undone (assuming it \emph{can} be undone) to return
from field operators to wavefunctions, the wavefunction dynamics is
required to preserve some extra structure.  This puts a linear
constraint on the allowed Hamiltonians.  A good viewpoint to adopt is
to attribute the extra invariant structure to the Hilbert space,
thereby turning it into a Nambu space.

It was conjectured some time ago \cite{az} that extending Dyson's
classification to the Nambu space setting, the relevant objects one is
led to consider are large families of \emph{symmetric spaces} of
compact type. Past understanding of the systematic nature of the
extended classification scheme relied on the mapping of disordered
fermion problems to field theories with supersymmetric target spaces
\cite{suprev} in combination with renormalization group ideas and the
classification theory of Lie superalgebras.

An extensive review of the mathematics and physics of symmetric
spaces, covering the wide range from the basic definitions to various
random-matrix applications, has recently been given in
\cite{magnea}. That work, however, offers no answers to the question
as to \emph{why} symmetric spaces are relevant for symmetry
classification, and under what assumptions the classification by
symmetric spaces is complete.

In the present paper, we get to the bottom of the subject and,
using a minimal set of tools from linear algebra, give a rigorous
answer to the classification problem for disordered fermions. The
rest of this introduction gives an overview over the mathematical
model to be studied and a statement of our main result.

We begin with a finite- or infinite-dimensional Hilbert space
$\mathcal{V}$ carrying a unitary representation of some compact
Lie group $G_0$ -- this is the group of unitary symmetries of the
disordered fermion system. We emphasize that $G_0$ need not be
connected; in fact, it might be just a finite group.

Let $\mathcal{W} = \mathcal{V} \oplus \mathcal{V}^*$, called the Nambu
space of fermionic field operators, be equipped with the induced
$G_0$-representation. This means that $\mathcal{ V}$ is equipped with
the given representation, and $g(f) : = f \circ g^{-1}$ for $f \in
\mathcal{V}^*$, $g \in G_0$. Let $C : \mathcal{W} \to \mathcal{W}$ be
the $\mathbb{C}$-antilinear involution determined by the Hermitian
scalar product $\langle \, , \, \rangle_\mathcal{V}$ on $\mathcal
{V}$. In physics this operator is called particle-hole
conjugation. Another canonical structure on $\mathcal{W}$ is the
symmetric complex bilinear form $b : \mathcal{W} \times \mathcal{W}
\to \mathbb{C}$ defined by
\begin{displaymath}
    b(v_1 + f_1 , v_2 + f_2) := f_1(v_2) + f_2(v_1) \;.
\end{displaymath}
It encodes the canonical anticommutation relations for fermions,
and is related to the unitary structure $\langle \, , \, \rangle$
of $\mathcal{W}$ by $b(w_1 , w_2) = \langle C w_1 , w_2 \rangle$
for all $w_1, w_2 \in \mathcal{W}$.

It is assumed that $G_0$ is contained in a group $G$ -- the total
symmetry group of the fermion system -- which is acting on
${\mathcal W}$ by transformations that are either unitary or
antiunitary.  An element $g \in G$ either stabilizes ${\mathcal
V}$ or exchanges ${\mathcal V}$ and ${\mathcal V}^*$.  In the
latter case we say that $g \in G$ mixes, and in the former case we
say that it is nonmixing.

The group $G$ is generated by $G_0$ and distinguished elements $g_T$
which act as anti\-unitary operators $T : {\mathcal W} \to {\mathcal
W}$.  These are referred to as distinguished `time-reversal'
symmetries, or $T$-symmetries for short.  The squares of the $g_T$ lie
in the center of the abstract group $G$; we therefore require that the
antiunitary operators $T$ representing them satisfy $T^2 = \pm
\mathrm{Id}$. The subgroup $G_0$ is defined as the set of all elements
of $G$ which are represented as unitary, nonmixing operators on
${\mathcal W}$.

If $T$ and $T_1$ are distinguished time-reversal operators, then $P:=
T T_1$ is a unitary symmetry.  $P$ may be mixing or nonmixing.  In the
latter case, $P$ is in $G_0$. Therefore, modulo $G_0$, there exist at
most two different $T$-symmetries.  If there are exactly two such
symmetries, we adopt the convention that $T$ is mixing and $T_1$ is
nonmixing.  Furthermore, it is assumed that $T$ and $T_1$ either
commute or anticommute, i.e., $T_1T=\pm TT_1$.

As explained throughout this article, all of these situations are well
motivated by physical considerations and examples. We note that
time-reversal symmetry (and all other $T$-symmetries) of the
disordered fermion system may also be broken; in this case $T$ and
$T_1$ are eliminated from the mathematical model and $G_0=G$.

Given $\mathcal{W}$ and the representation of $G$ on it, the object of
interest is the real vector space $\mathcal{H}$ of $\mathbb{C}$-linear
operators in $\mathrm{End}(\mathcal{W})$ that preserve the canonical
structures $b$ and $\langle \, , \, \rangle$ of $\mathcal{W}$ and
commute with the $G$-action. Physically speaking, $\mathcal{H}$ is the
space of `good' Hamiltonians: the field operator dynamics generated
by $H \in \mathcal{H}$ preserves both the canonical anticommutation
relations and the probability in Nambu space, and is compatible with
the prescribed symmetry group $G$.

When unitary symmetries are present, the space $\mathcal{H}$
decomposes by {\it blocks} associated with isomorphism classes of
$G_0$-subrepresentations occurring in $\mathcal{W}$. To formalize
this, recall that two unitary representations $\rho _i : G_0 \to
\mathrm{U}(V_i)$, $i = 1,2$, are equivalent if and only if there
exists a unitary $\mathbb{C }$-linear isomorphism $\varphi : V_1 \to
V_2$ so that $\rho_2(g) (\varphi (v)) = \varphi (\rho_1(g) (v))$ for
all $v \in V_1$ and for all $g \in G_0$. Let $\hat G_0$ denote the
space of equivalence classes of irreducible unitary representations of
$G_0$. An element $\lambda \in \hat{G}_0$ is called an isomorphism
class for short. By standard facts (recall that every representation
of a compact group is completely reducible) the unitary
$G_0$-representation on $\mathcal{V}$ decomposes as an orthogonal sum
over isomorphism classes:
\begin{displaymath}
    \mathcal{V} = \oplus_{\lambda} \mathcal{V}_\lambda \;.
\end{displaymath}
The subspaces $\mathcal{V}_\lambda$ are called the $G_0$-isotypic
components of $\mathcal{V}$.  Some of them may be zero. (Some of the
isomorphism classes of $G_0$ may just not be realized in
$\mathcal{V}$.)

For simplicity suppose now that there is only one distinguished
time-reversal symmetry $T$, and for any fixed $\lambda \in \hat{G}_0$
with $\mathcal{V}_ \lambda \not= 0$, consider the vector space $T
(\mathcal{V}_\lambda)$. If $T$ is nonmixing, i.e., $T : \mathcal{V}
\to \mathcal{V}$, then $T (\mathcal{V}_\lambda) \subset \mathcal{V}$
must coincide with the isotypic component for the same or some other
isomorphism class. (Since conjugation by $g_T$ is an automorphism of
$G_0$, the decomposition into $G_0$-isotypic components is preserved
by $T$.) If $T$ is mixing, i.e., $T : \mathcal{V} \to \mathcal{V}^
\ast$, then $T( \mathcal{V}_ \lambda) = \mathcal{V}_{\lambda
^\prime}^\ast$, still with some $\lambda^\prime \in \hat{G}_0$.

Now define the block $\mathcal{B}_\lambda $ to be the smallest
$G$-invariant space containing $\mathcal{V}_\lambda^{\vphantom{ \ast}}
\oplus \mathcal{V}_\lambda ^*$. Note that if we are in the situation
of nonmixing and $T(\mathcal{V}_\lambda) \not= \mathcal{ V}_{\lambda
}$, then
\begin{displaymath}
    \mathcal{B}_\lambda = \big( \mathcal{V}_\lambda \oplus
    T(\mathcal{V}_\lambda) \big) \oplus \big(\mathcal{V}_\lambda
    \oplus T(\mathcal{V}_\lambda) \big)^* \;.
\end{displaymath}
On the other hand, if we are in the situation of mixing and
$T(\mathcal{V}_\lambda) \not= \mathcal{V}_\lambda^*$, then
\begin{displaymath}
    \mathcal{B}_\lambda = \big(\mathcal{V}_\lambda^
    {\vphantom{\ast}} \oplus T(\mathcal{V}_\lambda^*) \big)
    \oplus \big( \mathcal{V}_\lambda^\ast \oplus
    T(\mathcal{V}_\lambda^{\vphantom{\ast}}) \big) \;.
\end{displaymath}
The block $\mathcal{B}_\lambda$ is halved if $T(\mathcal{V}_
\lambda) = \mathcal{V}_\lambda$ resp.\ $T(\mathcal{V}_\lambda) =
\mathcal{V}_\lambda^\ast$.

Note that if there are two distinguished $T$-symmetries, the above
discussion is only slightly more complicated.  In any case we now have
the basic $G$-invariant blocks $\mathcal{B}_\lambda$.

Because different blocks are built from representations of different
isomorphism classes, the good Hamiltonians do not mix blocks.  Thus
every $H \in \mathcal{H}$ is a direct sum over blocks, and the
structure analysis of $\mathcal{H}$ can be carried out for each block
$\mathcal{B}_ \lambda$ separately. If $\mathcal{V}_\lambda$ is
infinite-dimensional, then to have good mathematical control we
truncate to a finite-dimensional space $V_\lambda \subset
\mathcal{V}_\lambda$ and form the associated block $B_\lambda \subset
\mathcal{W}$.  The truncation is done in such a way that $B_\lambda$
is a $G$-representation space and is Nambu.

The goal now is to compute the space of Hermitian operators on
$B_\lambda$ which commute with the $G$-action and respect the
canonical symmetric $\mathbb{C}$-bilinear form $b$ induced from
that on $\mathcal{V} \oplus \mathcal{V}^*$; such a space of
operators realizes what is called a {\it symmetry class}.

For this, certain spaces of $G_0$-equivariant homomorphisms play
an essential role, i.e., linear maps $S : V_1 \to V_2$ between
$G_0$-representation spaces which satisfy
\begin{displaymath}
    \rho_2(g) \circ S = S\circ \rho_1(g)
\end{displaymath}
for all $g \in G_0$, where $\rho_i : G_0 \to \mathrm{U}(V_i)$, $i
= 1,2$, are the respective representations. If it is clear which
representations are at hand, we often simply write $g \circ S = S
\circ g$ or $S = g S g^{-1}$. Thus we regard the space $\mathrm{
Hom}_{G_0}(V_1,V_2)$ of equivariant homomorphisms as the space of
$G_0$-fixed vectors in the space $\mathrm{Hom}(V_1,V_2)$ of all
linear maps.  If $V_1 = V_2 = V$, then these spaces are denoted by
$\mathrm{End}_{G_0}(V)$ and $\mathrm{ End}(V)$ respectively.

Roughly speaking, there are two steps for computing the relevant
spaces of Hermitian operators. First, the block $B_\lambda$ is
replaced by an analogous block $H_\lambda $ of $G_0$-equi\-variant
homomorphisms from a fixed representation space $R_\lambda $ of
isomorphism class $\lambda$ and/or its dual $R_\lambda^*$ to
$B_\lambda $. The space $H_\lambda$ carries a canonical form
(called either $s$ or $a$) which is induced from $b$. As the
notation indicates, although the original bilinear form on
$B_\lambda$ is symmetric, this induced form is either symmetric or
alternating.

Change of parity occurs in the most interesting case when there is
a $G_0$-equivariant isomorphism $\psi : R_\lambda \to R_\lambda
^*$. In that case there exists a bilinear form $F_\psi : R_\lambda
\times R_\lambda \to \mathbb{C}$ defined by $F_\psi (r,t) = \psi(r
)(t)$, which is either symmetric or alternating. In a certain
sense the form $b$ is a product of $F_\psi$ and a canonical form
on $H_\lambda$. Thus, if $F_\psi$ is alternating, then the
canonical form on $H_\lambda$ must also be alternating.

After transferring to the space $H_\lambda$, in addition to the
canonical bilinear form $s$ or $a$ we have a unitary structure and
conjugation by one or two distinguished time-reversal symmetries. Such
a symmetry $T$ may be mixing or not, and both $T^2 = \mathrm{Id}$ and
$T^2 = - \mathrm{Id}$ are possible. The second main step of our work
is to understand these various cases, each of which is directly
related to a classical symmetric space of compact type.  Such are
given by a classical Lie algebra $\mathfrak{g}$ which is either
$\mathfrak{ su}_n$, $\mathfrak{usp}_{2n}$, or
$\mathfrak{so}_{n}(\mathbb R)$.

In the notation of symmetric spaces we have the following
situation. Let $\mathfrak{g}$ be the Lie algebra of \emph{
antihermitian} endomorphisms of $H_\lambda$ which are isometries (in
the sense of Lie algebra elements) of the induced complex bilinear
form $b = s$ or $b = a$.  This is of compact type, because it is the
intersection of the Lie algebra of the unitary group of $H_\lambda$
and the complex Lie algebra of the group of isometries of $b$.
Conjugation by the antiunitary mapping $T$ defines an involution
$\theta : \mathfrak{g}\to \mathfrak{g}$.

The good Hamiltonians (restricted to the reduced block $H_\lambda$)
are the \emph{Hermitian} operators $h \in \mathrm{i} \mathfrak{g}$
such that at the level of group action the one-parameter groups
$\mathrm{e}^{-\mathrm{i} th}$ satisfy $T \mathrm{ e}^{-\mathrm{i}t h}
= \mathrm{e}^{+\mathrm{i}t h} T$, i.e., $\mathrm{i}h \in \mathfrak{g}$
must anticommute with $T$.  Equivalently, if $\mathfrak{g} = \mathfrak
{k} \oplus \mathfrak{p}$ is the decomposition of $\mathfrak {g}$ into
$\theta$-eigenspaces, the space of operators which is to be computed
is the $(-1) $-eigenspace $\mathfrak{p}$.  The space of good
Hamiltonians restricted to $H_\lambda$ then is $\mathrm{i}
\mathfrak{p}$. Since the appropriate action of the Lie group
$\mathsf{K}$ (with Lie algebra $\mathfrak{k}$) on this space is just
conjugation, one identifies $\mathrm{i} \mathfrak{p}$ with the tangent
space $\mathfrak{g} / \mathfrak{k}$ of an associated symmetric space
${\mathsf G}/{\mathsf K}$ of compact type.

It should be underlined that there is more than one symmetric
space associated to a Cartan decomposition $\mathfrak{g} =
\mathfrak{k} \oplus \mathfrak{p}$. We are most interested in the
one consisting of the physical time-evolution operators
$\mathrm{e}^{-\mathrm{i}t h}$; if $\mathsf{G}$ (not to be confused
with the symmetry group $G$) is the semisimple and simply
connected Lie group with Lie algebra $\mathfrak{g}$, this is given
as the image of the compact symmetric space $\mathsf{G} /
\mathsf{K}$ under the Cartan embedding into $\mathsf{G}$ defined
by $g \mathsf{K} \mapsto g \theta(g) ^{-1}$, where $\theta :
\mathsf{G} \to \mathsf{G}$ is the induced group involution.

The following mathematical result is a conseqence of the detailed
classification work in Sects.\ \ref{sect: reduction},\ \ref{sect:
classify} and \ref{sect: classify two}.
\begin{thm}\label{rough statement}
The symmetric spaces which occur under these assumptions are
irreducible classical symmetric spaces ${\mathfrak g}/{\mathfrak
k}$ of compact type. Conversely, every irreducible classical
symmetric space of compact type occurs in this way.
\end{thm}
We emphasize that here the notion \emph{symmetric space} is
applied flexibly in the sense that depending on the circumstances
it could mean either the infinitesimal model $\mathfrak{g} /
\mathfrak{k}$ or the Cartan-embedded compact symmetric space
$\mathsf{G} / \mathsf{K}$.

Theorem \ref{rough statement} settles the question of symmetry
classes in disordered fermion systems; in fact every physics
example is handled by one of the situations above.

The paper is organized as follows.  In Sect.\ \ref{sect:
fermions}, starting from physical considerations we motivate and
develop the model that serves as the basis for subsequent
mathematical work. Sect.\ \ref{sect: reduction} proves a number of
results which are used to eliminate the group of unitary
symmetries $G_0$.  The main work of classification is given in
Sect.\ \ref{sect: classify} and Sect.\ \ref{sect: classify two}.
In Sect.\ \ref{sect: classify} we handle the case where at most
one distinguished time-reversal operator is present, and in Sect.\
\ref{sect: classify two} the case where there are two. There are
numerous situations that must be considered, and in each case we
precisely describe the symmetric space which occurs. Various
examples taken from the physics literature are listed in Sect.\
\ref{sect: realizations}, illustrating the general classification
theory.

\section{Disordered fermions with symmetries}\label{sect: fermions}

`Fermions' is the physics name for the elementary particles which
all matter is made of.  The goal of the present article is to
establish a symmetry classification of Hamiltonians which are
\emph{quadratic} in the fermion creation and annihilation
operators. To motivate this restriction, note that any Hamiltonian
for fermions at the fundamental level is of Dirac type; thus it is
always quadratic in the fermion operators, albeit with
time-dependent coefficients that are themselves operators. At the
nonrelativistic or effective level, quadratic Hamiltonians arise
in the Hartree-Fock mean-field approximation for metals and the
Hartree-Fock-Bogoliubov approximation for superconductors.  By the
Landau-Fermi liquid principle, such mean-field or noninteracting
Hamiltonians give an adequate description of physical reality at
very low temperatures.

In the present section, starting from a physical framework, we
develop the appropriate model that will serve as the basis for the
mathematical work done later on. Please be advised that
\emph{disorder}, though advertised in the title of the section and
in the title of paper, will play no explicit role here.
Nevertheless, disorder (and/or chaos) are the indispensable agents
that \emph{must be present} in order to remove specific and
nongeneric features from the physical system and make a
classification by basic symmetries meaningful. In other words,
what we carry out in this paper is the first step of a two-step
program. This first step is to identify in the total space of
Hamiltonians some linear subspaces that are relevant (in Dyson's
sense) from a symmetry perspective. The second step is to put
probability measures on these spaces and work out the disorder
averages and statistical correlation functions of interest.  It is
this latter step that ultimately justifies the first one and thus
determines the name of the game.

\subsection{The Nambu space model for fermions}
\label{sect: Nambu}

The starting point for our considerations is the formalism of
second quantization.  Its relevant aspects will now be reviewed so
as to introduce the key physical notions as well as the proper
mathematical language.

Let $i = 1, 2, \ldots$ label an orthonormal set of quantum states
for a single fermion. Second quantizing the many-fermion system
means to associate with each $i$ a pair of operators $c_i^\dagger$
and $c_i^{\vphantom{\dagger}}\,$, which are called fermion
creation and annihilation operators, respectively, and are related
to each other by an operation of Hermitian conjugation $\dagger :
c_i \mapsto c_i^\dagger$. They are subject to the \emph{canonical
anticommutation relations}
\begin{equation}\label{pCAR}
    c_i^\dagger c_j^\dagger + c_j^\dagger c_i^\dagger = 0\;, \quad
    c_i c_j + c_j c_i = 0 \;, \quad c_i^\dagger c_j^{\vphantom{
    \dagger}} + c_j^{\vphantom{\dagger}} c_i^\dagger = \delta_{ij} \;,
\end{equation}
for all $i, j$. They act in a Fock space, i.e., in a vector space
with a distinguished vector, called the `vacuum', which is
annihilated by all of the operators $c_i$ $(i = 1, 2,\ldots)$.
Applying $n$ creation operators to the vacuum one gets a state
vector for $n$ fermions. A \emph{field operator} $\psi$ is a
linear combination of creation and annihilation operators,
\begin{displaymath}
    \psi = \sum\nolimits_i \big( v_i^{\vphantom{\dagger}} \,
    c_i^\dagger + f_i^{\vphantom{\dagger}} \, c_i^{\vphantom{
    \dagger}} \big) \;,
\end{displaymath}
with complex coefficients $v_i$ and $f_i$.

To put this in mathematical terms, let $\mathcal{V}$ be the
complex Hilbert space of single-fermion states. (We do not worry
here about complications due to the dimension of $\mathcal{V}$
being infinite. Later rigorous work will be carried out in the
finite-dimensional setting.) Fock space then is the exterior
algebra
\begin{displaymath}
    \wedge \mathcal{V} = \mathbb{C} \oplus \mathcal{V} \oplus
    \wedge^2 \mathcal{V} \oplus \ldots \;,
\end{displaymath}
with the vacuum being the one-dimensional subspace of constants.
Creating a single fermion amounts to exterior multiplication by a
vector $v \in \mathcal{V}$ and is denoted by $\varepsilon(v) :
\wedge^n \mathcal{V} \to \wedge^{n+1} \mathcal{V}$. To annihilate
a fermion, one contracts with an element $f$ of the dual space
$\mathcal{V} ^\ast$.  In other words, one applies the
antiderivation $\iota(f) : \wedge^n \mathcal{V} \to \wedge^{n-1}
\mathcal{V}$ given by $\iota(f) \cdot 1 = 0$, $\iota(f)\, v =
f(v)$, $\iota(f)\, (v_1 \wedge v_2) = f(v_1)\, v_2 - f(v_2)\,
v_1$, etc.\

In that mathematical framework the canonical anticommutation
relations read
\begin{eqnarray}
    &&\varepsilon(v) \varepsilon(\tilde v) + \varepsilon(\tilde v)
    \varepsilon(v) = 0 \;, \nonumber \\ &&\iota(f) \iota(\tilde f) +
    \iota(\tilde f) \iota(f) = 0 \;, \label{mCAR} \\ &&\iota(f)
    \varepsilon(v) + \varepsilon(v) \iota(f) = f(v) \nonumber \;.
\end{eqnarray}
They can be viewed as the defining relations of an associative
algebra, the so-called \emph{Clifford algebra} $\mathcal{C}(
\mathcal{W})$, which is generated by the vector space $\mathcal{W}
:= \mathcal{V} \oplus \mathcal{V}^\ast$ over $\mathbb{C}$.  This
vector space $\mathcal{W}$ is sometimes referred to as \emph{Nambu
space} in physics.

Since we only consider Hamiltonians that are quadratic in the
creation and annihilation operators, we will be able to reduce the
second-quantized formulation on $\wedge \mathcal{V}$ to standard
single-particle quantum mechanics, albeit on the Nambu space
$\mathcal{W}$ carrying some extra structure.  Note that
$\mathcal{W}$ is isomorphic to the space of field operators
$\psi$.

On $\mathcal{W} = \mathcal{V} \oplus \mathcal{V}^\ast$ there
exists a canonical symmetric $\mathbb{C}$-bilinear form $b$
defined by
\begin{displaymath}
    b(v + f, \tilde v + \tilde f) = f(\tilde v) + \tilde{f}(v) =
    \sum\nolimits_i (f_i \, \tilde{v}_i + \tilde{f}_i \, v_i) \;.
\end{displaymath}
The significance of this bilinear form in the present context lies
in the fact that it encodes on $\mathcal{W}$ the canonical
anticommutation relations (\ref{pCAR}), or (\ref{mCAR}). Indeed,
we can view a field operator $\psi = \sum_i (
v_i^{\vphantom{\dagger}}\, c_i^\dagger + f_i^{\vphantom{
\dagger}}\, c_i^{\vphantom{\dagger}})$ either as a vector $\psi =
v + f \in \mathcal{V} \oplus \mathcal{V}^\ast$, or equivalently as
a degree-one operator $\psi = \varepsilon(v) + \iota(f)$ in the
Clifford algebra acting on $\wedge \mathcal{V}$. Adopting the
operator perspective, we get from (\ref{mCAR}) that
\begin{displaymath}
    \psi \tilde\psi + \tilde\psi \psi = f(\tilde{v}) +
    \tilde{f}(v) = \sum\nolimits_i \left( f_i\, \tilde{v}_i +
    \tilde{f}_i\, v_i \right) \;.
\end{displaymath}
Switching to the vector perspective we have the same answer from
$b(\psi,\tilde\psi)$.  Thus
\begin{displaymath}
    \psi \tilde\psi + \tilde\psi \psi = b(\psi , \tilde\psi) \;.
\end{displaymath}
\begin{defn}
In the Nambu space model for fermions one identifies the space of
field operators $\psi$ with the complex vector space $\mathcal{W}
= \mathcal{V} \oplus \mathcal {V}^\ast$ equipped with its
canonical unitary structure $\langle \, , \, \rangle$ and
canonical symmetric complex bilinear form $b$.
\end{defn}
\begin{rem}
Having already expounded the physical origin of the symmetric
bilinear form $b$, let us now specify the canonical unitary
structure of $\mathcal{W}$. The complex vector space
$\mathcal{V}$, being isomorphic to the Hilbert space of
single-particle states, comes with a Hermitian scalar product (or
unitary structure) $\langle \, , \, \rangle_\mathcal{V}$. Given
$\langle \, , \, \rangle_\mathcal {V}$ define a
$\mathbb{C}$-antilinear bijection $C : \mathcal{V} \to
\mathcal{V}^\ast$ by
\begin{displaymath}
    C v = \langle v , \cdot \rangle_\mathcal{V} \;,
\end{displaymath}
and extend this to an antilinear transformation $C : \mathcal{W}
\to \mathcal{W}$ by the requirement $C^2 = \mathrm{Id}$.  Thus $C
\vert_{\mathcal{V}^\ast} = \left( C \vert_\mathcal{V} \right)^
{-1}$. The operator $C$ is called \emph{particle-hole conjugation}
in physics. Using $C$, transfer the unitary structure from
$\mathcal{V}$ to $\mathcal{V}^\ast$ in the natural way:
\begin{displaymath}
    \langle f , \tilde{f} \rangle_{\mathcal{V}^\ast} :=
    \overline{\langle C f , C \tilde{f} \rangle}_{\mathcal{V}}
    = \langle C \tilde{f} , C f \rangle_{\mathcal{V}} \;.
\end{displaymath}
The canonical unitary structure of $\mathcal{W}$ is then given by
\begin{displaymath}
    \langle v + f , \tilde{v} + \tilde{f} \rangle = \langle v ,
    \tilde{v} \rangle_\mathcal{V} + \langle f , \tilde{f}
    \rangle_{\mathcal{V}^\ast} = \sum\nolimits_i \left( \bar{v}_i
    \, \tilde{v}_i + \bar{f}_i \, \tilde{f}_i \right) \;.
\end{displaymath}
Thus $\langle \, , \, \rangle$ is the orthogonal sum of the
Hermitian scalar products on $\mathcal{V}$ and $\mathcal{V}^\ast$.
\end{rem}
\begin{prop}\label{Crelates}
The canonical unitary structure and symmetric complex bilinear
form of $\mathcal{W}$ are related by
\begin{displaymath}
    \langle \psi, \tilde\psi \rangle = b(C\,\psi , \tilde\psi) \;.
\end{displaymath}
\end{prop}
\begin{proof}
Given an orthonormal basis $c_1^\dagger, c_1^{\vphantom{\dagger}}
, c_2^\dagger , c_2^{\vphantom{\dagger}}, \ldots$ this is
immediate from
\begin{displaymath}
    C \sum\nolimits_i ( v_i^{\vphantom{\dagger}}\, c_i^\dagger
    + f_i^{\vphantom{\dagger}}\, c_i^{\vphantom{\dagger}} ) =
    \sum\nolimits_i (\bar{v}_i^{\vphantom{\dagger}}\,
    c_i^{\vphantom{\dagger}} + \bar{f}_i^{\vphantom{\dagger}}
    \, c_i^\dagger)
\end{displaymath}
and the expressions for $\langle \, , \, \rangle$ and $b$ in
components.
\end{proof}
Returning to the physics way of telling the story, consider the
most general Hamiltonian $H$ which is quadratic in the
single-fermion creation and annihilation operators.  Assuming $H$
to be Hermitian, using the canonical anticommutation relations
(\ref{pCAR}), and omitting an additive constant (which is of no
consequence in physics) this has the form
\begin{displaymath}
    H = {\textstyle{\frac{1}{2}}} \sum\nolimits_{ij} A_{ij}
    \big( c_i^\dagger c_j^{\vphantom{\dagger}} - c_j^{\vphantom{
    \dagger}} c_i^\dagger \big) + {\textstyle{\frac{1}{2}}}
    \sum\nolimits_{ij} \big( B_{ij}\, c_i^\dagger c_j^\dagger +
    \bar{B}_{ij}\, c_j c_i \big) \;,
\end{displaymath}
where $A_{ij} = \bar{A}_{ji}$ (from $H = H^\dagger$) and $B_{ij} =
- B_{ji}$ (from $c_i c_j = - c_j c_i$). The Hamiltonians $H$ act
on the field operators $\psi$ by the commutator, $\psi \mapsto [H
, \psi] \equiv H \psi - \psi H$, and the time evolution is
determined by the Heisenberg equation of motion,
\begin{displaymath}
  - \mathrm{i} \hbar \frac{d \psi}{dt} = [H,\psi] \;,
\end{displaymath}
with $\hbar$ being Planck's constant. By the canonical
anticommutation relations, this dynamical equation is equivalent
to a system of linear differential equations for the coefficients
$v_i$ and $f_i\,$:
\begin{eqnarray*}
    - \mathrm{i} \hbar \dot v_i &=& \sum\nolimits_j \left( A_{ij}
    \, v_j + B_{ij}\, f_j \right) \;, \\ \mathrm{i} \hbar \dot
    f_i &=& \sum\nolimits_j \left( \bar B_{ij}\, v_j + {\bar A}_{ij}
    \, f_j \right) \;.
\end{eqnarray*}
If these are assembled into a column vector ${\bf v}$, the
evolution equation takes the form
\begin{displaymath}
    \dot {\bf v} = X {\bf v}\;, \quad X = \frac{\mathrm{i}}{\hbar}
    \begin{pmatrix} A &B \\ -\bar B &-\bar A \end{pmatrix} \;.
\end{displaymath}

To recast all this in concise terms, we need some further
mathematical background.  Notwithstanding the fact that in
practice we always consider the Fock space representation
$\mathcal{C}( \mathcal{W}) \to \mathrm{End}(\wedge \mathcal{V})$
by $w = v + f \mapsto \varepsilon(v) + \iota(f)$, it should be
stated that the primary (or universal) definition of the Clifford
algebra $\mathcal{C}( \mathcal{W})$ is as the associative algebra
generated by $\mathcal{W} \oplus \mathbb {C}$ with relations
\begin{equation}\label{uCAR}
    w_1 w_2 + w_2 w_1 = b(w_1 , w_2) \times \mathrm{Id} \qquad
    (w_1, w_2 \in \mathcal{W}) \;.
\end{equation}
The Clifford algebra is graded by
\begin{displaymath}
    \mathcal{C}(\mathcal{W}) = \mathcal{C}^0 (\mathcal{W})
    \oplus \mathcal{C}^1 (\mathcal{W}) \oplus \mathcal{C}^2
    (\mathcal{W}) \oplus \ldots \;,
\end{displaymath}
where $\mathcal{C}^0 (\mathcal{W}) \equiv \mathbb{C}$,
$\mathcal{C}^1(\mathcal{W}) \cong \mathcal{W}$, and $\mathcal{C}^n
(\mathcal{W})$ for $n \ge 2$ is the linear space of
skew-symmetrized degree-$n$ monomials in the elements of
$\mathcal{W}$.  In particular, $\mathcal{C}^2 ( \mathcal{W})$ is
the linear space of skew-symmetric quadratic monomials $w_1 w_2 -
w_2 w_1$ ($w_1, w_2 \in \mathcal{W}$).

From the Clifford algebra perspective, a quadratic Hamiltonian $H$
is viewed as an operator in the degree-two component $\mathcal
{C}^2 (\mathcal {W})$.  Let us therefore gather some standard
facts about $\mathcal{C}^2 (\mathcal{W})$.  First among these is
that $\mathcal{C}^2 (\mathcal{W})$ is a complex Lie algebra with
the commutator playing the role of the Lie bracket (an exposition
of this fact for the case of a Clifford algebra over $\mathbb{R}$
is found in \cite{BGV}; the complex case is no different).

Second, in addition to acting on itself by the commutator, the Lie
algebra $\mathcal{C}^2(\mathcal{W})$ acts (still by the
commutator) on all of the components $\mathcal{C}^k(\mathcal{W})$
of degree $k \ge 1$ of the Clifford algebra $\mathcal{C}(\mathcal
{W})$. In particular, $\mathcal {C}^2 (\mathcal{W})$ acts on
$\mathcal{C}^1 (\mathcal{W})$.

Third, $\mathcal{C}^2(\mathcal{W})$ turns out to be canonically
isomorphic to the complex orthogonal Lie algebra $\mathfrak{so}
(\mathcal {W},b)$ which is associated with the vector space
$\mathcal{W} = \mathcal {V} \oplus \mathcal{V}^\ast$ and its
canonical symmetric complex bilinear form $b$; this Lie algebra
$\mathfrak{so}( \mathcal{W},b)$ is defined to be the subspace of
elements $E \in \mathrm{End} (\mathcal{W})$ satisfying the
condition
\begin{displaymath}
    b(E w_1 , w_2) + b(w_1 , E w_2) = 0 \quad (\text{for all }\,
    w_1, w_2 \in \mathcal{W}) \;.
\end{displaymath}
The canonical isomorphism $\mathcal{C}^2(\mathcal{W}) \to
\mathfrak {so}(\mathcal{W},b)$ is given by the commutator action
of $\mathcal{C}^2(\mathcal{W})$ on $\mathcal{C}^1(\mathcal{W})
\cong \mathcal{W}$, i.e., by sending $a \in \mathcal{C}^2(\mathcal
{W})$ to $[a, \cdot] = E \in \mathrm{End}(\mathcal{W})$; the
latter indeed lies in $\mathfrak{so}(\mathcal{W},b)$ as follows
from the expression for $b(E w_1 , w_2) + b(w_1 , E w_2)$ given by
the canonical anticommutation relations (\ref{uCAR}), from the
Jacobi identity
\begin{displaymath}
    [ a , w_1 ] \, w_2 + w_2 \, [ a , w_1 ] + w_1\, [ a , w_2 ]
    + [ a , w_2 ]\, w_1 = [ a , w_1 w_2 + w_2 w_1 ] \;,
\end{displaymath}
and from the fact that $w_1 w_2 + w_2 w_1$ lies in the center of
the Clifford algebra.

To describe $\mathfrak{so}(\mathcal{W},b)$ explicitly, decompose
the endomorphisms $E \in \mathrm{End}(\mathcal{V} \oplus
\mathcal{V} ^\ast)$ into blocks as
\begin{displaymath}
    E = \begin{pmatrix} \mathsf{A} &\mathsf{B} \\
    \mathsf{C} &\mathsf{D} \end{pmatrix} \;,
\end{displaymath}
where $\mathsf{A} \in \mathrm{End}(\mathcal{V})$, $\mathsf{B} \in
\mathrm{Hom}(\mathcal{V}^\ast,\mathcal{V})$, $\mathsf{C} \in
\mathrm{Hom}(\mathcal{V},\mathcal{V}^\ast)$ and $\mathsf{D} \in
\mathrm{End}(\mathcal{V}^\ast)$. Let the adjoint (or transpose) of
$\mathsf{A} \in \mathrm{End}(\mathcal{V})$ be denoted by $\mathsf
{A}^\mathrm{t} \in \mathrm{End}(\mathcal{V}^\ast)$, and call an
element $\mathsf{C}$ in $\mathrm{Hom}(\mathcal{V},\mathcal{V}^\ast
)$ skew if $\mathsf{C}^\mathrm{t} = - \mathsf{C}$, i.e., if
$(\mathsf{C}v_1)(v_2) = - (\mathsf{C}\, v_2)(v_1)$.
\begin{prop}\label{when in so}
An endomorphism $E = \begin{pmatrix} \mathsf{A} &\mathsf{B}\\
\mathsf{C} &\mathsf{D} \end{pmatrix} \in \mathrm{End}(\mathcal{V}
\oplus \mathcal{V}^\ast)$ lies in the complex orthogonal Lie
algebra $\mathfrak{so}(\mathcal{V}\oplus \mathcal{V}^\ast ,b)$ if
and only if $\mathsf{B}, \mathsf{C}$ are skew and $\mathsf{D} = -
\mathsf{A}^\mathrm{t}$.
\end{prop}
\begin{proof}
Consider first the case $\mathsf{B} = \mathsf{C} = 0$, and let
$\mathsf{D} = - \mathsf{A}^\mathrm{t}$.  Then
\begin{eqnarray*}
    &&b \big( E (v+f) , \tilde{v} + \tilde{f} \big) =
    b(\mathsf{A}v - \mathsf{A}^\mathrm{t} f , \tilde{v} + \tilde{f})
    = \tilde{f} (\mathsf{A}v) - \mathsf{A}^\mathrm{t}f (\tilde v) \\
    &=& \mathsf{A}^\mathrm{t} \tilde{f} (v) -f(\mathsf{A} \tilde{v}) =
    - b(v + f , \mathsf{A} \tilde{v} - \mathsf{A}^\mathrm{t}\tilde{f})
    = - b \big( v+f , E (\tilde{v} + \tilde{f}) \big) \;.
\end{eqnarray*}
Using $\mathsf{B}^\mathrm{t} = - \mathsf{B}$ and $\mathsf{C}^
\mathrm{t} = - \mathsf{C}$, a similar calculation for the case
$\mathsf{A} = 0$ gives
\begin{eqnarray*}
    &&b\big( E(v+f) , \tilde{v} + \tilde{f} \big) =
    b(\mathsf{B}f + \mathsf{C}v , \tilde{v} + \tilde{f}) =
    \mathsf{C}v( \tilde{v} ) + \tilde{f} (\mathsf{B}f) \\
    &=& - f(\mathsf{B} \tilde{f}) - \mathsf{C} \tilde{v} (v) =
    - b(v + f , \mathsf{B} \tilde{f} + \mathsf{C} \tilde{v}) =
    - b \big( v + f , E (\tilde{v} + \tilde{f}) \big) \;.
\end{eqnarray*}
Since these two cases complement each other, we see that the
stated conditions on $E \in \mathrm{End}(\mathcal{W})$ are
sufficient in order for $E$ to be in $\mathfrak{so}(\mathcal{W}
,b)$. The calculation can equally well be read backwards; thus the
conditions are both sufficient and necessary.
\end{proof}
Let us now make the connection to physics, where $\mathcal{C}(
\mathcal{W})$ is represented on Fock space and the elements $v + f
= w \in \mathcal{W}$ become field operators $\psi = \varepsilon(v)
+ \iota(f)$.  Fixing orthonormal bases $c_1^\dagger, c_2^\dagger,
\ldots$ of $\mathcal{V}$ and $c_1, c_2, \ldots$ of $\mathcal{V}
^\ast$ as before, we assign matrices with matrix elements
$A_{ij}$, $B_{ij}$, $C_{ij}$ to the linear operators $\mathsf{A}$,
$\mathsf{B}$, $\mathsf{C}$. A straightforward computation using
the canonical anticommutation relations then yields:
\begin{prop}
The inverse of the Lie algebra automorphism $\mathcal{C}^2
(\mathcal{W}) \to \mathfrak{so}(\mathcal{W},b)$ is the
$\mathbb{C}$-linear mapping given by
\begin{displaymath}
    \begin{pmatrix} \mathsf{A} &\mathsf{B} \\ \mathsf{C}
     &-\mathsf{A}^\mathrm{t} \end{pmatrix} \mapsto
     {\textstyle{\frac{1}{2}}} \sum\nolimits_{ij} A_{ij}
    ( c_i^\dagger c_j^{\vphantom{\dagger}} - c_j^{\vphantom{
    \dagger}} c_i^\dagger ) + {\textstyle{\frac{1}{2}}}
    \sum\nolimits_{ij} ( B_{ij}\, c_i^\dagger c_j^\dagger +
    C_{ij}\, c_i c_j ) \;.
\end{displaymath}
\end{prop}
Now recall that $\mathcal {C}^2 (\mathcal{W})$ acts on the
degree-one component $\mathcal{C}^1 (\mathcal{W})$ by the
commutator. By the isomorphisms $\mathcal{C}^2 (\mathcal{W}) \cong
\mathfrak{so}(\mathcal{W},b)$ and $\mathcal {C}^1 (\mathcal{W})
\cong \mathcal{W}$, this action coincides with the fundamental
representation of $\mathfrak {so}(\mathcal{W},b)$ on its defining
vector space $\mathcal{W}$. In other words, taking the commutator
of the Hamiltonian $H \in \mathcal{C}^2( \mathcal {W})$ with a
field operator $\psi\in \mathcal{C}^1 (\mathcal{W})$ yields the
same answer as viewing $H$ as an element of $\mathfrak {so}
(\mathcal{W},b)$, then applying $H = \begin{pmatrix} \mathsf{A}
&\mathsf{B} \\ \mathsf{C} &- \mathsf{A} ^\mathrm{t}
\end{pmatrix}$ to the vector $\psi = v + f \in \mathcal{W}$ by
\begin{displaymath}
    H \cdot (v + f) = (\mathsf{A}v + \mathsf{B}f) +
    (\mathsf{C}v - \mathsf{A}^\mathrm{t}f) \;,
\end{displaymath}
and finally reinterpreting the result as a field operator in
$\mathcal{ C}^1 (\mathcal{W})$.

The closure relation $[ \mathcal{C}^2(\mathcal{W}) , \mathcal
{C}^1(\mathcal{W}) ] \subset \mathcal{C}^1(\mathcal{W})$ and the
isomorphism $\mathcal{C}^1 (\mathcal{W}) \cong \mathcal{W}$ make
it possible to reduce the dynamics of field operators to a
dynamics on the Nambu space $\mathcal{W}$. After reduction, as we
have seen, the generators $X \in \mathrm{End}( \mathcal{V} \oplus
\mathcal{V}^\ast)$ of time evolutions of the physical system are
of the special form
\begin{displaymath}
    X = \frac{\mathrm{i}}{\hbar} \begin{pmatrix}
    \mathsf{A} &\mathsf{B} \\ \mathsf{B}^\ast &-
    \mathsf{A}^\mathrm{t} \end{pmatrix} \;,
\end{displaymath}
where $\mathsf{B} \in \mathrm{Hom}(\mathcal{V}^\ast,\mathcal{V})$
is skew, and $\mathsf{A} = \mathsf{A}^\ast \in \mathrm{End}(
\mathcal{V})$ is self-adjoint w.r.t.\ $\langle\, ,\, \rangle_
\mathcal{V}$.
\begin{prop}\label{inv of struct}
The one-parameter groups of time evolutions $t \mapsto
\mathrm{e}^{tX}$ in the Nambu space model preserve both the
canonical unitary structure $\langle \, , \, \rangle$ and the
canonical symmetric complex bilinear form $b$ of $\mathcal{W} =
\mathcal{V} \oplus \mathcal{V}^\ast$.
\end{prop}
\begin{proof}
By Prop.\ \ref{when in so} the generator $X$ is an element of the
complex Lie algebra $\mathfrak{so}(\mathcal{W},b)$.  Hence the
exponential $U_t = \mathrm{e}^{tX}$ lies in the complex orthogonal
Lie group $\mathrm{SO}(\mathcal{W},b)$, which is defined to be the
set of solutions $g$ in $\mathrm{End}(\mathcal{W})$ of the
conditions
\begin{displaymath}
    b( g \psi , g \tilde\psi) = b( \psi , \tilde\psi) \;,
    \quad \text{and} \quad \mathrm{Det}(g) = 1 \;.
\end{displaymath}

Since $\mathsf{A} = \mathsf{A}^\ast$, and $\mathsf{B}^\ast \in
\mathrm{Hom}(\mathcal{V},\mathcal{V}^\ast)$ is the adjoint of
$\mathsf{B} \in \mathrm{Hom}(\mathcal{V}^\ast,\mathcal{V})$, the
generator $X$ is antihermitian with respect to the unitary
structure of $\mathcal{W}$. The exponentiated generator $U_t$
therefore lies in the unitary group $\mathrm{U}(\mathcal{W})$,
which is to say that
\begin{displaymath}
    \langle U_t \psi , U_t \tilde\psi \rangle = \langle
    \psi , \tilde\psi \rangle
\end{displaymath}
for all real $t$. Thus $U_t$ preserves both $b$ and $\langle \, ,
\, \rangle$.
\end{proof}
\begin{rem}
In physical language, the invariance of $b$ under time evolutions
means that the canonical anticommutation relations for fermionic
field operators do not change with time.  Invariance of $\langle
\, , \, \rangle$ means that probability in Nambu space is
conserved. (If the quadratic Hamiltonian $H$ arises as the
mean-field approximation to some many-fermion problem, the latter
conservation law holds as long as quasiparticles do not interact
and thereby are protected from decay into multi-particle states.)
\end{rem}
We now distill the essence of the information conveyed in this
section. The quantum theory of many-fermion systems is set up in a
Hilbert space called the fermionic Fock space in physics (or the
spinor representation in mathematics). The field operators of the
physical system span a vector space $\mathcal{W} = \mathcal{V}
\oplus \mathcal{V}^\ast$, which generates a Clifford algebra
$\mathcal{C}(\mathcal{W})$ whose defining relations are the
canonical anticommutation relations.

Since $[\mathcal{C}^2(\mathcal{W}) , \mathcal{C}^1(\mathcal{W})]
\subset \mathcal{C} ^1(\mathcal{W})$, the discussion of the field
operator dynamics for the important case of quadratic Hamiltonians
$H \in \mathcal{C}^2(\mathcal{W})$ can be reduced to a discussion
on the Nambu space $\mathcal{W} \cong \mathcal{C}^1 (\mathcal{W})
$. Via this reduction, the vector space $\mathcal{W}$ inherits two
natural structures: the canonical symmetric complex bilinear form
$b$ encoding the anticommutation relations, and a canonical
unitary structure $\langle \, , \, \rangle$ determined by the
Hermitian scalar product of $\mathcal{V}$.  Both of these
structures are invariant, i.e., are pre\-served by physical time
evolutions. Under the reduction to $\mathcal{W}$, the commutator
action of $\mathcal{C }^2(\mathcal{W})$ on $\mathcal{C}^1
(\mathcal{W})$ becomes the fundamental representation of
$\mathfrak{so} (\mathcal{W},b)$ on $\mathcal{W}$.

\subsection{Symmetry groups}\label{sect: symmetries}

Following Dyson, the classification of disordered fermion systems
will be carried out in a setting that prescribes two pieces of
data:
\begin{itemize}
\item[$\bullet$] One is given a Nambu space $\mathcal{W} =
\mathcal{V} \oplus \mathcal{V}^\ast$ equipped with its canonical
unitary structure $\langle \, , \, \rangle$ and canonical
symmetric $\mathbb{C }$-bilinear form $b$.
\item[$\bullet$] On $\mathcal{W}$ there acts a group $G$ of
unitary and antiunitary operators (the joint symmetry group of a
multi-parameter family of fermionic quantum systems).
\end{itemize}
Given this setup, one is interested in the linear space of
Hamiltonians $H$ with the property that they commute with the
$G$-action on $\mathcal{W}$, while preserving the invariant structures
$b$ and $\langle \, , \, \rangle$ of $\mathcal{W}$ under time
evolution by $\mathrm{e}^{-\mathrm{i} t H / \hbar}$.  Such a space of
Hamiltonians is of course reducible in general, i.e., the Hamiltonian
matrices decompose into blocks. The goal of classification is to
enumerate all the {\it symmetry classes}, i.e., all the types of
irreducible block which occur in this way.

In the present subsection we provide some information on what is
meant by unitary and antiunitary symmetries in the present
context. We begin by recalling the basic notion of a symmetry
group in quantum Hamiltonian systems.

In classical mechanics the symmetry group $G_0$ of a Hamiltonian
system is understood to be the group of symplectomorphisms that
commute with the phase flow of the system. Examples are the
rotation group for systems in a central field, and the group of
Euclidean motions for systems with Euclidean invariance.

In passing from classical to quantum mechanics, one replaces the
classical phase space by a complex Hilbert space $\mathcal{V}$,
and assigns to the symmetry group $G_0$ a (projective)
representation by unitary $\mathbb{C}$-linear operators on
$\mathcal{V}$. While the consequences due to one-parameter
continuous subgroups of $G_0$ are particularly clear from
Noether's theorem \cite{arnold}, the components of $G_0$
\emph{not} connected with the identity also play an important
role. A prominent example is provided by the operator for space
reflection. Its eigenspaces are the subspaces of states with
positive and negative parity, and they reduce the matrix of any
reflection-invariant Hamiltonian to two blocks.

Not all symmetries of a quantum mechanical system are of the
canonical, unitary kind: the prime counterexample is the operation
$g_T$ of inverting the time direction -- called time reversal for
short. In classical mechanics this operation reverses the sign of
the symplectic structure of phase space; in quantum mechanics its
algebraic properties reflect the fact that the time $t$ enters in
the Dirac, Pauli, or Schr\"odinger equation as ${\rm i}\hbar
d/dt$: there, time reversal $g_T$ is represented by an
\emph{antiunitary} operator $T$, which is to say that $T$ is
complex antilinear:
\begin{displaymath}
    T(z v) = {\bar z}\, T v \quad (z\in \mathbb{C},\;
    v \in \mathcal{V})\;,
\end{displaymath}
and preserves the Hermitian scalar product up to complex
conjugation:
\begin{displaymath}
    \langle v , \tilde{v} \rangle_\mathcal{V} =
    \overline{\langle T v , T \tilde{v} \rangle}_\mathcal{V} \;.
\end{displaymath}
Another example of such an operation is charge conjugation in
relativistic theories.  Further examples are provided by chiral
symmetry transformations (see Sect.\ \ref{sect: Dirac}).

By the symmetry group $G$ of a quantum mechanical system with
Hamiltonian $H$, one then means the group of all unitary and
antiunitary transformations $g$ of $\mathcal{V}$ that leave the
Hamiltonian invariant: $g H g^{-1} = H$. It should be noted that
finding the total symmetry group of a quantization of some
Hamiltonian system is not always straightforward. The reason is
that there may exist nonobvious quantum symmetries such as Hecke
symmetries, which are of number-theoretic origin and have no
classical limit. For our purposes, however, this complication will
not be an issue. We take the group $G$ and its action on the
Hilbert space to be \emph{fundamental and given}, and then ask
what is the linear space of Hamiltonians that commute with the
$G$-action.

For technical reasons, we assume the group $G_0$ to be compact;
this is an assumption that covers most (if not all) of the cases
of interest in physics. The noncompact group of space translations
can be incorporated, if necessary, by wrapping the system around a
torus, whereby translations are turned into compact torus
rotations.

What we have sketched -- a symmetry group $G$ acting on a Hilbert
space $\mathcal{V}$ -- is the framework underlying Dyson's
classification.  As was explained in Sect.\ \ref{sect: Nambu}, we wish
to enlarge it so as to capture all examples that arise in disordered
fermion physics.

For this, recall that in the Nambu space model for fermions, the
Hilbert space is not $\mathcal{V}$ but the space of field operators
$\mathcal{W} = \mathcal{V} \oplus \mathcal{V}^\ast$.  The given
$G$-representation on $\mathcal{V}$ therefore needs to be extended to
a representation on $\mathcal{W}$. This is done by the condition that
the pairing between $\mathcal{V}$ and $\mathcal{V} ^\ast$ (and thus
the pairing between fermion creation and annihilation operators) be
preserved. In other words, if $U : \mathcal{V} \to \mathcal{V}$ and $A
: \mathcal{V} \to \mathcal{V}$ are unitary resp.\ antiunitary
operators, their induced representations on $\mathcal{V}^\ast$ (which
we still denote by the same symbols) are defined by requiring that
\begin{displaymath}
    (U f)(U v) = f(v) = \overline{(A f)(A v)}
\end{displaymath}
for all $v \in \mathcal{V}$ and $f \in \mathcal{V}^\ast$. In
particular the $G_0$-representation on $\mathcal{V}^\ast$ is the
\emph{dual} one,
\begin{displaymath}
    U (f) = f \circ U^{-1} \;.
\end{displaymath}
Equivalently, the $G$-representation on $\mathcal{W}$ is defined
so as to be compatible with particle-hole conjugation $C :
\mathcal{W} \to \mathcal{W}$ in the sense that operations commute:
\begin{displaymath}
    C U = U C \;, \quad \text{and} \quad C A = A C \;.
\end{displaymath}
Indeed, if $f = C v$ then $f(\tilde{v}) = \langle v , \tilde{v}
\rangle$ and from the invariance of the pairing between $\mathcal{V}$
and $\mathcal{V}^\ast$ one infers the relations $\langle v , \tilde{v}
\rangle =(U f)(U \tilde{v}) = \langle U^{-1} C^{-1} U C v , \tilde{v}
\rangle$ and $\langle v , \tilde{v} \rangle = \overline{(A f)(A
\tilde{v})} = \langle A^{-1} C^{-1} A C v , \tilde{v} \rangle$.

While the framework so obtained is flexible enough to capture the
situations that arise in the nonrelativistic quasiparticle physics of
disordered metals, semiconductors and superconductors, it is still
slightly too narrow to accommodate some much studied examples that
have emerged from elementary particle physics. Let us explain this.

\subsection{The Euclidean Dirac operator}\label{sect: Dirac}

An important development in random-matrix physics over the last ten
years was the formulation \cite{vz} and study of the so-called chiral
ensembles, which model Dirac fermions in a random gauge field
background, and lie beyond Dyson's 3-way classification.  From the
viewpoint of applications, these random-matrix models have the merit
of capturing some universal features of the Dirac spectrum of quantum
chromodynamics (QCD) in the low-energy limit. In the present
subsection we will demonstrate that, but for one minor difference,
they fit naturally into our fermionic Nambu space model with
symmetries.

Let $M$ be a four-dimensional Euclidean space-time (more generally,
$M$ could be a Riemannian 4-manifold with spin structure), and
consider over $M$ a unitary spinor bundle $S$ twisted by a module $R$
for the action of some compact gauge group $K$. Denote by
$\mathcal{V}$ the Hilbert space of $L^2$-sections of the twisted
bundle $S \otimes R$.

Now let $D_A$ be a self-adjoint Dirac operator for $\mathcal{V}$
in a given gauge field background (or gauge connection) $A$.
Although $D_A$ is not a Hamiltonian in the strict sense of the
word, it has all the right mathematical attributes in the sense of
Sect.\ \ref{sect: Nambu}; in particular it determines a Hermitian
form, called the action functional, on differentiable sections
$\psi \in \mathcal{V}$. In physics notation this functional is written
\begin{displaymath}
    \psi \mapsto \int_M \bar\psi(x) \cdot \left( D_A \psi \right)(x)
    \, d^4x \;, \quad D_A = \mathrm{i} \gamma^\mu (\partial_\mu
    - A_\mu) \;,
\end{displaymath}
where $\gamma^\mu = \gamma(e^\mu)$ are the gamma matrices [i.e.,
the Clifford action $\gamma : T^\ast M \to \mathrm{End}(S)$
evaluated on the dual $e^\mu$ of an orthonormal coordinate frame
$e_\mu$ of $TM$], the operators $\partial_\mu$ are the partial
derivatives corresponding to the $e_\mu$, and $A_\mu(x) \in
\mathrm{Lie}(K)$ are the components of the gauge field.  If the
physical situation calls for a mass, then one adds a complex
number $\mathrm{i} m$ (times the unit operator on $\mathcal{V}$)
to the expression for $D_A$.

The Dirac operators of prime interest to low-energy QCD have zero
(or small) mass. To express the massless nature of $D_A$ one
introduces an object called the {\it chirality} operator $\Gamma$
in mathematics \cite{BGV}, or $\gamma_5 = \gamma^0 \gamma^1
\gamma^2 \gamma^3$ in physics. $\Gamma = \gamma_5$ is a section of
$\mathrm{End}(S)$ which is self-adjoint and involutory ($\Gamma^2
= \mathrm{Id}$) and anticommutes with the Clifford action ($\Gamma
\gamma^\mu + \gamma^\mu \Gamma = 0$). By the last property one has
\begin{displaymath}
    \Gamma D_A + D_A \Gamma = 0
\end{displaymath}
in the massless limit. This relation is called chiral symmetry in
physics. Note, however, that chiral `symmetry' \emph{is not a
symmetry} in the sense of the present paper. (Symmetries always
\emph{commute} with the Hamiltonian, never do they anticommute with
it!) Nonetheless, we shall now recognize chiral symmetry as being
equivalent to a true symmetry, by importing the Dirac operator into
the Nambu space model as follows.

As before, take Nambu space to be the sum $\mathcal{W} = \mathcal
{V} \oplus \mathcal{V}^\ast$ equipped with its canonical unitary
structure $\langle \, , \, \rangle$ and symmetric complex bilinear
form $b$. The antilinear bijection $C : \mathcal{V} \to \mathcal
{V}^\ast$ and $C : \mathcal{V}^\ast \to \mathcal{V}$ is still
defined by $\langle w_1 , w_2 \rangle = b(C w_1 , w_2)$.

Now extend the Dirac operator $D_A \in \mathrm{i} \mathfrak{u}(
\mathcal{V})$ to an operator $\mathcal{D}_A$ that acts diagonally on
$\mathcal{W} = \mathcal{V} \oplus \mathcal{V}^\ast$, by requiring
$\mathcal{D}_A$ to satisfy the commutation law $C\, \mathrm{i}
\mathcal{D }_A = \mathrm{i} \mathcal{D}_A C$, or equivalently $C
\mathcal{D }_A = - \mathcal{D}_A C$. Thus,
\begin{displaymath}
    \mathcal{D}_A \in \mathrm{End}(\mathcal{V}) \oplus \mathrm{End}
    (\mathcal{V}^\ast) \hookrightarrow \mathrm{End}(\mathcal{W}) \;,
\end{displaymath}
and $\mathcal{D}_A$ on $\mathrm{End}(\mathcal{V}^\ast)$ is given
by $- D_A^ \mathrm{t}$.  The diagonally extended operator
$\mathcal{D}_A$ lies in the intersection of $ \mathfrak{so}(
\mathcal{W},b)$ with $\mathrm{i} \mathfrak{u}(\mathcal{W})$ -- as
is required in order for the statement of Prop.\ \ref{inv of
struct} to carry over to the one-parameter group $t \mapsto
\mathrm{e}^{ \mathrm{i}t \mathcal {D}_A}$. The property that
$\mathcal{D}_A$ does not mix $\mathcal{V}$ and $\mathcal{V} ^\ast$
can be attributed to the existence of a $\mathrm{U}_1$ symmetry
group that has $\mathcal{V}$ and $\mathcal{V}^\ast$ as
inequivalent representation spaces.

To implement the chiral symmetry of the massless limit, extend the
chirality operator $\Gamma$ to a diagonally acting endomorphism in
$\mathrm{End}(\mathcal{V}) \oplus \mathrm{End}(\mathcal{V}^\ast)$
by $C \Gamma C^{-1} = \Gamma$.  The extended operators still
satisfy the chiral symmetry relation $\Gamma \mathcal{D}_A +
\mathcal{D}_A \Gamma = 0$. Then define an antiunitary operator $T$
by $T := C \Gamma$. Note that this is \emph{not} the operation of
reversing the time but will still be called the `time reversal'
for short.

Because $\mathcal{D}_A$ anticommutes with both $C$ and $\Gamma$,
one has
\begin{displaymath}
    T \mathcal{D}_A T^{-1} = \mathcal{D}_A \;.
\end{displaymath}
Thus $T$ is a true symmetry of the (extended) Dirac operator in
the massless limit.

Note that $CT = TC$ from $C \Gamma = \Gamma C$.  As was announced
above, the situation is the same as before but for one difference:
while the time reversal in Sect.\ \ref{sect: symmetries} was an
operator $T : \mathcal{V} \to \mathcal{V}$ and $T : \mathcal{V}
^\ast \to \mathcal{V}^\ast$, the present one is an operator $T :
\mathcal{V} \to \mathcal{V}^\ast$ and $T : \mathcal{V}^\ast \to
\mathcal{V}$. We refer to the latter type as \emph{mixing}, and
the former as \emph{nonmixing}.

To summarize, physical systems modelled by the Euclidean (or
positive signature) Dirac operator are naturally incorporated into
the framework of Sects.\ \ref{sect: Nambu} and \ref{sect:
symmetries}. The Hilbert space $\mathcal{V}$ here is the space of
$L^2$-sections of a twisted spinor bundle over Euclidean
space-time, and the role of the Hamiltonian is taken by the
quadratic action functional of the Dirac fermion theory.  When
transcribed into the Nambu space $\mathcal{W} = \mathcal{V} \oplus
\mathcal{V}^\ast$, the chiral `symmetry' of the massless theory
can be expressed as a true antiunitary symmetry $T$, with the only
new feature being that $T$ mixes $\mathcal{V}$ and $\mathcal{V}
^\ast$.

The most general situation occurring in physics may exhibit, beside
$T$, one or several other antiunitary symmetries.  In the example at
hand this happens if the representation space $R$ carries a complex
bilinear form which is invariant under gauge transformations (see
Sects.\ \ref{sect: BDI} and \ref{sect: CII} for the details). The
Dirac operator $\mathcal{D}_A$ then has one extra antiunitary
symmetry, say $T_1$, which is nonmixing. Forming the composition of
$T_1$ with $T$ we get a {\it mixing unitary symmetry} $P = T T_1 :
\mathcal{V} \leftrightarrow \mathcal{V} ^\ast$. This fact leads us to
adopt the final framework described in the next subsection.

\subsection{The mathematical model}

The following model is now well motivated.

We are given a Nambu space $(\mathcal{W}, b, \langle \, , \,
\rangle)$ carrying the action of a compact group $G$. The group
$G_0$ is defined to be the subgroup of $G$ which acts by canonical
unitary transformations, i.e., unitary transformations that
preserve the decomposition ${\mathcal W} = \mathcal{V} \oplus
\mathcal{V}^*$.  The full symmetry group $G$ is generated by $G_0$
and at most two distinguished antiunitary time-reversal operators.
If there is just one, we denote it by $T$, and if there are two,
by $T$ and $T_1$.  In the latter case we adopt the convention that
$T$ mixes, i.e., $T : \mathcal{V} \to \mathcal{V}^\ast$, while
$T_1$ is nonmixing. The distinguished time-reversal symmetries
always satisfy $T^2 = \pm \mathrm{Id}$ and $T_1^2 = \pm
\mathrm{Id}$.  In the case that there are two, it is assumed that
they commute or anticommute, i.e., $T_1T=\pm TT_1$.  Consequently
the unitary operator $P = T T_1$ (which mixes) also satisfies $P^2
= \pm \mathrm{Id}$. When $P$ is present we let $G_1$ denote the
$\mathbb {Z}_2$-extension of $G_0$ defined by $P$ and refer to it
as the full group of unitary symmetries.

We emphasize that the original action of $G_0$ on $\mathcal{V}$ has
been extended to $\mathcal{W}$ via its canonically induced action on
${\mathcal V}^*$. In other words, if $f\in \mathcal{V}^\ast$ then
$g(f) (v) = f(g^{-1}(v))$.  This is equivalent to requiring that a
unitary operator $U \in G_0$ commutes with particle-hole conjugation
$C : \mathcal{W} \to \mathcal{W}$. In fact we require that all
operators of $G$ commute with $C$.  Whereas the unitary operators
preserve the Hermitian scalar product $\langle \, , \, \rangle$, for
an antiunitary operator $A$ we have that $\langle A w_1 , A w_2\rangle
= \overline{\langle w_1 , w_2 \rangle}$ for all $w_1 , w_2 \in
{\mathcal W}$.

If $U$ is an operator coming from $G_0$ and $T$ is a distinguished
time-reversal symmetry, then $TUT^{-1}$ is unitary and nonmixing,
i.e., it is in $G_0$.  Thus, for the corresponding operator $g_T$
in $G$, we assume that $g_T$ normalizes $G_0$ and $g_T^2$ is in
the center of $G_0$.

According to Prop.\ \ref{inv of struct} the time evolutions of the
physical system leave the structure of Nambu space invariant. The
infinitesimal version of this statement is that the Hamiltonians
$H$ lie in the intersection of the complex orthogonal Lie algebra
$\mathfrak{so} (\mathcal{W},b)$ with $\mathrm{i} \mathfrak{u}(
\mathcal{W})$, the Hermitian operators on $\mathcal{W}$.

Let us summarize our situation in the language and notation
introduced above.

\begin{defn}\label{Nambu with syms}
The data in the Nambu space model for fermions with symmetries is
$(\mathcal{W}, b, \langle \, , \, \rangle; G)$, where the compact
group $G$ is called the symmetry group of the system. $G$ is
represented on $\mathcal{W} = \mathcal{V} \oplus \mathcal{V}^*$ by
unitary and antiunitary operators that preserve the structure of
$\mathcal{W}$; i.e., for every unitary $U$ and antiunitary $A$ one
has
\begin{displaymath}
    \langle \psi , \tilde\psi \rangle = \langle U \psi , U
    \tilde\psi \rangle = \overline{\langle A \psi , A \tilde\psi
    \rangle} \;, \quad b(\psi , \tilde\psi) = b(U \psi , U
    \tilde\psi) = \overline{ b(A \psi , A\tilde\psi)}
\end{displaymath}
for all $\psi, \tilde\psi \in \mathcal{W}$.  The space of `good'
Hamiltonians is the $\mathbb{R}$-vector space $\mathcal{H}$ of
operators $H$ in $\mathfrak{so} (\mathcal{W},b) \cap \mathrm{i}
\mathfrak{u}(\mathcal{W})$ that commute with the $G$-action:
\begin{displaymath}
    U H U^{-1} = H = A H A^{-1} \;.
\end{displaymath}
At the group level of time evolutions this means that
\begin{displaymath}
    U \mathrm{e}^{-\mathrm{i}t H / \hbar} =
    \mathrm{e}^{-\mathrm{i}t H / \hbar} U \;, \quad
    A \mathrm{e}^{-\mathrm{i}t H / \hbar} =
    \mathrm{e}^{+\mathrm{i}t H / \hbar} A \;,
\end{displaymath}
for all unitary $U$, antiunitary $A$, $H \in \mathcal{H}$, and $t \in
\mathbb{R}$.
\end{defn}
We remind the reader that the subgroup of unitary operators which
preserves the decomposition ${\mathcal W} = {\mathcal V}\oplus
{\mathcal V}^*$ is denoted by $G_0$, and the full group of unitaries
by $G_1$.

Several further remarks are in order. First, for a unitary $U \in G_1$
(resp.\ antiunitary $A$), the compatibility of $b$ with the $G$-action
is a consequence of Prop.\ \ref{Crelates} and the commutation law $CU
= UC$ and $CA = AC$. Second, it is possible that the fermion system
does not have any antiunitary symmetries and $G=G_0$. When some
antiunitary symmetries are present, $G$ is generated by $G_0$ and one
or at most two distinguished time-reversal symmetries as explained
above. Third, motivated by the prime physics example of time reversal,
we have assumed that the (one or two) distinguished time-reversal
symmetries $T$ satisfy $T^2 = \pm \mathrm{Id}$.  The reason for this
can be explained as follows.

The operator $T$ has been chosen to represent some kind of {\it
inversion} symmetry.  Since this means that conjugation by $T^2$
represents the unit operator, $T^2$ must be a unitary multiple of the
identity on any subspace of $\mathcal{W}$ which is irreducible under
time evolutions of the fermion system.  Thus for all practical
purposes we may assume that $T$ is a projective involution, i.e., $T^2
= z \times \mathrm{Id}$ with $z$ a complex number of unit modulus.
\begin{prop}
If a projective involution $T : \mathcal{W} \to \mathcal{W}$ of a
unitary vector space $\mathcal{W}$ is antiunitary, then either
$T^2 = +\mathrm{Id}_\mathcal{W}$ or $T^2 = -
\mathrm{Id}_\mathcal{W} $.
\end{prop}
\begin{proof}
A projective involution $T$ has square $T^2 = z \times \mathrm
{Id}$ with $z \in \mathbb{C} \setminus \{ 0 \}$.  Since $T$ is
antiunitary, $T^2$ is unitary, and hence $|z| = 1$.  But an
antiunitary operator is $\mathbb{C}$-antilinear, and therefore the
associative law $T^2 \cdot T = T \cdot T^2$ forces $z$ to be real,
leaving only the possibilities $T^2 = \pm \mathrm{Id}$.
\end{proof}
Since this work is meant to simultaneously handle symmetry at both
the Lie algebra and Lie group level, a final word should be said
about the notion that a bilinear form $F$ is respected by a
transformation $B$.  At the group level when $B$ is invertible and
is regarded as being in $\mathrm{GL}(W)$, where $W$ is the
underlying vector space of $F : W\times W \to \mathbb C$, this
means that $B$ is an isometry in the sense that $F(B w_1 , B w_2)
= F(w_1 , w_2)$ for all $w_1, w_2 \in W$. On the other hand, at
the Lie algebra level where $B \in \mathrm{End}(W)$, this means
that for all $w_1, w_2 \in W$ one has $\frac{d}{dt} F(\mathrm
{e}^{tB} w_1, \mathrm{e}^{tB} w_2) \vert_{t=0} = F(B w_1, w_2) +
F(w_1, B w_2) = 0$.

\section{Reduction to the case of $G_0 = \{ \mathrm{Id} \}$}
\label{sect: reduction}\setcounter{equation}{0}

Recall that our main goal, e.g., on the Lie algebra level, is to
describe the space of $G_0$-invariant endomorphisms which on a
block in Nambu space are compatible with the unitary structure,
time reversal and the symmetric $\mathbb{C}$-bilinear form.

Here we prove results which allow us to transfer this space to a
certain space of $G_0$-equivariant homomorphisms.  The unitary
structure, time reversal and the bilinear form are transferred
canonically, and as before, compatibility with these structures is
required. However, in the new setting $G_0$ acts trivially. This
is of course an essential simplification, and paves the way toward
our classification goal.

In this section $\lambda \in \hat G_0$ denotes a fixed isomorphism
class (i.e., an equivalence class of irreducible repre\-sentations
of $G_0$), and $\lambda ^*$ denotes its dual.  A \emph{block} is
determined by a choice of finite-dimensional $G_0$-invariant
subspace $V = V_\lambda$ (in the given Hilbert space $\mathcal
{V}$) such that all of its  irreducible subrepresentations have
isomorphism class $\lambda$. The full group $G$ of (unitary and
antiunitary) symmetries is generated by $G_0$ and at most two
distinguished time-reversal symmetries. Throughout this section
(and also in Sects.\ \ref{sect: symm space}, \ref{sect: typeI},
\ref{sect: smallblocks}) we assume that these time-reversal
operators $T$ stabilize the truncated subspace $W = V \oplus V^*$
of Nambu space:
\begin{displaymath}
    T W = W \;.
\end{displaymath}
The case where one or both time-reversal symmetries do not
stabilize $W$, i.e., where a larger block is generated, is handled
in Sects.\ \ref{sect: typeII} and \ref{sect: bigblocks}.

\subsection{Spaces of equivariant homomorphisms}
\label{sect: equihomo}

If $\langle \, , \, \rangle_V$ is the initial unitary structure on
$V$, one defines $C : V \to V^*$ by $C(v)(w) = \langle v,w
\rangle_V$. Taking $C \vert_{V^*}$ to be the inverse of this map,
one obtains the associated $\mathbb{C} $-antilinear isomorphism $C
: W \to W$.  All symmetries in $G$ are assumed to commute with
$C$.  We remind the reader that $G_0$ acts on $V^*$ by $g(f) =
f\circ g^{-1}$.

Let $R$ be a fixed irreducible $G_0$-representation space which is
in $\lambda$. Denote by $d$ its dimension.  Of course $R^*$ is a
representative of $\lambda^*$.  We fix an antilinear bijection
\begin{displaymath}
    \iota : R \to R^* \;,
\end{displaymath}
which is defined by a $G_0$-invariant unitary structure $\langle
\, , \, \rangle_R$ on $R$.  (Note the change of meaning of the
symbol $\iota$ as compared to Sect.\ \ref{sect: Nambu}.)

In the sequel we will often make use of the following consequence
of Schur's Lemma.  (Note the change of meaning of the symbol
$\psi$ as compared to Sect.\ \ref{sect: Nambu}.)
\begin{prop} \label{Schur}
If two irreducible $G_0$-representation spaces $R_1$ and $R_2$ are
equi\-variantly isomorphic by $\psi : R_1 \to R_2$, then
$\mathrm{Hom}_{G_0}(R_1,R_2) = \mathbb{C} \cdot \psi$, i.e., the
linear space of $G_0$-equivariant homomorphisms from $R_1$ to
$R_2$ has complex dimension one and every operator in it is some
multiple of $\psi$.
\end{prop}
The following related statement was essential to Dyson's
classification and will play a similarly important role in the
present article.
\begin{lem} \label{sym/alt}
If an irreducible $G_0$-representation space $R$ is equivariantly
isomorphic to its dual $R^\ast$ by an isomorphism $\psi : R \to
R^*$, then $\psi$ is either symmetric or alternating, i.e., either
$\psi(r)(t) = \psi(t)(r)$ or $\psi(r) (t) = - \psi(t)(r)$ for all
$r,t \in R$.
\end{lem}
\begin{proof}
It is convenient to think of $\psi$ as defining an invariant bilinear
form $B(r,t) = \psi(r)(t)$ on $R$.  We then decompose $B$ into its
symmetric and alternating parts, $B = S + A$, where
\begin{displaymath}
    S(r,t) = {\textstyle{\frac{1}{2}}} \big( B(r,t) + B(t,r)
    \big) \quad \text{and} \quad A(r,t) = {\textstyle{
    \frac{1}{2}}} \big( B(r,t) - B(t,r) \big) \;.
\end{displaymath}
Both are $G_0$-invariant, and consequently their degeneracy subspaces
are invariant. Since the representation space $R$ is irreducible, it
follows that each is either nondegenerate or vanishes identically. But
both being nondegenerate would violate the fact that up to a constant
multiple there is only one equivariant isomorphism in $\mathrm{End}
(R)$.  Therefore $B$ is either symmetric or alternating as claimed.
\end{proof}
Now let $H := \mathrm{Hom}_{G_0}(R,V)$ be the space of
$G_0$-equivariant linear mappings from $R$ to $V$. Its dual space
is $H^* = \mathrm{Hom}_{G_0}(R^*,V^*)$.  The key space for our
first considerations is $(H \otimes R) \oplus (H^*\otimes R^*)$.
(Here, and throughout this paper, tensor products are understood
to be tensor products over the field of complex numbers.) Note
that $G_0$ acts on it by
\begin{displaymath}
    g(h \otimes r + f \otimes t) =
    h \otimes g(r) + f \otimes g(t) \;.
\end{displaymath}
We can apply $h \in H$ to $r \in R$ to form $h(r) \in V$. Since
$h$ is $G_0$-equivariant we have $g \cdot h(r) = h( g(r) )$. The
same goes for the corresponding objects on the dual side. Thus in
our finite-dimensional setting the following is immediate. (Once
again, note the change of meaning of the symbol $\varepsilon$ as
compared to Sect.\ \ref{sect: Nambu}.)
\begin{prop}
If $H = \mathrm{Hom}_{G_0} (R,V)$ and $H^\ast = \mathrm{Hom}_{G_0}
(R^\ast, V^\ast)$ the map
\begin{eqnarray*}
    \varepsilon : (H\otimes R)\oplus (H^*\otimes R^*) &\to&
    V\oplus V^* = W\;, \\ h \otimes r + f \otimes t
    &\mapsto& h(r) + f(t) \;,
\end{eqnarray*}
is a $G_0$-equivariant isomorphism.
\end{prop}
Transferring the unitary structure from $W$ to $(H\otimes R)\oplus
(H^*\otimes R^*)$ induces a unitary structure on $H\oplus H^*$.
For this, note for example that for $h_1 \otimes r_1$ and $h_2
\otimes r_2$ in $H\otimes R$ we have
\begin{displaymath}
    \langle h_1\otimes r_1, h_2 \otimes r_2 \rangle_{H \otimes R}
    := \langle h_1(r_1), h_2(r_2) \rangle_V \;.
\end{displaymath}
Observe that for $h_1$ and $h_2$ fixed, the right-hand side of
this equality defines a $G_0$-invariant unitary structure on $R$
which is unique up to a multiplicative constant. Thus we define
$\langle \, , \, \rangle_H$ by
\begin{displaymath}
    \langle h_1\otimes r_1, h_2 \otimes r_2 \rangle_{H
    \otimes R} = \langle h_1, h_2 \rangle_H\cdot
    \langle r_1, r_2 \rangle_R \;.
\end{displaymath}
Given the fixed choice of $\langle \, , \, \rangle_R$ this
definition is canonical.

We will in fact transfer all of our considerations for $V\oplus
V^*$ to the space $H \oplus H^*$, the latter being equipped with
the unitary structure defined as above.  One of the key points for
this is to understand how to express a $G_0$-invariant
endomorphism
\begin{displaymath}
    S\in \mathrm{End}_{G_0} (V\oplus V^*) ~ \cong_\varepsilon
    ~ \mathrm{End}_{G_0} (H \otimes R \oplus H^*\otimes R^*)
\end{displaymath}
as an element of $\mathrm{End}(H \oplus H^*)$.  Also, we must
understand the role of time reversal.

In this regard the two cases $\lambda \not=\lambda ^*$ and
$\lambda =\lambda ^*$ pose slightly different problems.  Before
going into these in the next sections, we note several facts which
are independent of the case.

First, let $V_1$ and $V_2$ be vector spaces where $G_0$ acts
trivially, and let $R_1$ and $R_2$ be arbitrary
$G_0$-representation spaces.
\begin{prop} \label{equivariantHom}
\begin{displaymath}
    \mathrm{Hom}_{G_0}(V_1\otimes R_1,V_2\otimes R_2) =
    \mathrm{Hom}(V_1,V_2)\otimes \mathrm{Hom}_{G_0}(R_1,R_2) \;.
\end{displaymath}
\end{prop}
\begin{proof}
Note that $\mathrm{Hom}(V_1\otimes R_1,V_2\otimes R_2) =
\mathrm{Hom}(V_1,V_2) \otimes \mathrm{Hom}(R_1,R_2)$, and let
$(\varphi_1, \ldots, \varphi_m)$ be a basis of $\mathrm{Hom} (V_1,
V_2)$. Then for every element $S$ of $\mathrm{Hom}(V_1, V_2)
\otimes \mathrm{Hom}(R_1,R_2)$ there are unique elements $\psi_1,
\ldots , \psi_m$ so that $S = \sum \varphi_i\otimes \psi_i$. If
$S$ is $G_0$-equivariant, then
\begin{displaymath}
    S = g \circ S \circ g^{-1} = \sum \varphi_i \otimes
    (g \circ \psi_i \circ g^{-1}) \;,
\end{displaymath}
and the desired result follows from the uniqueness statement.
\end{proof}
Our second general remark concerns the way in which a distinguished
time-reversal symmetry $T$ is transferred to an antilinear
endomorphism of $H \otimes R \oplus H^* \otimes R^*$. Let us consider
for example the case of mixing where it is sufficient to understand $T
: H \otimes R \to H^* \otimes R^*$. For that purpose we view
$\mathrm{End} (H \otimes R)$ as $\mathrm{End}(H) \otimes \mathrm
{End}(R)$, let $(\varphi_1, \ldots, \varphi_m)$ be a basis of
$\mathrm{End}(H)$ and write
\begin{displaymath}
    \Gamma  = CT = \sum \varphi_i \otimes \psi_i
\end{displaymath}
for $\psi_1, \ldots, \psi_m \in \mathrm{End}(R)$.  Now $T$ is
equivariant in the sense that $T \circ g = a(g) \circ T$, where $a$ is
the automorphism of $G_0$ determined by conjugation with $g_T$. Thus,
since the $\mathbb{C}$-antilinear operator $C$ intertwines
$G_0$-actions, the $\mathbb{C}$-linear mapping $\Gamma = CT$ is
invariant with respect to the twisted conjugation $\Gamma \mapsto a(g)
\Gamma g^{-1}$. Consequently, every $\psi_i$ is invariant with respect
to this conjugation.

This means that the $\psi_i : R \to R$ are equivariant with
respect to the original $G_0$-representation on the domain space
and the new $G_0$-action, $v \mapsto a(g) (v)$, on the image
space. But by Prop.\ \ref{Schur}, up to a constant multiple there
is only one such element of $\mathrm{End}(R)$, i.e., we may assume
that
\begin{displaymath}
    \Gamma = \varphi \otimes \psi \;,
\end{displaymath}
where $\psi$ is unique up to a multiplicative constant.

Note further that $C$ is also of this factorized form.  Indeed, we
have
\begin{displaymath}
    \langle h \otimes r, \cdot \rangle_{H \otimes R} = \langle
    h ,\cdot \rangle_H \, \langle r , \cdot \rangle_R \;,
\end{displaymath}
and if $\gamma : H \to H^*$ is defined by $h \mapsto \langle h , \cdot
\rangle_H$, then $C = \gamma \otimes \iota$. Furthermore, since
$\Gamma $ and $C$ are pure tensors, so is $T = C\Gamma = T_H \otimes
T_R$ , with the factors being antilinear mappings $T_H = \gamma\circ
\varphi : H \to H^\ast$ and $T_R = \iota \circ \psi : R \to R^\ast$.

Of course we have only considered a piece of $T$, and that only in
the case of mixing.  However, exactly the same arguments apply to
the other piece and also in the case of nonmixing. Thus we have
the following observation.
\begin{prop}\label{PureTensors}
The induced map
\begin{displaymath}
    T : (H \otimes R) \oplus (H^*\otimes R^*) \to
    (H \otimes R) \oplus (H^* \otimes R^*) \;,
\end{displaymath}
is the sum $T = A_1 \otimes B_1 + A_2 \otimes B_2$ of pure
tensors.
\end{prop}
In the case of mixing this means that $A_1 \otimes B_1$ is an
antilinear mapping from  $H \otimes R$ to $H^* \otimes R^*$ and
vice versa for $A_2 \otimes B_2$.

If $T$ doesn't mix, then $A_1\otimes B_1 : H \otimes R \to H \otimes
R$ and $A_2 \otimes B_2 : H^* \otimes R^* \to H^\ast \otimes R^\ast$.
In this case we impose the natural condition that the $A_i$ and $B_i$
be antiunitary.  For later purposes we note that this condition
determines the factors only up to multiplication by a complex number
of unit modulus. Using the formula $C = \gamma \otimes \iota$ and the
fact that $C$ commutes with $T$, one immediately computes $A_2 \otimes
B_2$ from $A_1 \otimes B_1$ (or vice versa).

The involutory property $T^2 = \pm \mathrm{ Id}$ also adds strong
restrictions. Of course there may be two distinguished time
reversals, $T$ and $T_1$, and we require that they commute with
$C$ and $T_1T=\pm TT_1$.  These properties are automatically
transferred at this level, because the transfer process from $(H
\otimes R) \oplus (H^* \otimes R^*)$ to $V \oplus V^*$ is an
isomorphism.

Finally, we prove an identity which is essential for transferring the
complex bilinear form.  For this we begin with
\begin{displaymath}
    h \otimes r + f \otimes t \in
    (H \otimes R) \oplus (H^* \otimes R^*) \;,
\end{displaymath}
apply $\varepsilon$ to obtain $h(r) + f(t)$, and then apply the linear
function $f(t)\in V^\ast$ to the vector $h(r) \in V$. The result $f(t)
(h(r))$ is to be compared to the product $f(h) \, t(r)$.  Recall that
the dimension of the vector space $R$ is denoted by $d$.
\begin{prop} \label{BilinearForm}
\begin{displaymath}
    f(t)(h(r)) = d^{-1} \, f(h) \, t(r) \;.
\end{displaymath}
\end{prop}
Before beginning the proof, which uses bases for the various
spaces, we set the notation and prove a preliminary lemma. Let $m$
denote the multiplicity of the component $V$ and fix an
identification
\begin{displaymath}
    V\oplus V^* = R\oplus \ldots \oplus R \oplus
    R^* \oplus \ldots \oplus R^*
\end{displaymath}
with $m$ summands of $R$ and $R^*$.  Let $(e_1, \ldots, e_d)$ be a
basis of $R$ and $(\vartheta_1, \ldots, \vartheta_d)$ be its dual
basis. These define bases $(e_1^k, \ldots, e_d^k)$ and
$(\vartheta_1^k, \ldots, \vartheta_d^k)$ of the corresponding
$k$-th summands above.  Let $I^k_R$ and $I^k_{R^*}$ be the
respective identity mappings.
\begin{lem}
\begin{displaymath}
    I^\ell_{R^*}(I^k_R) = \delta _{k\ell} \, d \;.
\end{displaymath}
\end{lem}
\begin{proof}
Expressing the operators in the bases, i.e.,
\begin{displaymath}
   I^k_R = \sum\nolimits_i \vartheta^k_i\otimes e^k_i \quad
   \text{and} \quad I^\ell_{R^*} = \sum\nolimits_j
   e^\ell_j \otimes \vartheta^\ell_j \;,
\end{displaymath}
one has
\begin{displaymath}
    I^\ell_{R*}(I^k_R) = \sum\nolimits_{i,j} \vartheta^k_i(
    e^\ell_j)\, \vartheta^\ell_j(e^k_i) = \sum\nolimits_{i,j}
    \delta^{k\ell}_{ij} = \delta_{k\ell} \, d \;,
\end{displaymath}
which is the statement of the lemma.
\end{proof}
\noindent{\it Proof of Prop.\ \ref{BilinearForm}}. --- We expand
$h \in H = \mathrm{Hom}_{G_0}(R,V)$ as $h = \sum h_k I^k_R$ , and
$f \in H^\ast = \mathrm{Hom}_{G_0}(R^\ast, V^\ast)$ as $f = \sum
f_\ell I^\ell _{R^*}$ . If $r = \sum r_i e_i$ and $t = \sum t_j
\vartheta_j$ , then
\begin{displaymath}
    h(r) = \sum\nolimits_{i,k} h_k r_i \, e^k_i \quad
    \text{and} \quad f(t) = \sum\nolimits_{j,\ell}
    f_\ell t_j \, \vartheta^\ell_j \;.
\end{displaymath}
Thus
\begin{displaymath}
    f(t)(h(r)) = \sum\nolimits_{i j k \ell}
    \delta^{k\ell}_{ij} f_\ell h_k t_j r_i =
    \left( \sum\nolimits_k f_k h_k \right) t(r) \;.
\end{displaymath}
Prop.\ \ref{BilinearForm} now follows from the above lemma which
implies that $f(h) = d \, \sum f_k h_k$.

\subsection{The case where $\lambda \not =\lambda ^*$}

Recall that our goal is to canonically transfer the data on
$V\oplus V^*$ to $H \oplus H^*$, thus removing $G_0$ from the
picture.  In the case where $\lambda \not =\lambda ^*$ this is a
particularly simple task.

First, we apply Prop.\ \ref{equivariantHom} to transfer elements
of $\mathrm{End}_{G_0}(V\oplus V^*)$. In the case at hand
$\mathrm{Hom}_{G_0}(R,R^*)$ and $\mathrm{Hom}_{G_0}(R^*,R)$ are
both zero, and both $\mathrm{End}_{G_0}(R)$ and $\mathrm{End}_{
G_0}(R^*)$ are isomorphic to $\mathbb C$.  Thus it follows from
Prop.\ \ref{equivariantHom} that
\begin{eqnarray*}
    \mathrm{End}_{G_0}(V\oplus V^*) &\cong& \mathrm{End}_{G_0}
    (H\otimes R\oplus H^*\otimes R^*) \\ &\cong& \mathrm{End}(H)
    \oplus \mathrm{End}(H^*)\hookrightarrow \mathrm{End}(H
    \oplus H^*) \;.
\end{eqnarray*}
We always normalize operators in $\mathrm{End}_{G_0}(H\otimes R)$
to the form $\varphi \otimes \mathrm{Id}_R$ and normalize
operators in $\mathrm{End}_{G_0}(H^*\otimes R^*)$ in a similar
way.  Thus we identify $\mathrm{End}_{G_0}(V\oplus V^*)$ with
$\mathrm{End}(H) \oplus \mathrm{End}(H^*)$ as a subspace of
$\mathrm{End} (H \oplus H^*)$ and have the following result.
\begin{prop}
The condition that an operator in $\mathrm{End}_{G_0}(V\oplus
V^*)$ respects the unitary structure on $V\oplus V^*$ is
equivalent to the canonically transferred operator in
$\mathrm{End}(H \oplus H^*)$ respecting the canonically
transferred unitary structure on $H\oplus H^*$.
\end{prop}
Now let us turn to the condition of compatibility with a
transferred time-reversal operator $T : H \otimes R \oplus
H^*\otimes R^* \to H \otimes R \oplus H^*\otimes R^*$.  There are
a number of cases, depending on whether or not $T$ mixes and which
of the conditions $T^2 = -\mathrm{Id}$ or $T^2 = \mathrm{Id}$ are
satisfied. The arguments are essentially the same in every case.
Let us first go through the details in one of them, the mixing
case where $T^2 = - \mathrm{Id}$. To be consistent with the
slightly more complicated discussion in the case where $\lambda =
\lambda^*$, let us write this in matrix notation.

For $A\in \mathrm{End}(H)$ and $D\in \mathrm{End}(H^*)$, we regard
\begin{displaymath}
    M = \begin{pmatrix} A\otimes \mathrm{Id}_R & 0\\
    0 & D\otimes \mathrm{Id}_{R^*} \end{pmatrix}
\end{displaymath}
as the associated transformation in $\mathrm{End}_{G_0}(H \otimes
R \oplus H^*\otimes R^*)$. To construct the transferred
time-reversal operator recall the statement of Prop.\
\ref{PureTensors}. In the setting under consideration $T$ squares
to minus the identity; it is therefore expressed as
\begin{displaymath}
    T = \begin{pmatrix} 0 & - \alpha^{-1} \otimes \beta^{-1}
    \\ \alpha \otimes \beta &0 \end{pmatrix} \;,
\end{displaymath}
where $\alpha : H \to H^\ast$ and $\beta : R \to R^\ast$ are
complex antilinear. Note that since $\alpha \otimes \beta = z \,
\alpha \otimes z^{-1} \beta$, the mappings $\alpha$ and $\beta$
are determined only up to a common multiplicative constant $z \in
\mathbb{C} \setminus \{ 0 \}$. Conjugation of $M$ in $\mathrm{
End}_{G_0}(H\otimes R \oplus H^*\otimes R^*)$ by $T$ yields
\begin{displaymath}
    T M T^{-1} = \begin{pmatrix} \alpha^{-1} D\, \alpha
    \otimes \mathrm{Id}_R & 0\\ 0 & \alpha A \alpha^{-1}
    \otimes \mathrm{Id}_{R^*} \end{pmatrix} \;.
\end{displaymath}
Clearly, compatibility of $M$ with $T$ here means that $D = \alpha
A \alpha^{-1}$.

Formulating this in a less detailed way gives the appropriate
statement: conjugation of $M$ in $\mathrm{End}_{G_0} (H \otimes R
\oplus H^*\otimes R^*)$ by $T$ yields the same compatibility
condition as conjugating
\begin{displaymath}
    \begin{pmatrix} A & 0\\ 0 & D \end{pmatrix} \quad
    \mbox{by} \quad \begin{pmatrix} 0 & \mp \alpha^{-1}\\
    \alpha & 0 \end{pmatrix} \;.
\end{displaymath}
Here the sign in front of $\alpha^{-1}$ is arbitrary. For
definiteness we choose it in such a way that the transferred
time-reversal operator has the same involutory property $T^2 = -
\mathrm{Id}$ or $T^2 = \mathrm{Id}$ as the original operator; in
the case under consideration this means that we choose the minus
sign.
\begin{prop}\label{transfer T}
There is a transferred time-reversal operator $T : H
\oplus H^* \to H\oplus H^*$ which satisfies either $T^2 = -
\mathrm{Id}$ or $T^2 = \mathrm{Id}$. It mixes if and only if the
original operator mixes, and a canonically transferred mapping in
$\mathrm{End}(H \oplus H^*)$ commutes with it if and only if the
original mapping in $\mathrm{End}_{G_0} (V\oplus V^*)$ commutes
with the original time-reversal operator.
\end{prop}
\begin{proof}
It only remains to handle the case of nonmixing, e.g., when $T^2 =
- \mathrm{Id}$.  As we have seen, $T : H \otimes R \to H\otimes R$
is a pure tensor:
\begin{displaymath}
    T \vert_{H \otimes R} = \alpha \otimes \beta \;,
\end{displaymath}
which gives $T^2 \vert_{H \otimes R} = \alpha^2 \otimes \beta^2 =
- \mathrm{Id}_H \otimes \mathrm{Id}_R$ in the case at hand. Since
the induced map $\beta : R \to R$ is antiunitary by convention, we
have $\beta^2 = z \times \mathrm{Id}_R$ with $|z| = 1$.
Associativity ($\beta^2 \cdot \beta = \beta \cdot \beta^2$) then
implies $z = \pm 1$. Unlike the case of mixing, $\beta$ now plays
a role through its parity.  If $\beta^2 = + \mathrm{Id}_R$, the
transferred time-reversal operator $\alpha$ on $H$ still satisfies
$\alpha^2 = - \mathrm{Id}_H$.  On the other hand, if $\beta^2 = -
\mathrm{Id}_R$ we have $\alpha^2 = + \mathrm{Id}_H$ instead.  Thus
the involutory property $T^2 = \pm \mathrm{Id}$ is passed on to
the transferred time-reversal operator, but depending on the
involutory character of $\beta$ the parity may change.
\end{proof}
We remind the reader that two distinguished time-reversal symmetries
may be present.  The above shows that both can be transferred with
appropriate involutory properties.  Further, it must be shown that
they can be transferred (along with $C$) so that $TC = CT$, $T_1 C = C
T_1$, and $T_1 T = \pm T T_1$ still hold. Even if there is just one such
operator, it must be shown that the transferred operator can be chosen
to satisfy $T C = C T$. Since the discussion for this is the same as
in the case where $\lambda = \lambda ^*$, we postpone it to Sect.\
\ref{precise T-transfer}.

Finally, we turn to the problem of transferring the complex bilinear
form on $V \oplus V^*$ to $H \oplus H^*$. If $b$ denotes the pullback
by $\varepsilon$ of the canonical symmetric bilinear form on $V\oplus
V^*$, then by Prop.\ \ref{BilinearForm}
\begin{displaymath}
    b(h_1 \otimes r_1 + f_1 \otimes t_1, h_2 \otimes r_2 + f_2
    \otimes t_2 ) = d^{-1} (f_1(h_2) t_1(r_2) + f_2(h_1) t_2(r_1)) \;.
\end{displaymath}
Now in this case, i.e., where $\lambda \not= \lambda^*$, the
$G_0$-invariant endomorphisms are acting on $H \otimes R \oplus
H^*\otimes R^*$ by $\begin{pmatrix} A\otimes \mathrm{Id}_R & 0\\
0 & D\otimes \mathrm{Id}_{R^*} \end{pmatrix}$, where
\begin{displaymath}
    A\oplus D \in \mathrm{End}(H) \oplus \mathrm{End}(H^*)
    \hookrightarrow \mathrm{End} (H\oplus H^*) \;.
\end{displaymath}
Inserting the operator $A \oplus D$ into the above expression for
$b$ we have the following fact involving the canonical symmetric
bilinear form $s$ on $H \oplus H^*$,
\begin{displaymath}
    s(h_1 + f_1, h_2 + f_2) = f_1(h_2) + f_2(h_1) \;.
\end{displaymath}
\begin{prop}
A map in $\mathrm{End}_{G_0}(V \oplus V^*)$ respects the canonical
symmetric bilinear form if and only if the transferred map in
$\mathrm{End}(H) \oplus \mathrm{End}(H^*) \hookrightarrow
\mathrm{End} (H \oplus H^*)$ respects the canonical symmetric
bilinear form $s$ on $H\oplus H^*$.
\end{prop}
In summary, we have shown that if $\lambda \not=\lambda ^*$, then
all relevant structures on $V\oplus V^*$ transfer to data of
essentially the same type on $H\oplus H^*$ (the only exception
being that the parity of the transferred time-reversal operator
may be reversed). In this case $\mathrm{End}_{G_0}(V \oplus V^*)$
is canonically isomorphic to $\mathrm{End}(H) \oplus
\mathrm{End}(H^*) \hookrightarrow \mathrm{End}(H \oplus H^*)$. An
operator in $\mathrm{End}_{G_0}(V \oplus V^*)$ respects the
original structures if and only if the corresponding operator in
$\mathrm{End}(H\oplus H^*)$ respects the transferred structures on
$H\oplus H^*$.  The latter are the transferred unitary structure,
induced time reversal and the symmetric bilinear form $s$.

\subsection{The case where $\lambda = \lambda^*$}
\label{sect: selfdual}

Throughout this section it is assumed that $\lambda = \lambda^*$, and
$\psi : R\to R^*$ is a $G_0$-equivariant isomorphism.  Thus we have
the identification
\begin{eqnarray*}
    H\otimes R \oplus H^*\otimes R &\cong& H\otimes R\oplus
    H^*\otimes R^*, \\ h \otimes r + f \otimes t
    &\mapsto& h \otimes r + f \otimes \psi(t) \;.
\end{eqnarray*}
Applying Prop.\ \ref{equivariantHom} to each component of an
operator in $\mathrm{End}_{G_0}(H \otimes R \oplus H^*\otimes R)$
it follows that
\begin{displaymath}
    \mathrm{End}_{G_0} (H\otimes R\oplus H^*\otimes R^*)
    \cong \mathrm{End}(H\oplus H^*) \;.
\end{displaymath}
We therefore identify $\mathrm{End}(H\oplus H^*)$ with $\mathrm
{End}_{G_0}(H\otimes R \oplus H^*\otimes R^*) = \mathrm{End}_{G_0}
(V \oplus V^*)$ by the mapping
\begin{gather*}
    M = \begin{pmatrix} \mathsf{A} & \mathsf{B}\\ \mathsf{C}
    & \mathsf{D} \end{pmatrix} \mapsto \begin{pmatrix} \mathsf{A}
    \otimes \mathrm{Id}_R & \mathsf{B}\otimes \psi^{-1}\\
    \mathsf{C}\otimes \psi & \mathsf{D}\otimes \mathrm{Id}_{R^\ast}
    \end{pmatrix} \;.
\end{gather*}

Recall the induced unitary structure which is defined, e.g., on $H
\otimes R$ by
\begin{displaymath}
    \langle h_1 \otimes r_1, h_2 \otimes r_2 \rangle_{H\otimes R}
    := \langle h_1(r_1), h_2(r_2) \rangle_V = \langle h_1 , h_2
    \rangle_H \, \langle r_1 , r_2 \rangle_R \;.
\end{displaymath}
It is easy to verify that this defines a unitary structure on $H
\oplus H^*$ with the desired property: a map in $\mathrm{End}_
{G_0} (V\oplus V^*)$ preserves the given unitary structure on
$V\oplus V^*$ if and only if the transferred map $M$ preserves the
induced unitary structure on $H\oplus H^*$.

Now let us consider time reversal.  For example, take the case of
nonmixing where $T_1 : H \otimes R \to H \otimes R$.  Using Prop.\
\ref{PureTensors} we have
\begin{gather*}
    T_1 = \begin{pmatrix} \alpha \otimes \beta & 0\\
    0 & \tilde \alpha \otimes \tilde \beta \end{pmatrix} \;,
\end{gather*}
and conjugating the transformation $\begin{pmatrix} \mathsf{A}
\otimes \mathrm{Id}_R & \mathsf{B}\otimes \psi^{-1}\\ \mathsf{C}
\otimes \psi & \mathsf{D}\otimes \mathrm{Id}_{R^*} \end{pmatrix}$
at the level of operators on $H \otimes R \oplus H^* \otimes R^*$
yields
\begin{gather*}
    \begin{pmatrix} \alpha \mathsf{A} \alpha^{-1} \otimes
    \mathrm{Id}_R &\alpha \mathsf{B} \tilde\alpha^{-1} \otimes
    \beta \psi^{-1} \tilde\beta^{-1}\\ \tilde\alpha \mathsf{C}
    \alpha^{-1} \otimes \tilde\beta \psi \beta^{-1} &\tilde\alpha
    \mathsf{D} \tilde \alpha^{-1}\otimes \mathrm{Id}_{R^*}
    \end{pmatrix} \;.
\end{gather*}
Now, as has been mentioned in Sect.\ \ref{sect: equihomo}, the
equivariant antiunitary maps $\beta$ and $\tilde\beta$ are only
unique up to multiplicative constants of unit modulus.  They will
be chosen in the next subsection so that the distinguished
time-reversal operator(s) and the unitary structure $C$ commute.
These choices having been made, we make a compatible choice of
$\psi$ so that $\tilde \beta \psi \beta ^{-1} = \psi $.  In this
way, in the case where $T_1$ is nonmixing as above, conjugation of
the matrix $M$ by $T_1$ is given by
\begin{gather}\label{nonmixing conjugation}
  \begin{pmatrix}\mathsf{A} & \mathsf{B}\\ \mathsf{C}
    &\mathsf{D} \end{pmatrix} \mapsto \begin{pmatrix}
    \alpha \mathsf{A} \alpha ^{-1} & \alpha \mathsf{B}
    \tilde\alpha ^{-1} \\ \tilde\alpha \mathsf{C}\alpha^{-1}
    & \tilde\alpha \mathsf{D} \tilde\alpha ^{-1}\end{pmatrix}
  \;.
\end{gather}
Thus the transferred time-reversal operator is simply given by $T_1 =
\alpha \oplus \tilde \alpha $ on $H\oplus H^\ast$.

Consider now the case of a mixing time-reversal symmetry $T$ where
\begin{displaymath}
  T = \begin{pmatrix} 0 & \alpha ^{-1}\otimes \beta ^{-1} \\
    \varepsilon_T \, \alpha \otimes \beta & 0 \end{pmatrix}
\end{displaymath}
with $\varepsilon_T = \pm 1$. In this case the compatibility
condition on $\psi$ is $\beta \psi^{-1}\beta = \varepsilon_\beta
\psi$, with $\varepsilon_\beta = \pm 1$.  If this holds,
conjugation of $M$ by $T$ is given by
\begin{gather}\label{mixing conjugation}
  \begin{pmatrix} \mathsf{A} & \mathsf{B}\\ \mathsf{C}
    &\mathsf{D} \end{pmatrix} \mapsto \begin{pmatrix}
    \alpha^{-1} \mathsf{D} \alpha & \varepsilon_\alpha \,
    \alpha^{-1} \mathsf{C} \alpha^{-1} \\ \varepsilon_\alpha
    \, \alpha \mathsf{B}\alpha &\alpha \mathsf{A}
    \alpha^{-1} \end{pmatrix}
\end{gather}
with $\varepsilon_\alpha = \varepsilon_\beta \varepsilon_T$. In this
case the appropriate transferred operator is given by
\begin{gather*}
  T = \begin{pmatrix} 0 &\alpha^{-1}\\ \varepsilon_\alpha
    \, \alpha &0 \end{pmatrix} \;.
\end{gather*}
Given the (essentially unique) choices of the tensor-product
representations of $T$, $T_1$ and $C$ which are defined by $T_1 T
= \pm T T_1$ and by the conditions that $T$ and $T_1$ commute with
$C$, we show in Sect.\ \ref{precise T-transfer} that there is a
unique choice of $\psi$ so that both of these compatibility
conditions (from $T_1$ and $T$) on $\psi$ hold.

If we are in the nonmixing case $\beta : R \to R$, and it so
happens that $\beta$ is $G_0$-invariant, then the two alternatives
for the involutory property of $T$ (actually, $T_1$) can be
distinguished by the type of the unitary representation $R$ as
follows. Defining $\iota : R \to R^\ast$ by $r \mapsto \langle r ,
\cdot \rangle_R$ as before, consider the unitary mapping $\psi : R
\to R^\ast$ given as the composition $\psi = \iota \circ \beta$.
Since $\beta$ is $G_0$-invariant, $\psi$ is $G_0$-equivariant, and
the statement of Lemma \ref{sym/alt} applies. Using the
antiunitarity of $\beta$ one has
\begin{displaymath}
    \psi(r)(t) = \langle \beta r , t \rangle_R = \overline{
    \langle \beta^2 r , \beta t \rangle}_R = \psi(t)(\beta^2 r)
    \;,
\end{displaymath}
and therefore the following statement is immediate.
\begin{lem}\label{parity of beta}
The parity of an antiunitary and $G_0$-invariant mapping $\beta :
R \to R$ is determined by the parity of the irreducible
$G_0$-representation space $R$; i.e., $\beta$ satisfies $\beta^2 =
\mathrm{Id}_R$ resp.\ $\beta^2 = - \mathrm {Id}_R$ if $R$ carries
an invariant $\mathbb{C}$-bilinear form which is symmetric resp.\
alternating.
\end{lem}
If $\beta^2 = \mathrm{Id}_R$, the transferred time reversal
satisfies $T^2 = - \mathrm{Id}$ or $T^2 = \mathrm{Id}$ if the
original time reversal has these properties. On the other hand, if
$\beta^2 = - \mathrm{ Id}_R$, then the properties are reversed;
e.g., if $T^2 = - \mathrm{Id}$ on the original space, then
transferred time reversal satisfies $T^2 = \mathrm{Id}$.  We again
remind the reader that we must check that the transferred
time-reversal operator(s) and $C$ can be chosen compatibly.  It
turns out that there is in fact just enough freedom in the choice
of the constants to achieve this (see Sect.\ \ref{precise
T-transfer}).
\begin{ex}
An example of particular importance in physics is the transfer of
the (true) time reversal $T$ in the case where all spin rotations
are symmetries.  On fundamental grounds, $T$ is a (nonmixing)
operator which commutes with the spin-rotation group $\mathrm
{SU}_2$ and satisfies $T^2 = (-1)^n \, \mathrm{Id}$ on quantum
mechanical states with spin $S = n/2$.

Let $V = H \otimes \mathbb C^{n+1}$ be the tensor product of a
vector space $H$ with the spin $n/2$ re\-pre\-sentation space of
$\mathrm{SU}_2$. For simplicity assume that there are no further
symmetries.

Our Nambu space is already in the form $V \oplus V^* = (H \otimes
R) \oplus (H^* \otimes R^*)$. Thus the reduced space is $H \oplus
H^*$. Let the time-reversal operator on $V = H \otimes \mathbb{C}
^{n+1}$ be written $T = \alpha \otimes \beta$. The $\mathrm{SU}
_2$-representation space $\mathbb{C}^{n+1}$ is known to have
parity $+1$ (symmetric invariant form) for $n$ even, and $-1$
(alternating invariant form) for $n$ odd. By Lemma \ref{parity of
beta} this implies $\beta^2 = (-1)^n \, \mathrm{Id}$. The
situation on the dual space $V^\ast$ is the same. Thus in this
case, since $T^2 = (-1)^n\, \mathrm{Id}$, the transferred
time-reversal operator $\alpha : H \oplus H^\ast \to H \oplus
H^\ast$ always satisfies $\alpha^2 = + \mathrm{Id}_{ H \oplus
H^\ast}$, independent of the spin. \hfill $\square $
\end{ex}
Now let us turn to the problem of transferring the complex
bilinear form.  For this Lemma \ref{sym/alt} is an essential fact.
Earlier we identified $H \otimes R \oplus H^* \otimes R^\ast$ with
$V \oplus V^*$ by the map $\varepsilon : h \otimes r + f \otimes t
\mapsto h(r) + f(t)$. Using this along with Prop.\
\ref{BilinearForm} we now transfer the canonical symmetric
bilinear form on $V \oplus V^*$ to $H \oplus H^*$.  For this let
$s$ (resp.\ $a$) denote the canonical symmetric (resp.\
alternating) form on $H \oplus H^*$.
\begin{prop}\label{transfer b}
Depending on $\psi$ being symmetric or alternating, a transferred
map in $\mathrm{End}(H \oplus H^*)$ respects the canonical
symmetric form $s$ or alternating form $a$ if and only if the
original endomorphism in $\mathrm{End}_{G_0}(V \oplus V^*)$
respects the canonical symmetric complex bilinear form on $V
\oplus V^*$.
\end{prop}
\begin{proof}
We give the proof for the case where $\psi$ is alternating.  The
proof in the symmetric case is completely analogous.

Let $M = \begin{pmatrix} \mathsf{A} & \mathsf{B}\\ \mathsf{C} &
\mathsf{D} \end{pmatrix} \in \mathrm{ End} (H \oplus H^*)$ act as
a $G_0$-invariant operator
\begin{displaymath}
    \begin{pmatrix} \mathsf{A} \otimes \mathrm{Id}_R &\mathsf{B}
    \otimes \psi^{-1} \\ \mathsf{C}\otimes \psi &\mathsf{D}
    \otimes \mathrm{Id}_{R^\ast}\end{pmatrix}
\end{displaymath}
on $H \otimes R \oplus H^* \otimes R^\ast$ and let $b$ be the
symmetric complex bilinear form on this space which is induced
from the canonical symmetric form on $V \oplus V^*$.  We assume
that $M\in \mathrm{GL}(H\oplus H^*)$ and give the proof in terms
of the isometry property $b(Mv,Mw) = b(v,w)$.  Let us do this in a
series of cases. First, for $h_1 \otimes r_1$ and $h_2\otimes r_2$
in $H\otimes R$,
\begin{eqnarray*}
    &\phantom{=}& b(M(h_1 \otimes r_1), M(h_2 \otimes r_2)) \\
    &=& b(\mathsf{A} h_1\otimes r_1 + \mathsf{C} h_1 \otimes
    \psi(r_1), \mathsf{A} h_2 \otimes r_2 + \mathsf{C} h_2
    \otimes \psi (r_2)) \\ &=& \mathsf{C} h_1 (\mathsf{A} h_2)\,
    \psi(r_1)(r_2) / d + \mathsf{C} h_2(\mathsf{A} h_1)\,
    \psi(r_2)(r_1) / d \\ &=& a(\mathsf{A} h_1 + \mathsf{C} h_1,
    \mathsf{A} h_2 + \mathsf{C} h_2)\, \psi(r_1)(r_2) / d \;.
\end{eqnarray*}
When $M$ is the identity this becomes
\begin{displaymath}
    b(h_1 \otimes r_1, h_2 \otimes r_2) =
    a(h_1, h_2)\, \psi(r_1) (r_2) / d \;.
\end{displaymath}
Therefore $b (h_1\otimes r_1, h_2\otimes r_2) = b(M(h_1 \otimes
r_1), M(h_2 \otimes r_2))$ if and only if $a(h_1, h_2) = a(M(h_1)
, M(h_2))$. For $f_1 \otimes t_1 , f_2 \otimes t_2 \in H^*\otimes
R^\ast$ the discussion is analogous.

For $h \otimes r\in H\otimes R$ and $f \otimes t \in H^*\otimes
R^\ast$ we have a similar calculation:
\begin{eqnarray*}
    &\phantom{=}& b(M(f \otimes t), M(h \otimes r)) \\ &=&
    b(\mathsf{B} f\otimes \psi^{-1}(t) + \mathsf{D} f \otimes t
    , \mathsf{A} h\otimes r + \mathsf{C} h \otimes \psi(r)) \\
    &=& \mathsf{D} f (\mathsf{A} h)\, t(r) / d +
    \mathsf{C} h (\mathsf{B} f)\, \psi(r)( \psi^{-1}(t)) / d
    \\ &=& a(M(f) , M(h))\, t(r) / d \;.
\end{eqnarray*}
Of course the analogous identity holds for $b(M(h \otimes r), M(f
\otimes t))$.
\end{proof}
\begin{rem}
To avoid making sign errors and misidentifications in later
computations, we find it helpful to transfer the particle-hole
conjugation operator $C$ along with the complex bilinear form.
This is done by insisting that the statement of Lemma
\ref{Crelates} remains true after the transfer.  Thus the relation
$b(C w_1,w_2) = \langle w_1 , w_2 \rangle$ continues to hold in
all cases. By an almost identical variant of the computation that
led to Lemma \ref{parity of beta}, the transferred operator $C$
has parity $C^2 = +\mathrm{Id}$ or $C^2 = - \mathrm{Id}$ depending
on whether the transferred bilinear form is symmetric or
alternating.
\end{rem}

\subsection {Precise choice of time-reversal transfer}
\label{precise T-transfer}

Recalling the situation of this section, we have assumed that the
distinguished time-reversal operator(s) stabilize the initial block $V
\oplus V^*$, and we have transferred all structures to the space $(H
\otimes R) \oplus (H^* \otimes R^*)$ which is isomorphic to $V \oplus
V^*$.

The time-reversal operator(s) $T$ and the operator $C$ are given
by $(2\times 2)$-matrices of pure tensors on this space. The space
of endomorphisms $B$ that commute with the $G_0$-action is
identified with $\mathrm{End} (H\oplus H^*)$ or $\mathrm{End}(H)
\oplus \mathrm {End} (H^*)$ depending on whether or not $\lambda =
\lambda^*$.  The good Hamiltonians $B$ anticommute with $C$, and
commute with the time-reversal operator(s) $T$. If the matrix of
pure tensors representing the antiunitary operator $C$ (resp.\
$T$) has entries $\gamma \otimes \delta$, this means that $B$
anticommutes (resp.\ commutes) with the matrices defined by the
operators $\gamma$. Although the pure tensor decomposition is not
unique, this statement is independent of that decomposition.

It has been shown above that the transferred operators $T_1$, $T$,
and $C$ on $H\oplus H^*$, i.e., those defined by the operators
$\gamma$, can be chosen with the desired involutory properties. It
will now be shown that there is just enough freedom to insure that
\begin{displaymath}
TC = CT \;, \quad T_1 C = C T_1 \;, \quad T_1 T = \pm T T_1 \;,
\end{displaymath}
still hold after transferral.  After these conditions have
been met, we show as promised that $\psi : R \to R^*$ can be
chosen in a unique way so that the compatibility conditions of
Sect.\ \ref{sect: selfdual} hold, i.e., so that it makes sense to
define the transferred operators by the first factors of the
tensor-product representations.

We carry this out in the case where $\lambda = \lambda ^*$ and two
distinguished time-reversal operators are present.  All other cases
are either subcases of this or are much simpler.

The operator $C$ always mixes.  We will always choose it to be of the
form $C = \gamma \otimes \iota : H \otimes R \to H^* \otimes R^*$ and
$C = \gamma ^{-1} \otimes \iota^{-1} : H^* \otimes R^* \to H \otimes
R$.  Of course this is in the case where $b$ is symmetric.  If $b$ is
alternating, then we have $C^2 = - \mathrm{Id}$, and we make the
necessary sign change.

Here we restrict to the case where $T^2 = T_1^2 = \mathrm{Id}$.  The
various other involutory properties make no difference in the
argument. Just as in the case of $C$ we choose $T = \alpha \otimes
\beta : H \otimes R \to H^* \otimes R^*$ and $T = \alpha ^{-1} \otimes
\beta ^{-1}: H^* \otimes R^* \to H\otimes R$.  Similarly, we choose
$T_1 = \alpha _1 \otimes \beta_1 : H \otimes R \to H \otimes R$ and
$T_1 = \alpha _2 \otimes \beta_2 : H^* \otimes R^* \to H^* \otimes
R^*$.

On $(H \otimes R) \oplus (H^* \otimes R^*)$, the operators $T$ and
$T_1$ commute with $C$, and we have $T_1 T = \pm T T_1$.  We now
choose the tensor representations so that the same relations hold for
the induced operators on the first factors.

If $\alpha$, $\alpha_1$, $\alpha_2$, and $\gamma$ are any choices
for the first factors of the tensor-product representations of
$T$, $T_1$ and $C$, then there exist constants $c_1$, $c_2$ and
$c_3$ so that $\alpha_2 \alpha = c_1 \alpha \alpha_1$ (from $T T_1
= \pm T_1 T$), $\gamma \alpha_1 = c_2 \alpha_2 \gamma$ (from $C
T_1 = T_1 C$), and $\gamma \alpha^{-1} \gamma = c_3 \alpha$ ($CT =
TC$).

Let $\tilde\alpha = \xi\alpha$, $\tilde\gamma = \eta \gamma$, and
$\tilde\alpha_i = z_i \alpha$ (for $i = 1,2$), where $\xi$, $\eta$ and
$z_i$ are complex numbers yet to be determined.  Just as the $c_i$,
these constants are of modulus one.

The scaled operators satisfy $\tilde\alpha_2 \tilde\alpha = \chi_1 c_1
\tilde\alpha \tilde\alpha_1$, $\tilde\gamma \tilde\alpha_1 = \chi_2
c_2 \tilde\alpha_2 \tilde\gamma$, and $\tilde\gamma \tilde\alpha^{-1}
\tilde\gamma = \chi_3 c_3 \tilde\alpha$, where $\chi_1 = \xi^{-2} z_1
z_2$, $\chi_2 = \eta^2 (z_1 z_2)^{-1}$, and $\chi_3 = \xi^{-2}
\eta^2$. Observe that the characters $\chi_i$ satisfy the relation
$\chi_1 \chi_2 = \chi_3$, and that, e.g., $\chi_2$ and $\chi_3$ are
independent.

The constants $c_i$ satisfy an analogous relation.  For this first use
$\gamma \alpha_1 \gamma^{-1} = c_2 \alpha _2$ and $\gamma \alpha^{-1}
\gamma = c_3 \alpha $ to obtain $\gamma \alpha_1 \alpha^{-1} \gamma =
(c_2 / c_3) \alpha_2 \alpha $.  Then compose both sides of this
equation with the inverse of $\alpha_1$ on the right and use the
relation $\alpha_2 \alpha \alpha_1^{-1} = c_1 \alpha$ to obtain
$\gamma \alpha_1 \alpha^{-1} \gamma \alpha_1^{-1} = (c_1 c_2 / c_3)
\alpha $. Now $\gamma \alpha_1^{-1} = (c_2 \alpha_2)^{-1} \gamma$.
Thus
\begin{displaymath}
  \gamma \alpha_1 \alpha^{-1} \gamma \alpha_1^{-1} = \gamma \alpha_1
  \alpha^{-1} \alpha_2^{-1} c_2^{-1} \gamma = c_2^{-1} \gamma
  c_1 \alpha^{-1} \gamma = (c_1 c_2)^{-1} c_3 \alpha \;,
\end{displaymath}
and hence $c_1 c_2 / c_3 = c_3 / c_1 c_2$, i.e., $c_1^2 c_2^2 =
c_3^{2}$.

Since $\chi_2$ and $\chi_3$ are independent, we can choose the scaling
numbers so that $c_2 = c_3 = 1$, thereby arranging that $CT = TC$ and
$C T_1 = T_1 C$ still hold after transferral.  To preserve these
relations we must now keep $\chi_2$ and $\chi_3$ fixed at unity, which
from $\chi_1 \chi_2 = \chi_3$ implies that $\chi_1 = 1$.  Since
$c_3^{2} = c_1^{2} c_2^{2}$, we then conclude that $c_1$ takes one of
the two values $\pm 1$, and further scaling does not change this
constant.

In summary we have the following result.
\begin{prop}
The transferred operators $T_1$, $T$ and $C$ can be chosen so that
$T_1 C = C T_1$, $T C = C T$, and $T_1 T =\pm T T_1$. Assuming that
the time-reversal operators have been transferred to commute with $C$
in this way, the relation $T_1 T = \pm T T_1$ is automatic and further
scaling does not change the sign. Furthermore, the $\mathbb{C}$-linear
isomorphism $\psi : R \to R^*$ can be chosen to meet the compatibility
conditions which determine the conjugation rules (\ref{nonmixing
conjugation}) and (\ref{mixing conjugation}).
\end{prop}
\begin{proof}
It remains to prove that $\psi$ can be chosen as stated.  For the
nonmixing operator $T_1$ the compatibility condition is $\beta_2 \psi
\beta_1^{-1} = \psi$.  Given some choice of $\psi$ (which we will
modify) there is a constant $c \in \mathbb{C}$ so that $\beta_2 \psi
\beta_1^{-1} = c \psi$.  This constant $c$ is unimodular since
$\beta_1$ and $\beta_2$ are antiunitary. To satisfy the compatibility
condition, replace $\psi$ by $\xi \psi$, where $\bar\xi \xi^{-1} c =
1$.  Note that this choice of $\xi$ only determines its argument.

Turning to the compatibility condition $\beta \psi^{-1} \beta =
\varepsilon_\beta \psi$ for the mixing operator $T$, we start from $c
\psi = \beta \psi^{-1}\beta$ for some other $c \in \mathbb{C}$, and
use the $\mathbb {C}$-antilinearity of $\beta$ to deduce $\psi^{-1} =
\bar{c} \, \beta^{-1} \psi \beta^{-1}$.  Multiplying expressions gives
$c = \bar c \in \mathbb{R}$. Then, rescaling $\psi$ to $\xi \psi$, the
compatibility condition is achieved by setting $\varepsilon_\beta := c
/ \vert c \vert$ and solving $\vert \xi \vert^2 = \vert c \vert$.
Since this rescaling (with $\xi \in \mathbb{R}$) does not affect the
compatibility condition for the nonmixing operator, we have determined
the desired isomorphism $\psi$.
\end{proof}
Finally, since $C$ is a pure tensor, it follows from our
representation of the transferred bilinear form $b$ that $c b(C h_1 ,
h_2) = \langle h_1 , h_2 \rangle$ for some constant $c$.  Thus we
replace $b$ by $cb$ and obtain the following final transferred setup
on $H\oplus H^*$:
\begin{itemize}
\item[$\bullet$] The canonical bilinear form $b$ which is either
symmetric or alternating.
\item[$\bullet$] A unitary structure $\langle \, , \,\rangle$ which is
compatible with $b$ in the sense that $b(C h_1 , h_2) = \langle h_1 ,
h_2 \rangle$.  The operator $C : H \leftrightarrow H^\ast$ satisfies
either $C^2 = \mathrm{Id}$ or $C^2 = - \mathrm{Id}$, depending on $b$
being symmetric or alternating.
\item [$\bullet$] Either zero, one, or two time-reversal operators.
They are antiunitary and commute with $C$. In the case of two, $T$ is
mixing and $T_1$ is nonmixing. In the case of one, both mixing and
nonmixing are allowed. The same involutory properties hold as before
transfer, but signs might change, i.e., if $T^2 = \mathrm{Id}$ holds
before transfer, then it is possible that $T^2 = - \mathrm{Id}$
afterwards.  Furthermore, $T_1 T = \pm T T_1$, and consequently the
unitary product $P := T T_1$ satisfies $P^2 = \pm \mathrm{Id}$.
\end{itemize}
In the following sections all of the symmetric spaces which occur in
our basic model will be described, using the transferred setup.  This
means that we describe the subspace of Hermitian operators in
$\mathrm{End}(H) \oplus \mathrm{End}(H^*)$ or $\mathrm{End}(H \oplus
H^*)$ which are compatible with $b$ and the $T$-symmetries. We first
handle the case of one or no time-reversal operator (Sect.\ \ref{sect:
classify}), and then carry out the classification when both $T$ and
$T_1$ are present (Sect.\ \ref{sect: classify two}).  The final
classification result, Theorem \ref{rough statement}, then follows.

\section{Classification: at most one distinguished time reversal}
\label{sect: classify}\setcounter{equation}{0}

This section is devoted to giving a precise statement of Theorem
\ref{rough statement} and its proof in the case where at most one
distinguished time-reversal symmetry is present. Combining this with
the results of Sect.\ \ref{sect: reduction}, we obtain a precise
description of the blocks that occur in the model motivated and
described in Sects.\ \ref{sect: intro} and \ref{sect: fermions}.

\subsection{Statement of the main result}
\label{sect: main statement}

Throughout this section, $V$ denotes a finite-dimensional unitary
vector space. The associated space $W = V \oplus V^*$ is equipped
with the canonically induced unitary structure $\langle \, , \,
\rangle$ and $\mathbb{C}$-antilinear map $C : V \to V^\ast$, $v
\mapsto \langle v , \cdot \rangle$. The results of the previous
section allow us to completely eliminate $G_0$ from the discussion
so that it is only necessary to consider the following data:
\begin{itemize}
\item[$\bullet$] \emph{The relevant space $E$ of endomorphisms.}  This
is either the full space $\mathrm{End}(W)$ or $\mathrm{End} (V) \oplus
\mathrm{End}(V^*)$ embedded as usual in $\mathrm{End}(W)$.
\item[$\bullet$] \emph{The canonical complex bilinear form} $b : W
\times W \to \mathbb{C}$. This is either the symmetric form $s$
which is given by
\begin{displaymath}
    s(v_1 + f_1, v_2 + f_2) = f_1(v_2) + f_2(v_1) \;,
\end{displaymath}
or the alternating form $a$ which is given by
\begin{displaymath}
    a(v_1 + f_1, v_2 + f_2) = f_1(v_2) - f_2(v_1) \;.
\end{displaymath}
Equivalently, $C : V \to V^\ast$ is extended to a
$\mathbb{C}$-antilinear mapping $C : W \to W$ by $C^2 = +
\mathrm{Id}$ resp.\ $C^2 = - \mathrm{Id}$, and $b(Cw_1 , w_2) =
\langle w_1 , w_2 \rangle$ holds in all cases.
\item[$\bullet$] \emph{The antiunitary mapping} $T : W \to W$, which
satisfies either $T^2 = -\mathrm{Id}$ or $T^2 = \mathrm {Id}$. We say
that $T$ is nonmixing if $T \vert_V: V \to V$ and $T \vert_{V^*} : V^*
\to V^*$. If $T\vert_V : V \to V^*$, then we refer to $T$ as
mixing. In all cases $T$ commutes with $C$. We also include the case
where $T$ is not present.
\end{itemize}
Fixing one of these properties each, we refer to $(V,E,b,T)$ as
{\it block data}; e.g., $E = \mathrm{End} (W)$, $b = s$, $T^2 = -
\mathrm{Id}$ and $T$ being nonmixing would be such a choice.

Our main result describes the symmetric spaces associated to given
block data. Let us state this at the Lie algebra level, where for
convenience of formulation we only consider the case of trace-free
operators.  In order to state this result, it is necessary to
introduce some notation.

Given block data $(V,E,b,T)$, let $\mathfrak{g}$ be the subspace
of $E$ of antihermitian operators $A$ which are compatible with
$b$ in the sense that
\begin{displaymath}
    b(A w_1 , w_2) + b(w_1 ,A w_2) = 0
\end{displaymath}
for all $w_1 , w_2 \in W$.  It will be shown that $\mathfrak{g}$
is a Lie subalgebra of $E$ which is invariant under conjugation $A
\mapsto T A T^{-1}$ with $T$.  This defines a Lie algebra
automorphism
\begin{displaymath}
    \theta :{\mathfrak g}\to {\mathfrak g} \;, \quad
    A \mapsto T A T^{-1} \;,
\end{displaymath}
which is usually called a {\it Cartan involution}. If $\mathfrak{
k} := \mathrm{Fix}(\theta ) = \{ A \in \mathfrak{g} : \theta(A) =
A \}$ and $\mathfrak{p}$ is the space $\{ A \in \mathfrak{g} :
\theta(A) = - A \}$ of antifixed points, then
\begin{displaymath}
    \mathfrak{g} = \mathfrak{k} \oplus \mathfrak{p}
\end{displaymath}
is called the associated {\it Cartan decomposition}.

The space $\mathcal{H} = \mathcal{H}(V,E,b,T)$ of Hermitian
operators which are compatible with the block data is $\mathrm{i}
\mathfrak{p}$, which is identified with the infinitesimal version
$\mathfrak{p} = \mathfrak{g} / \mathfrak{k}$.

In order to give a smooth statement of our classification result, we
recall that the Lie algebras $\mathfrak{su}_n$, $\mathfrak{usp}_{2n}$,
and $\mathfrak{so}_{2n}$ are commonly referred to as being of type
$A$, $C$, and $D$, respectively.  By an irreducible $ACD$-symmetric
space of compact type one means an (irreducible) compact symmetric
space of any of these Lie algebras.  With a slight exaggeration we use
the same terminology in Theorem \ref{precise statement} below. The
exaggeration is that the case $\mathfrak{so} _{2n} / (\mathfrak{so}_p
\oplus \mathfrak {so}_q)$ with $p$ and $q$ odd must be excluded in
order for that theorem to be true.  For the overall statement of
Theorem \ref{rough statement} there is no danger of misinterpretation,
as the case where $p$ and $q$ are odd does occur in the situation
where two distinguished time-reversal symmetries are present (see
Sect.\ \ref{sect: classify two}).
\begin{thm}\label{precise statement}
Given block data $(V,E,b,T)$, the space $\mathcal{H} = \mathcal{
H} (V,E,b,T) \cong \mathfrak{g} / \mathfrak{k}$ is the
infinitesimal version of an irreducible $ACD$-symmetric space of
compact type. Conversely, the infinitesimal version of any
irreducible $ACD$-symmetric space of compact type can be
constructed in this way.
\end{thm}
There are several remarks which should be made concerning this
statement. First, as we have already noted, in order to give a
smooth formulation, we have reduced to trace-free operators.  As
will be seen in the proof, there are several cases where without
this assumption $\mathfrak{g}$ would have a one-dimensional
center.

Secondly, recall that one of the important cases of a compact
symmetric space is that of a compact Lie group $K$ with the
geodesic inversion symmetry at the identity being defined by $k
\mapsto k^{-1}$. Usually one equips $K$ with the action of $G = K
\times K$ defined by left- and right-multiplication, and views the
symmetric space as $G/K$, where the isotropy group $K$ is
diagonally embedded in $G$. The infinitesimal version is then
$(\mathfrak{k} \oplus \mathfrak{k}) / \mathfrak{k}$, and the
automorphism $\theta : \mathfrak{g} \to \mathfrak{g}$ is defined
by $(X_1,X_2) \mapsto (X_2,X_1)$.  In this setting one speaks of
symmetric spaces of type II.

In our case the classical compact Lie algebras do indeed arise
from appropriate block data, but in the situation where $T$ does
not leave the original space $W$ invariant. In that setting, $T$
maps $W = W_1^{\vphantom{*}} = V_1^{\vphantom{*}} \oplus V_1^*$ to
$W_2^{\vphantom{*}} = V_2^{\vphantom{*}} \oplus V_2^*$, which has
different $G_0$-representations from those in $W_1$. Thus the
relevant block is $W_1 \oplus W_2$. Using the results of the
previous section, in this case we also remove $G_0$ from the
picture.

Nevertheless, we are left with a situation where the block is $W_1
\oplus W_2$ and $T : W_1 \to W_2$. Thus we wish to allow situations of
this type, i.e., where $V \oplus V^*$ is not $T$-invariant, to be
allowed block data. These cases are treated separately in Sect.\
\ref{sect: typeII}.

The case where the symmetric space is just the compact group
associated to $\mathfrak{g}$ also arises when $T$ is not present,
i.e., when there is no condition which creates isotropy.

Finally, as has already been indicated in Sect.\ \ref{sect:
intro}, the appropriate homogeneous space version of Theorem
\ref{precise statement} is given by replacing the infinitesimal
symmetric space $\mathfrak{g} / \mathfrak{k}$ by the
Cartan-embedded symmetric space $M \cong G / K$. Here $G$ is the
simply connected group associated to $\mathfrak{g}$, a mapping
$\theta : G \to G$ is defined as the Lie group automorphism whose
derivative at the identity is the Cartan involution of the Lie
algebra, and $M$ is the orbit of $e \in G$ of the twisted
$G$-action given by $x \mapsto g x \, \theta(g)^{-1}$.

\subsection{The associated symmetric space}
\label{sect: symm space}

In this and the next subsection we work in the context of simple
block data $(V,E,b,T)$ where $W = V\oplus V^*$ is $T$-invariant.

In the present subsection we prove the first half of Theorem
\ref{precise statement}, namely that $\mathcal{H} \cong
\mathfrak{g} / \mathfrak{k}$ is an infinitesimal version of a
classical symmetric space of compact type.  This essentially
amounts to showing that all the involutions which are involved
commute.

Let $\sigma : E \to E$ be the $\mathbb{C}$-antilinear Lie-algebra
involution that fixes the Lie algebra of the unitary group in $E$.
If the adjoint operation $A\mapsto A^*$ is defined by
\begin{displaymath}
    \langle A w_1 , w_2 \rangle = \langle w_1 , A^* w_2 \rangle \;,
\end{displaymath}
then $\sigma(A) = -A^*$. The transformations $S \in E$ which are
isometries of the canonical bilinear form satisfy
\begin{displaymath}
    b(S w_1 , S w_2) = b(w_1 , w_2)
\end{displaymath}
for all $w_1, w_2 \in W$.  Thus the appropriate Lie algebra
involution is the $\mathbb C$-linear automorphism
\begin{displaymath}
    \tau : E \to E \;, \quad A \mapsto -A^\mathrm{t} \;,
\end{displaymath}
where $A \mapsto A^\mathrm{t}$ is the adjoint operation defined by
\begin{displaymath}
    b(A w_1 , w_2) = b(w_1, A^\mathrm{t} w_2) \;.
\end{displaymath}
Finally, let $\theta : E \to E$ be the $\mathbb{C}$-antilinear map
defined by $A \mapsto T A T^{-1}$.
\begin{prop} \label{adjoint formulas}
The operations $A \mapsto A^\ast$ and $A \mapsto A^\mathrm{t}$ are
related by $A^\ast = C A^\mathrm{t} C^{-1}$.
\end{prop}
\begin{proof}
From $b(C w_1 , w_2) = \langle w_1 , w_2 \rangle$ and the
definition of $A \mapsto A^\ast$ we have
\begin{displaymath}
    b(A w_1 , w_2) =  \langle C^{-1} A w_1 , w_2 \rangle =
    \langle C^{-1} w_1 , (C^{-1} A C)^\ast w_2 \rangle =
    b(w_1 , C^{-1} A^* C w_2) \;,
\end{displaymath}
i.e., $A^\mathrm{t} = C^{-1} A^* C$, independent of the case $b =
s$ or $b = a$.
\end{proof}
\begin{prop}
The involutions $\sigma$, $\tau$ and $\theta$ commute.
\end{prop}
\begin{proof}
Using $\langle C^{-1} A^* C w_1 , w_2 \rangle = \langle w_1 ,
C^{-1} A C w_2 \rangle$ along with $A^\mathrm{t} = C^{-1} A^\ast
C$ we have
\begin{displaymath}
    (A^\mathrm{t})^* = C^{-1} A C = (A^*)^\mathrm{t} \;,
\end{displaymath}
and consequently $\sigma\tau = \tau\sigma $.

Since $T$ is antiunitary, one immediately shows from the
definition of $A^*$ that
\begin{displaymath}
    \langle w_1 , T A^* T^{-1} w_2 \rangle
    = \langle T A T^{-1} w_1 , w_2 \rangle \;.
\end{displaymath}
In other words,
\begin{displaymath}
    \theta(\sigma(A)) = - T A^* T^{-1} =
    - (TAT^{-1})^* = \sigma(\theta(A)) \;.
\end{displaymath}

Finally, since $\theta(A) = T A T^{-1}$ and $T$ commutes with $C$,
it follows that $\theta\tau = \tau\theta$.
\end{proof}
Let $\mathfrak{s} := \mathrm{Fix}(\tau)$. Since $\theta$ and
$\sigma$ commute with $\tau$, it follows that they restrict to
$\mathbb{C}$-antilinear involutions of the complex Lie algebra
$\mathfrak{s}$.  We denote these restrictions by the same letters.
For future reference let us summarize the relevant formulas.
\begin{prop} \label{involution formulas}
For $A \in \mathfrak{s}$ it follows that
\begin{displaymath}
    \sigma(A) = C A C^{-1} \quad \text{and}
    \quad \theta(A) = T A T^{-1}.
\end{displaymath}
The parity of $C$ is $C^2 = + \mathrm{Id}$ for $b = s$ symmetric,
and $C^2 = - \mathrm{Id}$ for $b = a$ alternating.
\end{prop}
The space $\mathfrak{g}$ of antihermitian operators in $E$ that
respect $b$ is therefore the Lie algebra of $\sigma$-fixed points
in $\mathfrak{s}$.  Since $\sigma$ defines the unitary Lie algebra
in $E$, it follows that $\mathfrak{g}$ is a compact real form of
$\mathfrak{s}$. Let us explicitly describe $\mathfrak{s}$ and
$\mathfrak{g}$.

If $E = \mathrm{End}(W)$ and $b = s$ is symmetric, then $\mathfrak
{s}$ is the complex orthogonal Lie algebra {$\mathfrak{ so}(W,s)
\cong \mathfrak{so}_{2n}(\mathbb{C})$. If $E = \mathrm{End} (W)$
and $b = a$ is alternating, then $\mathfrak{s}$ is the complex
symplectic Lie algebra $\mathfrak{sp}(W,a) \cong \mathfrak{sp}_
{2n} (\mathbb C)$. If $E = \mathrm{End}(V) \oplus \mathrm {End}
(V^*)$, then in both cases for $b$ it follows that its isometry
group is $\mathrm{SL}_\mathbb{C} (V)$ acting diagonally by its
defining representation on $V$ and its dual representation on
$V^*$.  In this case we have $\mathfrak{s} = \mathfrak{sl}(V)
\cong \mathfrak{sl}_n (\mathbb{C})$. Note that this is a situation
where we have used the trace-free condition to eliminate the
one-dimensional center.

For the discussion of $\mathfrak{g}$ it is important to note that
since $\sigma(A) = C A C^{-1}$,  it follows that $\mathfrak{g}$
just consists of the elements of $\mathfrak{s}$ which commute with
$C$.

In the symmetric case $b = s$, where $C$ defines a real structure on
$W$, it is appropriate to consider the set of real points
$W_\mathbb{R} = \{v + C v : v \in V \}$. Thinking in terms of
isometries, we regard $G = \exp(\mathfrak{g})$ as being the group of
$\mathbb {R}$-linear isometries of the restriction of $b = s$ to
$W_\mathbb{R}$ which are extended complex linearly to $W$. Note that
in this case $b\vert_{W_\mathbb{R}} = 2\, \mathrm{Re}\, \langle \, ,
\, \rangle$, and that every $\mathbb R$-linear transformation of
$W_\mathbb{R}$ which preserves $\mathrm{Re}\, \langle \, ,\, \rangle$
extends $\mathbb{C}$-linearly to a unitary transformation of
$W$. Thus, if $E = \mathrm{End}(W)$ and $b = s$, then $\mathfrak{g}$
is naturally identified with $\mathfrak{so} (W_\mathbb{R},
s\vert_{W_\mathbb{R}}) \cong \mathfrak{so}_{2n} (\mathbb{R})$.

In the alternating case $b = a$, if $E = \mathrm{End}(W)$, then as
in the previous case, since $\sigma$ defines $\mathfrak{u}(W)
\subset E$, it follows that its set $\mathfrak{g}$ of fixed points
in $\mathfrak{s}$ is a compact real form of $\mathfrak{s}$. Since
$\mathfrak{s}$ is the complex symplectic Lie algebra $\mathfrak{
sp} (W,a)\cong \mathfrak{sp}_{2n}(\mathbb{C})$, it follows that
$\mathfrak{g}$ is isomorphic to the Lie algebra $\mathfrak{usp
}_{2n}$ of the unitary symplectic group.

It is perhaps worth mentioning that $C$ for $b = a$ defines a
quaternionic structure on the complex vector space $W$. Thus the
condition $A = C A C^{-1}$ defines the subalgebra $\mathfrak{gl}_{
n} (\mathbb{H})$ in $\mathrm{End}(W)$. The further condition $A =
- A^*$ shows that $\mathfrak{g}$ can be identified with the
algebra of quaternionic isometries, another way of seeing that
$\mathfrak{g} \cong \mathfrak{usp}_{2n}$.

Finally, in the case where $E = \mathrm{End}(V) \oplus
\mathrm{End}(V^*)$ we have already noted that $\mathfrak{s} =
\mathfrak{sl}(V)$ which is acting diagonally.  It is then immediate
that in both the symmetric and alternating cases $\mathfrak{g} =
\mathfrak{su}(V) \cong \mathfrak{su}_n$. Of course $\mathfrak{g}$ acts
diagonally as well.

Let us summarize these results.
\begin{prop}\label{g/k}
In the case where $E = \mathrm{End}(W)$ the following hold:
\begin{itemize}
\item[$\bullet$] If $b = s$ is symmetric, then $\mathfrak{g} \cong
\mathfrak{so}_{2n}(\mathbb{R})$.
\item[$\bullet$] If $b = a$ is alternating, then $\mathfrak{g}
\cong \mathfrak{usp}_{2n}$.
\end{itemize}
If $E = \mathrm{End}(V) \oplus \mathrm{End}(V^*)$, then
$\mathfrak{g}$ is isomorphic to $\mathfrak{su}_{n}$ and acts
diagonally.
\end{prop}
Since $\theta$ commutes with $\sigma$, it stabilizes
$\mathfrak{g}$. Hence, $\theta \vert_\mathfrak{g}$ is a Cartan
involution which defines a Cartan decomposition
\begin{displaymath}
    \mathfrak{g} = \mathfrak{k} \oplus \mathfrak{p}
\end{displaymath}
of $\mathfrak{g}$ into its $(\pm 1)$-eigenspaces. The fixed
subspace $\mathfrak{k} = \{ A \in \mathfrak{g}: \theta(A) = A \}$
is a subalgebra and $\mathfrak{g} / \mathfrak{k}$ is the
infinitesimal version of a symmetric space of compact type.

Recall that, given block data $(V,E,b,T)$, the associated space
\begin{displaymath}
    \mathcal{H} = \mathcal{H}(V,E,b,T)
    \cong \mathrm{i} \mathfrak{p}
\end{displaymath}
of structure-preserving Hamiltonians has been identified with
$\mathfrak{p} = \mathfrak{g} / \mathfrak{k}$. Thus we have proved
the first part of Theorem \ref{precise statement}. The second part
is proved in the next section by going through the possibilities
in Prop.\ \ref{g/k} along with the various possibilities for $T$.

It should be noted that if $T = \pm C$, then $\mathfrak{g} =
\mathfrak{k}$, i.e., the symmetric space is just a point. Such a
degenerate situation, where the set of Hamiltonians is trivial
(consisting only of the zero Hamiltonian), never occurs in a
well-posed physics setting.

\subsection{Concrete description: symmetric spaces of type I}
\label{sect: typeI}

Here we describe the possibilities for each set of block data
$(V,E,b,T)$ under the assumption that $W = V \oplus V^*$ is
$T$-invariant. The results are stated in terms of the
$ACD$-symmetric spaces, with $n := \mathrm{dim}_\mathbb{C} V$. The
methods of proof of showing which symmetric spaces arise also show
how to explicitly construct them. In the present subsection, all
of these are compact irreducible classical symmetric spaces of
type I in the notation of \cite{helgason}.

\subsubsection{The case $E = \mathrm{End}(V) \oplus
\mathrm{End}(V^*)$}\label{sect: 4.3.1}

Under the assumption $E = \mathrm{End}(V) \oplus \mathrm{End}
(V^*)$ it follows that $\mathfrak{g}$ is just the unitary Lie
algebra $\mathfrak{su}(V) \cong \mathfrak{su}_n$ which is acting
diagonally on $W = V \oplus V^*$. This is independent of $b$ being
symmetric or alternating. Thus we need only consider the various
possibilities for $T$. If $T$ is not present, the symmetric space
is $\mathfrak{g} = \mathfrak{su}_n$.
\begin{enumerate}
\item $T^2 = - \mathrm{Id}$, nonmixing:
$\mathfrak{su}_{n} / \mathfrak{usp}_n$. \\
Since $T$ is nonmixing and satifies $T^2 = - \mathrm{Id}$, it
follows that $\langle Tv_1 , v_2\rangle = a(v_1,v_2)$ is a
$\mathbb{C} $-linear symplectic structure on $V$ which is
compatible with $\langle v_1 , v_2 \rangle$. Thus the dimension
$n$ of $V$ must be even here. The facts that $\mathfrak{g}$ is
acting diagonally as $\mathfrak{su} (V)$ and that the elements of
$\mathfrak{k}$ are precisely those which commute with $T$, imply
that $\mathfrak{k} = \mathfrak{usp}_n$ as announced.
\hfill\vspace{0.2cm}
\item $T^2 = \mathrm{Id}$, nonmixing:
$\mathfrak{su}_n / \mathfrak{so}_n$. \\
Since $T$ and $\mathfrak{g}$ are acting diagonally, as in the
previous case it is enough to only discuss the matter on $V$. In
this case $T$ defines a real structure on $V$ with $V_\mathbb{R} =
\{v + Tv : v \in V \}$, and the unitary isometries which commute
with $T$ are just those transformations which stabilize $V_
\mathbb{R}$ and preserve the restriction of $\langle \, , \,
\rangle$. Since $\langle x , y \rangle_{V_\mathbb{R}} = \mathrm{Re
} \, \langle x , y \rangle_V$ for $x,y \in V_\mathbb{R}$, it
follows that $\mathfrak{k} = \mathfrak{so}(V_\mathbb{R}) \cong
\mathfrak{so}_n(\mathbb{R})$. \hfill\vspace{0.2cm}
\item $T^2 = \pm \mathrm{Id}$, mixing: $\mathfrak{su}_n /
\mathfrak{s} (\mathfrak{u}_{p} \oplus \mathfrak{u}_{q})$.\\
Here it is convenient to introduce the unitary operator $P = CT$,
which satisfies $P^2 = \mathrm{Id}$ or $P^2 = - \mathrm{Id}$,
depending on the parity of $T$. Denote the eigenvalues of $P$ by
$u$ and $-u$. Since $P$ does not mix, the condition that a
diagonally acting unitary operator commutes with $T$ (or
equivalently, with $P$) is just that it preserves the
$P$-eigenspace decomposition $V = V_u \oplus V_{-u}$. Since the
two eigenspaces $V_u$ and $V_{-u}$ are $\langle \, , \, \rangle
$-orthogonal, we have $\mathfrak{k} = \mathfrak{s} \left(
\mathfrak{u} (V_u) \oplus \mathfrak{u} (V_{-u}) \right)$, and the
desired result follows with $p = \mathrm{dim}\, V_u$ and $q =
\mathrm{dim}\, V_{-u}$. \\
In the case $P^2 = - \mathrm{Id}$, if there existed a subspace
$V_\mathbb{R}$ of real points that was stabilized by $P$, then $P$
would be a complex structure of $V_\mathbb{R}$ and the dimensions
of $V_u$ and $V_{-u}$ would have to be equal. In general, however,
no such space $V_\mathbb{R}$ exists and the dimensions $p$ and $q$
are arbitrary. \hfill\vspace{0.2cm}
\end{enumerate}

\subsubsection{The case $E = \mathrm{End}(W)$, $b = s$}
\label{sect: 4.3.2}

In this case we have the advantage that we may restrict the entire
discussion to the set of real points
\begin{displaymath}
    W_\mathbb{R} = \mathrm{Fix}(C) = \{v + C v : v \in V \} \;.
\end{displaymath}
Thus $\mathfrak{k}$ is translated to being the Lie algebra of the
group of isometries of $2\, \mathrm{Re}\, \langle \, , \, \rangle$
on $V$. Here the Lie algebra $\mathfrak{g}$ is
$\mathfrak{so}(W_\mathbb{R})$. Thus in the case where $T$ is not
present, the symmetric space is $\mathfrak{so}_{2n}(\mathbb{R})$.
\begin{enumerate}
\item $T^2 = - \mathrm{Id}$, nonmixing or mixing:
$\mathfrak{so}_{2n}(\mathbb{R}) / \mathfrak{u}_n$. \\
Independent of whether or not it mixes, $T\vert_{W_\mathbb{R}} :
W_\mathbb{R} \to W_\mathbb{R}$ is a complex structure on
$W_\mathbb{R}$. A transformation in $\mathrm{SO}(W_\mathbb{R})$
commutes with $T$ if and only if it is holomorphic. Since
$\mathrm{Re}\, \langle \, , \, \rangle$ is $T$-invariant, this
condition defines the unitary subalgebra $\mathfrak{k} \cong
\mathfrak{u}_n$ in $\mathfrak{g} \cong \mathfrak{so}_{2n}(
\mathbb{R})$. \hfill\vspace{0.2cm}
\item $T^2 = \mathrm{Id}$, nonmixing: $\mathfrak{so}_{2n}
(\mathbb{R}) / (\mathfrak{so}_n(\mathbb{R}) \oplus \mathfrak{so}_n
(\mathbb{R}))$.\\
Since $T\vert_{W_\mathbb{R}} : W_\mathbb{R} \to W_\mathbb{R}$, we
have the decomposition $W_\mathbb{R} = W^+_\mathbb{R} \oplus
W^-_\mathbb{R}$ into the $(\pm 1)$-eigenspaces of $T$.  We still
identify $\mathfrak{g}$ with the Lie algebra of the group of
isometries of $W_\mathbb{R}$ equipped with the restricted form
$\mathrm{Re}\, \langle \, , \, \rangle$.  The subalgebra
$\mathfrak{k}$, which is fixed by $\theta : X \mapsto T X T^{-1}$,
is the stabilizer $\mathfrak{so}(W^+_\mathbb{R}) \oplus
\mathfrak{so} (W^-_\mathbb{ R})$ of the above decomposition. Now
let us compute the dimensions of the eigenspaces. In the case at
hand $T$ defines a real structure on both $V$ and $V^*$.  Since
$C$ commutes with $T$, it follows that $\mathrm{Fix}(T) =
V_\mathbb{R}^{\vphantom{*}} \oplus V^*_\mathbb{R}$ is
$C$-invariant. Thus $W_\mathbb{R}^+ = \{ v + Cv : v \in
V_\mathbb{R} \}$.  A similar argument shows that $W_\mathbb{R}^- =
\{ v + Cv : v \in \mathrm{i} V_\mathbb{R}\}$. \hfill\vspace{0.2cm}
\item $T^2 = \mathrm{Id}$, mixing: $\mathfrak{so}_{2n}(\mathbb{R})
/ (\mathfrak{so}_{2p} (\mathbb{R}) \oplus \mathfrak{so}_{2q}
(\mathbb{R}))$. \\
The exact same argument as above shows that $\mathfrak{k} =
\mathfrak{so}(W^+_\mathbb{R}) \oplus \mathfrak{so}(W^-_\mathbb{R}
)$. It only remains to show that the eigenspaces are
even-dimensional. For this we consider the unitary operator $P =
CT$ which leaves both $V$ and $V^*$ invariant. Its
$(+1)$-eigenspace $W_{+1}$ is just the complexification of
$W^+_\mathbb{R}$.  The intersections of $W_{+1}$ with $V$ and
$V^*$ are interchanged by $C$, and therefore $\mathrm{dim}_
\mathbb{C} W_{+1} = :2p$ is even.  Of course the same argument
holds for $W_{-1}$.
\end{enumerate}

\subsubsection{The case $E = \mathrm{End}(W)$, $b = a$}
\label{sect: 4.3.3}

Since in this case $\mathfrak{g}$ is the Lie algebra of
antihermitian endomorphisms which respect the alternating form $a$
on $W$, it follows that $\mathfrak{g} \cong \mathfrak{usp}_{2n}$.
Thus if $T$ is not present the associated symmetric space is
$\mathfrak{usp}_{2n}$.

If $T$ is present, we let $P := CT$.  The unitary operator $P$
always commutes with $T$, and from $a(w_1 , w_2) = \langle C^{-1}
w_1 , w_2 \rangle$ one infers that $a(P w_1, P w_2) = a(w_1 ,
w_2)$ in all cases, independent of $T$ being mixing or not.

The classification spelled out below follows from the fact that
commutation with $T$ is equivalent to preservation of the
$P$-eigenspace decomposition of $W$.
\begin{enumerate}
\item $T^2 = - \mathrm{Id}$, nonmixing: $\mathfrak{usp}_{2n} /
(\mathfrak{usp}_n \oplus \mathfrak{usp}_n)$. \\
In this case $P^2 = \mathrm{Id}$, and $T^2 = - \mathrm{Id}$ forces
$n$ to be even. Let $W$ be decomposed into $P$-eigen\-spaces as $W
= W_{+1} \oplus W_{-1}$. If $w_1 \in W_{+1}$ and $w_2 \in W_{-1}$,
then
\begin{displaymath}
    a(w_1 , w_2) = a(P w_1 , P w_2) = - a(w_1 , w_2) = 0 \;,
\end{displaymath}
and we see that $W_{+1}$ and $W_{-1}$ are $a$-orthogonal. The
mixing operator $P$ is traceless. Therefore the dimensions of
$W_{+1}$ and $W_{-1}$ are equal, and both of them are symplectic
subspaces of $W$. The fact that the decomposition $W = W_{+1}
\oplus W_{-1}$ is also $\langle \, , \, \rangle$-orthogonal
therefore implies that $\mathfrak{k} = \mathfrak{usp}(W_{+1})
\oplus \mathfrak{usp}(W_{-1})$. \hfill\vspace{0.2cm}
\item $T^2 = -\mathrm{Id}$, mixing: $\mathfrak{usp}_{2n} /
(\mathfrak{usp}_{2p} \oplus \mathfrak{usp}_{2q})$. \\
Here, using the same argument as in the previous case, one shows
that the $P$-eigenspace decomposition $W = W_{+1} \oplus W_{-1}$
still is a direct sum of $a$-orthogonal, complex symplectic
subspaces. Since these are also $\langle\,,\,\rangle$-orthogonal,
it follows that $\mathfrak{k} = \mathfrak{usp}(W_{+1}) \oplus
\mathfrak{usp} (W_{-1})$. Note that in the present case the
nonmixing operator $P$ stabilizes the decomposition $W = V \oplus
V^*$. Thus, since $P$ commutes with $C$, it follows that $W_{+1}^{
\vphantom{*}} = V_{+1}^{\vphantom {*}} \oplus V^*_{+1}$ and
$W_{-1}^{\vphantom{*}} = V_{-1}^{ \vphantom{*}} \oplus V^*_{-1}$.
\hfill\vspace{0.2cm}
\item $T^2 = \mathrm{Id}$, mixing or nonmixing:
$\mathfrak{usp}_{2n} / \mathfrak{u}_n$. \\
In this case $P^2 = -\mathrm{Id}$. Here $a(P w_1 ,P w_2) = a(w_1 ,
w_2)$ implies that the $P$-eigenspace decomposition $W = W_{+
\mathrm{i}} \oplus W_{- \mathrm{i}}$ is Lagrangian. (This means in
particular $\mathrm{dim}\, W_{+\mathrm{i}} = \mathrm{dim}\,
W_{-\mathrm{i}}$.) Thus its stabilizer in $\mathfrak{sp }(W)$ is
the diagonally acting $\mathfrak{gl} (W_{+\mathrm{i}})$. Since the
decomposition is $\langle \, , \, \rangle$-orthogonal, it follows
that $\mathfrak{k} = \mathfrak{u}(W_{+\mathrm{i}}) \cong
\mathfrak{u}_n$. \hfill\vspace{0.2cm}
\end{enumerate}

\subsection{Concrete description: symmetric spaces of type II}
\label{sect: typeII}

Recall the original situation where the symmetry group $G_0$ is
still in the picture.  As described in Sect.\ \ref{sect: intro} we
select from the given Hilbert space a basic finite-dimensional
$G_0$-invariant subspace $V$ which is composed of irreducible
subrepresentations all of which are equivalent to a fixed
irreducible representation $R$.

Although the initial block of interest is $W = V \oplus V^*$, it
is possible that it is not $T$-invariant and that it must be
expanded. Let us formalize this situation by denoting the initial
block by $W_1^{\vphantom{*}} = V_1^{\vphantom{*}} \oplus V_1^*$.
We then let $P = CT$ and regard this as a unitary isomorphism $P :
W_1 \to W_2$, where $W_2^{\vphantom{*}} = V_2^{\vphantom{*}}
\oplus V_2^*$ is another initial block.

For $i \in \{ 1, 2\}$, let $R_i$ be the irreducible
$G_0$-representation on $V_i$ which induces the representation on
$W_i$. The map $P$ is equivariant, but only with respect to the
automorphism $a$ of $G_0$ which is defined by $g_T$-conjugation:
$P \circ g = a(g) \circ P$.

As a brief interlude, let us investigate the consequences of this
automorphism $a$ being inner versus outer.  If $a$ is inner there
exists $A \in G_0$ such that $a(g) = A^{-1} g A$ and hence $AP
\circ g = g \circ AP$ for all $g \in G_0$.  Thus $AP : W_1 \to
W_2$ is a $G_0$-equivariant isomorphism and we have either $R_1
\cong R_2$ or $R_1^{\vphantom{\ast}} \cong R_2^\ast$ depending on
whether $T$ is mixing or not.  In either case we may build a new
block $W = V \oplus V^*$ which \emph{is} $T$-invariant so that the
results of the previous section can be applied: if $R_1\cong R_2$,
then we let $V := V_1 \oplus V_2$ and if $R_1^{\vphantom{*}} \cong
R_2^*$, then $V := V_1^{\vphantom{*}} \oplus V_2^*$.

If the $G_0$-automorphism $g \mapsto a(g)$ is outer, it may still
happen that $R_1 \cong R_2$ or $R_1^{\vphantom{\ast}} \cong
R_2^\ast$, and then we may still build a new block $W = V \oplus
V^\ast$ and apply the previous results.

We assume now that neither $R_1\cong R_2$ nor $R_1^{\vphantom{*}}
\cong R_2^*$, and consider the expanded block $W = W_1\oplus W_2$.
Recall that $W_1$ and $W_2$ are in the Nambu space $\mathcal{W}$
which decomposes as a direct sum of nonisomorphic representation
spaces that are orthogonal with respect to both the unitary
structure and the canonical symmetric form. Thus the decomposition
$W = W_1 \oplus W_2$ is orthogonal with respect to both of these
structures.

Under the assumption at hand it is immediate that
\begin{displaymath}
    \mathrm{End}_{G_0}(W) = \mathrm{End}_{G_0}(W_1) \oplus
    \mathrm{End}_{G_0}(W_2) \;.
\end{displaymath}
Thus we are in a position to apply the results of Sect.\
\ref{sect: reduction}.

To do so in the case where $R_1^{\vphantom{*}} \cong R_1^*$, we
let $\psi_1^{\vphantom{*}} : R_1^{\vphantom{*}} \to R_1^*$ denote
an equivariant isomorphism, and organize the notation so that $P :
V_1 \to V_2$. Of course $R_1$ and $R_2$ are abstract
representations, but we now choose realizations of them in $V_1$
and $V_2$ so that $\psi_2 := P\, \psi_1 P^{-1} : R_2^{
\vphantom{*}} \to R_2^\ast$ makes sense. Since
\begin{eqnarray*}
    &&P\, \psi_1 P^{-1} (g(v_2)) = P(\psi_1(a^{-1}(g) P^{-1}
    (v_2))) \\ &=& P(a^{-1}(g) \psi_1 (P^{-1}(v_2))) = g (P\,
    \psi_1 P^{-1} (v_2)) \;,
\end{eqnarray*}
it follows that $\psi_2^{\vphantom{*}} : R_2^{\vphantom{*}} \to
R_2^*$ is a $G_0$-equivariant isomorphism.

Assume for simplicity that $\psi_1$ is symmetric, i.e., that
$\psi_1 (v_1) (\tilde{v}_1) = \psi_1(\tilde{v}_1)(v_1)$.  Then
\begin{eqnarray*}
    &&\psi_2(v_2)(\tilde{v}_2) = P\,\psi_1 P^{-1}(v_2)(\tilde{v}_2)
    = \psi_1(P^{-1}(v_2))(P^{-1}(\tilde{v}_2))\\ &=& \psi_1(P^{-1}
    (\tilde{v}_2))(P^{-1}(v_2)) = P\,\psi_1 P^{-1}(\tilde{v}_2)(v_2)
    = \psi_2(\tilde{v}_2)(v_2) \;.
\end{eqnarray*}
The computation in the case where $\psi_1$ is odd is the same
except for a sign change.  Thus $\psi_1$ and $\psi_2$ have the
same parity.

Now let $E_i$ (for $i = 1, 2$) be the relevant space of
endomorphisms that was produced by our analysis of $W_i$ in Sect.\
\ref{sect: reduction}. Recall that this is either the space
$\mathrm{End}(H_i) \oplus \mathrm{End}(H^*_i)$ or $\mathrm{End}
(H_i^{\vphantom{*}} \oplus H_i^*)$.  Let $\mathfrak{g}_i$ be the
Lie algebra of the group of unitary transformations which preserve
$b_i$. The key points now are that the unitary structure on $E :=
E_1 \oplus E_2$ is the direct sum structure, the complex bilinear
form on $E$ is $b = b_1 \oplus b_2$, and the parity of $b_1$ is
the same as that of $b_2$.  Thus $\mathfrak{g}_1 \cong \mathfrak{g
}_2$.

For the statement of our main result in this case, let us recall
that the infinitesimal versions of symmetric spaces of type II are
of the form $\mathfrak{g} \oplus \mathfrak{g} / \mathfrak{g}$,
where the isotropy algebra is embedded diagonally.
\begin{prop} \label{type II classification}
If $R_1$ is neither isomorphic to $R_2$ nor to $R_2^*$, then the
infinitesimal symmetric space associated to the $T$-invariant
block data is a type-II $ACD$-symmetric space of compact type.
Specifically, the classical Lie algebras $\mathfrak{su}_n$,
$\mathfrak{so}_{2n}(\mathbb{R})$, and $\mathfrak{usp}_{2n}$ arise
in this way.
\end{prop}
\begin{proof}
Identify $\mathfrak{g}_1$ and $\mathfrak{g}_2$ by the isomorphism
$P$.  Call the resulting Lie algebra $\mathfrak{g}$. The
transformations that commute with $T$ are those in the diagonal in
$\mathfrak{g} \oplus \mathfrak{g}$. Thus the associated
infinitesimal version of the symmetric space is of type II.  The
fact that the only Lie algebras which occur are those in the
statement has been proved in \ref{sect: symm space}.
\end{proof}
This completes the proof of Theorem \ref{precise statement}. In
closing we underline that under the assumptions of Prop.\
\ref{type II classification} the odd-dimensional orthogonal Lie
algebra does not appear as a type-II space; only the
even-dimensional one does.

\section {Classification: two distinguished time-reversal symmetries}
\label{sect: classify two}\setcounter{equation}{0}

Here we describe in detail the situation where both of the
distinguished time-reversal operators $T$ and $T_1$ are present.  As
would be expected, there are quite a few cases.  The work will be
carried out in a way which is analogous to our treatment of the case
where only one time-reversal operator was present. In the first part
(Sect.\ \ref{sect: smallblocks}) we operate under the assumption that
the initial truncated space $V \oplus V^*$ is invariant under both of
the distinguished operators. In the second part (Sect.\ \ref{sect:
bigblocks}) we handle the general case where bigger blocks must be
considered.

\subsection{The case where $V \oplus V^\ast$ is $G$-invariant}
\label{sect: smallblocks}

Throughout, $T$ is mixing, $T_1$ is nonmixing and $P := T T_1$. Our
strategy in Sects.\ \ref{sect: 5.1.3} and \ref{sect: 5.1.4} will be to
first compute the operators which are $b$-isometries, are unitary and
commute with $P$.  This determines the Lie algebra $\mathfrak{g}$ and
its action on $V \oplus V^*$.  Then $\mathfrak{k}$ is determined as
the subalgebra of operators which commute with $T$ or $T_1$, whichever
is most convenient for the proof. The space of Hamiltonians is
identified with $\mathfrak{g} / \mathfrak{k}$ as before.

In the case of $E = \mathrm{End}(V) \oplus \mathrm{End}(V^*)$, where
$\mathfrak{g}$ acts diagonally, the answer for $\mathfrak{g} /
\mathfrak{k}$ does not depend on the involutory properties of $C$,
$T$, and $T_1$ individually, but only on those of the nonmixing
operators $CP = C T T_1$ and $T_1$.  The pertinent Sects.\ \ref{sect:
5.1.1} and \ref{sect: 5.1.2} are organized accordingly.

\subsubsection{The case $E = \mathrm{End}(V) \oplus \mathrm{End}(V^*)$,
$(CP)^2 = \mathrm{Id}$}\label{sect: 5.1.1}

Recall that in the case of $E = \mathrm{End}(V) \oplus \mathrm{End}
(V^*)$ it follows that the $b$-isometry group is $\mathrm{SL}_
\mathbb{C}(V)$ acting diagonally. Thus the Lie algebra $\mathfrak{g}$
consists of those elements of the unitary algebra $\mathfrak{su}(V)$
which commute with the mixing unitary symmetry $P$.  Equivalently,
$\mathfrak{g}$ is the subalgebra of $\mathfrak{su}(V)$ defined by
commutation with the antiunitary operator $CP$.

In the present case $CP$ defines a real structure on $V$, and we have
the $\mathfrak{g}$-invariant decomposition $V = V_\mathbb{R} \oplus
\mathrm{i} V_\mathbb{R}$. Since the unitary structure $\langle \, , \,
\rangle $ is compatible with this real structure, it follows that
$\mathfrak{g} = \mathfrak{so} (V_\mathbb{R})$.  Our argumentation is
based around $T_1$.  If it anticommutes with $P$, then we replace $P$
by $\mathrm{i} P$ so that it commutes.  Of course this has the effect
of changing to the case $(CP)^2 = - \mathrm{Id}$ which is, however,
handled below. Hence, in both cases we may assume that $P$ and $T_1$
commute.
\begin{enumerate}
\item $T_1^2 = \mathrm{Id}$: $\mathfrak{so}_n / (\mathfrak{so}_p
\oplus \mathfrak{so}_q)$. \\
The space of $CP$-real points $V_\mathbb {R}$ is $T_1$-invariant and
splits into a sum $V^+_\mathbb{R} \oplus V^-_\mathbb{R}$ of $T_1$-
eigenspaces.  The Lie algebra $\mathfrak{k}$ is the stabilizer of this
decomposition, which is $\langle \, , \, \rangle $-orthogonal.  Thus
$\mathfrak{k} = \mathfrak{so}(V^+_\mathbb{R}) \oplus \mathfrak{so}
(V^-_\mathbb{R})$. \\
Observe that in this case $n$ can be any even or odd number and that
$p$ and $q$ are arbitrary with the condition that $n = p + q$.
\hfill\vspace{0.2cm}
\item $T_1^2 = -\mathrm{Id}$: $\mathfrak{so}_{2n} / \mathfrak{u}_n$. \\
In this case $T_1$ is a complex structure on $V_\mathbb{R}$ which is
compatible with the unitary structure.  Thus $\mathfrak{k} =
\mathfrak{u}(V_\mathbb R, T_1)$ and the desired result follows with
$2n = \mathrm{dim}_\mathbb{C} V$.
\end{enumerate}

\subsubsection{The case $E = \mathrm{End}(V) \oplus \mathrm{End}(V^*)$,
$(CP)^2 = -\mathrm{Id}$}\label{sect: 5.1.2}

The first remarks made at the beginning of Sect.\ \ref{sect: 5.1.1}
still apply: $\mathfrak{g}$ is the subalgebra of the diagonally acting
$\mathfrak{su}(V)$ which commutes with the antiunitary operator $CP$.
But now $CP$ defines a $\mathbb{C}$-bilinear symplectic structure on
$W = V \oplus V^*$ by $a(w_1,w_2) := \langle C P w_1 , w_2 \rangle$.
Actually $CP$ is already defined on $V$ and transported to $V^*$ by
$C$.  Thus $\mathfrak{g} = \mathfrak{usp}(V)$.
\begin{enumerate}
\item $T_1^2 = -\mathrm{Id}$: $\mathfrak{usp}_{2n} / (\mathfrak
{usp}_{2p} \oplus \mathfrak{usp}_{2q})$. \\
In this case $\Gamma := CT : V \to V$ is a unitary operator which
satisfies $\Gamma^2 = \mathrm{Id}$, and which defines the eigenspace
decomposition $V = V^{+} \oplus V^{-}$.  This decomposition is both
$a$- and $\langle \, , \, \rangle$-orthogonal, and consequently
$\mathfrak{k} = \mathfrak{usp}(V^{+}) \oplus \mathfrak{usp}
(V^{-})$. \\
Note that there is no condition on $p$ and $q$ other than $p + q = n$.
\hfill\vspace{0.2cm}
\item $T_1^2 = \mathrm{Id}$: $\mathfrak{usp}_{2n} / \mathfrak{u}_n$.\\
Let $V_\mathbb R$ be the $T_1$-real points of $V$. Then $\mathfrak{k}$
is the stabilizer of $V_\mathbb{R}$ in $\mathfrak{g} = \mathfrak{usp}
(V)$. Here the symplectic structure $a$ on $V$ restricts to a real
symplectic structure $a_\mathbb{R}$ on $V_\mathbb{R}$.  Since the
unitary structure $\langle \, , \, \rangle $ is compatible with this
structure, $\mathfrak{k}$ is the maximal compact subalgebra
$\mathfrak{u}_n$ of the associated real symplectic algebra.
\end{enumerate}

\subsubsection{The case $E = \mathrm{End}(V \oplus V^*)$, $b = s$}
\label{sect: 5.1.3}

Recall that in this case $C^2 = \mathrm{Id}$, and the $b$-isometry
group of $W = V \oplus V^\ast$ is $\mathrm{SO}(W)$.  Before going into
the various cases, let us remark on the relevance of whether or not
time-reversal operators commute with $P$.

If $P^2 = u^2 \mathrm{Id}$, where either $u = \pm 1$ or $u = \pm
\mathrm{i}$, we consider the $P$-eigenspace decomposition $W = W_{u}
\oplus W_{-u}$. Note $\mathrm{dim}\, W_u = \mathrm{dim}\, W_{-u}$ from
$\mathrm{Tr}\, P = 0$.  The Lie algebra $\mathfrak{g}\subset\mathfrak
{so}(W)$ of operators which preserve $b = s$ and commute with $P$ is
$\mathfrak{so}_\mathbb{R} (W_{u}) \oplus \mathfrak{so}_\mathbb{R}
(W_{-u})$.

An antiunitary operator which commutes with $P$ preserves the
decomposition $W = W_u \oplus W_{-u}$ if $u = \pm 1$, and exchanges
the summands if $u = \pm \mathrm{i}$. Similarly, if it anticommutes
with $P$, then it exchanges the summands in $W = W_{+1} \oplus W_{-1}$
and preserves the decomposition $W = W_{+ \mathrm{i}} \oplus W_{-
\mathrm{i}}$.  For this reason, as will be clear from the first case
below, the sign of $T T_1 = \pm T_1 T$ has no bearing on our
classification.
\begin{enumerate}
\item $T^2 = T_1^2 = \mathrm{Id}$: $(\mathfrak{so}_n / (\mathfrak{
so}_p \oplus \mathfrak{so}_q)) \oplus (\mathfrak{so}_n /
(\mathfrak{so}_p \oplus \mathfrak{so}_q ))$. \\
Suppose first that $P^2 = \mathrm{Id}$, giving the $P$-eigenspace
decomposition $W = W_{+1} \oplus W_{-1}$.  Each of the time-reversal
operators commutes with $P$.  To determine $\mathfrak{k}$ we consider
the unitary operator $\Gamma = C T_1$ which is a mixing $b$-isometry
satisfying $\Gamma P = P \Gamma$ and $\Gamma^2 = \mathrm{Id}$.  Thus
$W_{+1}$ further decomposes into a direct sum $W_{+1} = W_{+1}^{+1}
\oplus W_{+1}^{-1}$ of $\Gamma$-eigenspaces, which are orthogonal with
respect to both $b$ and $\langle \, , \, \rangle$. The same discussion
holds for $W_{-1}$. The stabilizer of this refined decomposition is
$\mathfrak{k} = \big( \mathfrak{so}_\mathbb{R} (W_{+1}^{+1}) \oplus
\mathfrak{so}_\mathbb{R} (W_{+1}^{-1}) \big) \oplus \big( \mathfrak
{so}_\mathbb{R}(W_{-1}^{+1}) \oplus \mathfrak {so}_\mathbb{R}
(W_{-1}^{-1}) \big)$.  From $\mathrm{Tr}\, P = \mathrm{Tr}\, \Gamma =
0$ one infers $\mathrm{dim}\, W_{+1}^{+1} = \mathrm{dim}\, W_{-1}^{
-1} = p$ and $\mathrm{dim}\, W_{+1}^{-1} = \mathrm{dim}\, W_{-1}^{+1}
= q$. \\
Now consider the case where $P^2 = - \mathrm{Id}$ but the
time-reversal operators \emph{anti}com\-mute with each other and hence
with $P$.  In this situation the $P$-eigenspace decomposition $W =
W_{+ \mathrm{i}} \oplus W_{- \mathrm{i}}$ is still $T$-invariant.
Therefore we are in exactly the same situation as above, and of course
obtain the same result. \\
This happens in all cases below.  Thus, for the remainder of this
section we assume that the time-reversal operators commute with $P$.
\hfill\vspace{0.2cm}
\item $T^2 = T_1^2 = -\mathrm{Id}$: $(\mathfrak{so}_{2n}/ (\mathfrak{
so}_n \oplus \mathfrak{so}_n )) \oplus (\mathfrak{so}_{2n} /
(\mathfrak{so}_n \oplus \mathfrak{so}_n ))$. \\
The situation is exactly the same as that above, except that $\Gamma =
C T_1$ now satisfies $\Gamma^2 = - \mathrm{Id}$.  Since $\Gamma$
preserves the sets of $C$-real points of $W_{+1}$ and $W_{-1}$,
$\Gamma$ defines a complex structure of these real vector spaces.
Therefore we have the additional condition $\mathrm{dim}\, W_{+1}^{+
\mathrm{i}} = \mathrm{dim}\, W_{+1}^{- \mathrm{i}}$ on the dimensions
of the $\Gamma$-eigenspaces. \hfill\vspace{0.2cm}
\item $T^2 = -T_1^2$: $(\mathfrak{so}_n \oplus \mathfrak{so}_n) /
\mathfrak{so}_n$. \\
The argument to be given is true independent of whether $T^2 =
\mathrm{Id}$ or $T^2 = - \mathrm{Id}$. \\
As usual we consider the $P$-eigenspace decomposition $W = W_{+
\mathrm {i}} \oplus W_{-\mathrm{i}}$.  Since $P$ is an isometry of
both $b$ and $\langle \, , \, \rangle $, the decomposition is $b$- and
$\langle \, , \,\rangle $-orthogonal.  Thus $\mathfrak{g} = \mathfrak
{so}_\mathbb{R} (W_{+ \mathrm{i}}) \oplus \mathfrak{so}_ \mathbb{R}
(W_{-\mathrm{i}})$.  Now $T$ is antilinear and commutes with $P$.
Thus it permutes the $P$-eigenspaces, i.e., $T:W_{+ \mathrm{i}} \to
W_{- \mathrm{i}}$.  Since $\mathfrak{k}$ consists of those operators
in $\mathfrak{g}$ that commute with $T$, and $T$ is compatible with
both the unitary structure and the bilinear form $b$, it follows that
$(A,B) \in \mathfrak{g}$ is in $\mathfrak{k}$ if and only if $B = T A
T^{-1}$.  In other words, after applying the obvious automorphism,
$\mathfrak{k}$ is the diagonal in $\mathfrak{g} \cong \mathfrak{so}_n
\oplus \mathfrak{so}_n$.
\end{enumerate}

\subsubsection {The case $E = \mathrm{End} (V \oplus V^*)$, $b = a$}
\label{sect: 5.1.4}

Recall that in this case $C^2 = -\mathrm{Id}$, and the $b$-isometry
group of $W = V \oplus V^\ast$ is $\mathrm{Sp}(W)$.  For the same
reasons as indicated above we may assume that the time-reversal
operators commute with $P$.
\begin{enumerate}
\item $T^2 = T^2_1 = \mathrm{Id}$: $(\mathfrak{usp}_{2n}/\mathfrak{
u}_n)\oplus (\mathfrak{usp}_{2n} /\mathfrak{u}_n )$. \\
Observe that the $P$-eigenspace decomposition $W = W_{+1} \oplus
W_{-1}$ is $a$- and $\langle \, , \, \rangle $-orthogonal and that
therefore $\mathfrak{g} = \mathfrak{usp}(W_{+1}) \oplus \mathfrak
{usp}(W_{-1})$.  Let the dimension be denoted by $\mathrm{dim}
_\mathbb{C}(W_{+1}) = \mathrm{dim}_\mathbb{C}(W_{-1}) = 2 n$.  \\
Now $T$ defines real structures on $W_{+1}$ and $W_{-1}$, and these
are compatible with $a$.  Hence in both cases the restriction
$a_\mathbb{R}$ to the set $W_{\pm 1}^\mathbb{R}$ of fixed points of
$T$ is a real symplectic structure.  The algebra $\mathfrak{k}$
consists of the pairs $(A,B)$ of operators in $\mathfrak{g}$ which
stabilize $W_{+1}^\mathbb{R} \oplus W_{-1}^\mathbb{R}$. This means
that $A$, e.g., is in the maximal compact subalgebra of the real
symplectic Lie algebra determined by $a_\mathbb{R}$ on $W_{+1}
^\mathbb{R}$, i.e., in a unitary Lie algebra isomorphic to $\mathfrak
{u}_n$.  A similar statement holds for $B$.  \hfill\vspace{0.2cm}
\item $T^2 = T^2_1 = -\mathrm{Id}$: $(\mathfrak{usp}_{2n}/
(\mathfrak{usp}_{2p}\oplus \mathfrak{usp}_{2q})\oplus (\mathfrak{
usp}_{2n}/(\mathfrak{usp}_{2p}\oplus \mathfrak{usp}_{2q}))$. \\
The argument made above still shows that $\mathfrak{g} = \mathfrak{
usp}(W_{+1})\oplus \mathfrak{usp}(W_{-1})$. \\
Now, to determine $\mathfrak{k}$ we consider the operator $\Gamma := C
T_1$ which stabilizes this decomposition and satisfies $\Gamma^2 =
\mathrm{Id}$. Thus the further condition to be satisfied in order for
an operator to be in $\mathfrak{k}$ is that the $\Gamma $-eigenspace
decomposition of each summand must be stabilized, i.e., $\mathfrak{k}
= \oplus_{\varepsilon, \delta = \pm 1} \mathfrak{usp} (W_{\varepsilon}
^{\delta})$.  The dimensions must match pairwise because $\mathrm{Tr}
\, P = \mathrm{Tr}\, \Gamma = 0$. \hfill\vspace{0.2cm}
\item $T^2 = - T_1^2$: $\mathfrak{su}_n / \mathfrak{so}_n$. \\
The answer for $\mathfrak{g} / \mathfrak{k}$ is the same for the two
cases $T^2 = \mathrm{Id}$ or $T^2 = - \mathrm {Id}$. \\
In either case it follows from $a(w_1, w_2) = a(P w_1, P w_2)$ that
the summands of the $P$-decomposition $W = W_{+ \mathrm{i}} \oplus
W_{- \mathrm{i}}$ are $a$-Lagrangian.  Thus an $a$-isometry stabilizes
the decomposition if and only if it is a $\mathbb{C}$-linear
transformation acting diagonally, and consequently $\mathfrak{g} =
\mathfrak{su}(W_{+ \mathrm{i}})$ (which is acting diagonally as
well). \\
Without loss of generality we may assume that $T^2 = \mathrm{Id}$ (or
else we replace $T$ by $T_1$).  Then $T$ is a real structure which
permutes the $P$-eigenspaces.  Thus the diagonal action $(w^+ , w^-)
\mapsto (B w^+ , B w^-)$ commutes with $T$ if and only if $T B T^{-1}
= B$.  Since $T$ is compatible with the initial unitary structure, if
follows that $B$ is in the associated real orthogonal group.  For
example, if unitary coordinates are chosen so that $T$ is given by
$(z,w) \mapsto (\bar w,\bar z)$, then $T B T^{-1} = B$ simply means
that $B = \bar B$.
\end{enumerate}

\subsection{Building bigger blocks}
\label{sect: bigblocks}

Before $G_0$-reduction we must determine the basic block associated to
the $G_0$-repre\-sentation space $V$.  This has been adequately
discussed in all cases with the exception of the one where there are
two time-reversal operators.  Here we handle that case by reducing it
to the situation where there is only one.

Write the initial block as $V_1^{\vphantom{*}} \oplus V_1^*$ and
build a diagram consisting of the four spaces $V_i^{\vphantom{*}}
\oplus V_i^*$, $i = 1, \ldots, 4$, with the maps $T$, $T_1$, and
$P$ emanating from each of them.  To be concrete, $T :
V_1^{\vphantom{*}} \oplus V_1^*\to V_2^{\vphantom{*}} \oplus
V_2^*$ defines $V_2$, and $T_1 : V_1^{\vphantom{*}} \oplus V_1^*
\to V_3^{\vphantom{*}} \oplus V_3^*$ defines $V_3$, and $T_1 :
V_2^{\vphantom{*}} \oplus V_2^* \to V_4^{\vphantom{*}} \oplus
V_4^*$ defines $V_4$.  The relation $P = T T_1$ defines the
remaining maps. At this point there is no need to discuss mixing.

We also underline that, by the nature of the basic model, any two
spaces $V_i^{\vphantom{*}} \oplus V_i^*$ and $V_j^{\vphantom{*}}
\oplus V_j^*$ are either disjoint in the big Nambu space or are equal.

Let us now complete the proof of our classification result, Theorem
\ref{rough statement}, by running through the various cases which
occur in the present setting where the initial block must be
extended. We only sketch this, because given how the extended block
case was handled in the setting of one distinguished time-reversal
symmetry (Sect.\ \ref{sect: typeII}) and the detailed classification
results above, the proof requires no new ideas or methods.

\medskip\noindent{\it 1) $V_1^{\vphantom{\ast}} \oplus V_1^*$ is
  $T$-invariant and is not $T_1$-invariant}.  ---
Here it is only necessary to consider $P : W_1 = V_1^{\vphantom{\ast}}
\oplus V_1^* \to V_3^{\vphantom{\ast}} \oplus V_3^\ast = W_3$.  If
$\mathfrak{g}$ is the Lie algebra of unitary operators which commute
with the $G_0$-action and respect the $b$-structure on $V_1^{\vphantom
{\ast}} \oplus V_1^*$, then the further condition of compatibility
with $P$ means that the algebra in the present case is $\mathfrak{g}$
acting diagonally via $P$ on $W_1 \oplus W_3$.  Thus we have reduced
to the case of only one time-reversal operator on $W_1$, which has
been classified above.

Note that this argument has nothing to do with whether or not $T$ is
mixing.  Hence, in this and all of the following cases there is no
need to differentiate between $T$ and $T_1$.

\medskip\noindent{\it 2) $V_1^{\vphantom{\ast}} \oplus V_1^*$ is
neither $T$- nor $T_1$-invariant}. ---
Consider the diagram introduced above where all the spaces $W_i =
V_i^{\vphantom{\ast}} \oplus V_i^*$ occur.  If any of the $W_i$ is
invariant by either $T$ or $T_1$, then we change our perspective,
replace $W_1$ by that space and apply the above argument.  Thus we may
assume that no $W_i$ is stabilized by either $T$ or $T_1$.  It is
still possible, however, that $W_1 = W_4$, and in that case it follows
that $W_2 = W_3$.

\smallskip\noindent{\it 2.1) $W_1=W_4$}. ---
Here both $W_1$ and $W_4$ are $P$-invariant.  We leave it to the
reader to check that $P$ can be transferred to the level of $\mathrm
{End}(H)\oplus \mathrm{End}(H^*)$ or $\mathrm{End}(H\oplus H^*)$ just
as we transferred the time-reversal operators.  Thus, e.g., it is
enough to know the Lie algebra of operators $\mathfrak{g}$ on $W_1$
which are compatible with the unitary structure, are $b$-isometries
and are compatible with $P$. This has been computed in Sect.\
\ref{sect: smallblocks}.  Of course we did this in the case where $V
\oplus V^*$ is $T$- and $T_1$-invariant, but the compatibility with
$P$ had nothing to do with time reversal.

In the present case both $T$ and $T_1$ exchange $W_1$ and $W_2$.  Thus
our symmetric space is $(\mathfrak{g}\oplus \mathfrak{g})/\mathfrak{
g}$.

\smallskip\noindent{\it 2.2) The spaces $W_i$ are pairwise disjoint}. ---
Here we will go through a number of subcases, depending on whether or
not there exist (equivariant) isomorphisms between various spaces.
Such an isomorphism is of course assumed to be unitary and to commute
with $C$; in particular it is a $b$-isometry.

\smallskip\noindent{\it 2.2.1) $W_1\cong W_4$}. ---
If $\varphi $ is the isomorphism which does this, then $T\varphi
T^{-1}=:\psi $ is an isomorphism of $W_2$ and $W_3$.  Using these
isomorphisms, we build $W:=W_1\oplus W_4$ and $\tilde W:=W_2\oplus
W_3$ which are of our initial type; they are stabilized by $P$ and
exchanged by $T$.  Thus, as in {\it 2.1)}, if ${\mathfrak g}$ is the
Lie algebra of operators on $W$ which are compatible with the unitary
structure, are $b$-isometries and are compatible with $P$, then our
symmetric space is $(\mathfrak{g}\oplus \mathfrak{g})/\mathfrak{g}$.

\smallskip\noindent{\it 2.2.2) $W_1\cong W_2$}. ---
For the reasons given above, $W_3\cong W_4$ and we build $W$ and
$\tilde W$ as in that case.  In the present situation $P$ exchanges
$W$ and $\tilde W$.  We must then consider two subcases during our
procedure for identifying $\mathfrak{g}$.

The simplest case is where $W$ and $\tilde W$ are not isomorphic.  In
that setting the Lie algebra $\mathfrak{g}$ of unitary operators on
$W$ which commute with the $G_0$-action and are compatible with $b$
acts diagonally on $W\oplus \tilde W$. This is exactly our algebra of
interest.

Thus in this case we can forget $\tilde W$, and regard $\mathfrak{g}$
as acting on $W$.  Here $T$ stabilizes $W$ and thus the associated
symmetric space is $\mathfrak{g}/\mathfrak{k}$, where $\mathfrak{k}$
consists of the operators in $\mathfrak{g}$ which commute with $T$.
This situation has been classified above; in particular, only
classical irreducible symmetric spaces of compact type occur.

Our final case occurs under the assumption $W_1\cong W_2$ in the
situation where $W$ and $\tilde W$ are isomorphic.  Here we view
an operator which commutes with the $G_0$-action as a matrix
\begin{displaymath}
  \begin{pmatrix} \mathsf{A} &\mathsf{B}\\
  \mathsf{C} &\mathsf{D} \end{pmatrix} \;.
\end{displaymath}
Compatibility with $P$ can then be interpreted as $\mathsf{B}$ and
${\mathsf D}$ being determined from $\mathsf{A}$ and $\mathsf{C}$ by
$P$-conjugation. In this notation $\mathsf{A} : W \to W$ and
$\mathsf{C} : W \to \tilde W$.  But we may also regard $\mathsf{C}$ as
an operator on $W$ which is transferred to a map from $W$ to $\tilde
W$ by the isomorphism at hand.  Therefore the Lie algebra of interest
can be identified with the set of pairs $(\mathsf{A}, \mathsf{C})$ of
operators on $W$ which are compatible with the unitary and
$b$-structures and commute with the $G_0$-action on $W$.  Hence the
associated symmetric space is the direct sum $\mathfrak{g} /
\mathfrak{k} \oplus \mathfrak{g} / \mathfrak{k}$ where $\mathfrak{k}$
is determined by compatibility with $T : W \to W$, i.e., a direct sum
of two copies of an arbitrary example that occurs with only one
$T$-symmetry.

\section{Physical realizations}\label{sect: realizations}
\setcounter{equation}{0}

We now illustrate Theorem \ref{rough statement} by the two large
sets of examples that were already referred to in Sect.\
\ref{sect: fermions}: (i) fermionic quasiparticle excitations in
disordered normal- and superconducting systems, and (ii) Dirac
fermions in a stochastic gauge field background. In each case we
fix a specific Nambu space $\mathcal{W}$, and show how a variety
of symmetric spaces (each corresponding to a symmetry class) is
realized by \emph{varying the group of unitary and antiunitary
symmetries}, $G$.

The invariable nature of $\mathcal{W}$ is a principle imposed by
physics: electrons, e.g., have electric charge $e = -1$ and spin
$S = 1/2$ and these properties cannot ever be changed.  What can
be changed, however, by varying the experimental conditions, are
the symmetries of the Hamiltonian governing the specific situation
at hand.  For example, turning on an external magnetic field
breaks time-reversal symmetry, adding spin-orbit scatterers to the
system breaks spin-rotation symmetry, lowering the temperature
enhances the pairing forces that may lead to a spontaneous
breakdown of the global $\mathrm{U}_1$ charge symmetry, and so
on.

\subsection{Quasiparticles in metals and superconductors}

The setting here is the one already described in Sect.\ \ref{sect:
Nambu}: given the complex Hilbert space $\mathcal{V}$ of
single-electron states, we form the Nambu space $\mathcal{W} =
\mathcal{V} \oplus \mathcal{V}^\ast$ of electron field operators.
On $\mathcal{W}$ we then have the canonical symmetric bilinear
form $b$, the particle-hole conjugation operator $C : \mathcal{W}
\to \mathcal{W}$, and the canonical unitary structure $\langle \,
, \, \rangle$.

The complex Hilbert spaces $\mathcal{V}$ and $\mathcal{V}^\ast$ are to
be viewed as representation spaces of a $\mathrm{U}_1$ group, which is
the global $\mathrm{U}_1$ gauge degree of freedom of
electrodynamics. Indeed, creating or annihilating one electron amounts
to adding one unit of negative or positive electric charge to the
fermion system. In representation-theoretic terms, this means that
$\mathcal{V}$ carries the fundamental representation of the
$\mathrm{U}_1$ gauge group while $\mathcal{V}^\ast$ carries the
antifundamental one. Thus $z \in \mathrm{U}_1$ here acts on
$\mathcal{V}$ by multiplication with $z$, and on $\mathcal{V}^\ast$ by
multiplication with $\bar z$.

Extra structure arises from the fact that electrons carry spin
1/2, which implies that $\mathcal{V}$ is a tensor product of
spinor space, $\mathbb{C}^2$, with the Hilbert space $X$ for the
orbital motion in real space. The spin-rotation group
$\mathrm{Spin}_3 = \mathrm{SU }_2$ acts trivially on $X$ and by
the spinor representation on the factor $\mathbb{C}^2$.  (In a
framework more comprehensive than is of relevance to the
disordered systems setting developed here, the spinor
representation would enter as a projective representation of the
rotation group $\mathrm{SO}_3$, and $\mathrm{SO}_3$ would act on
the factor $X$ by rotations in the three-dimensional Euclidean
space.) On physical grounds, spin rotations must preserve the
canonical anticommutation relations as well as the unitary
structure of $\mathcal{V}$.  Therefore, by Prop.\ \ref{Crelates}
spin rotations commute with the particle-hole conjugation operator
$C$.

Another symmetry operation of importance for present purposes is
time reversal. As always in quantum mechanics, time reversal is
implemented as an antiunitary operator $T$ on the single-electron
Hilbert space $\mathcal{V}$.  Its algebraic properties are
influenced by the spin 1/2 nature of the electron: fundamental
physics considerations dictate $T^2 = - \mathrm{Id}$. A closely
related condition is that time reversal commutes with spin
rotations. $T$ extends to an operation on $\mathcal{W}$ by $C T =
T C$.

In physics one uses the word \emph{quasiparticle} for the
excitations that are created by acting with a fermionic field
operator on a many-fermion ground state.

\subsubsection{Class $D$}

In the general context of quasiparticle excitations in metals and
superconductors, this is the fundamental class where \emph{no}
symmetries are present.

A concrete realization takes place in superconductors where the order
parameter transforms under spin rotations as a spin triplet, $S = 1$
(i.e., the adjoint representation of $\mathrm{SU}_2$), and transforms
under $\mathrm{SO}_2$-rotations of two-dimensional space as a $p$-wave
(the fundamental representation of $\mathrm{SO}_2$). A recent
candidate for a quasi-2d (or layered) spin-triplet $p$-wave
superconductor is the compound $\mathrm{Sr}_2 \mathrm{Ru}\,
\mathrm{O}_4$ \cite{mm,annalen}. (A noncharged analog is the $A$-phase
of superfluid ${}^3 \mathrm{He}$ \cite{vw}.) Time-reversal symmetry in
such a system may be broken spontaneously, or else can be broken by an
external magnetic field creating vortices in the superconductor.
Further realizations proposed in the recent literature include
double-layer fractional quantum Hall systems at half filling \cite{rg}
(more precisely, a mean-field description for the composite fermions
of such systems), and a network model for the random-bond Ising model
\cite{sf}.

The time-evolution operators $U = {\rm e}^{-{\rm i}t H / \hbar}$
in this class are constrained only by the requirement that they
preserve both the unitary structure and the symmetric bilinear
form of $\mathcal{W}$.  If $\mathcal{W}_\mathbb{R}$ is the set of
real points $\{ v + Cv : v \in \mathcal{V} \}$, we know from
Prop.\ \ref{g/k} that the space of time evolutions is a real
orthogonal group $\mathrm{SO} (\mathcal{W}_\mathbb{R})$.  In
Cartan's notation this is called a symmetric space of the $D$
family. The Hamiltonians $H$ are such that $\mathrm{i}H \in
\mathfrak{so} (\mathcal{W}_ \mathbb{R} )$; this means that the
Hamiltonian matrices are imaginary skew in a suitably chosen basis
(called Majorana fermions in physics).

Note that since $\mathcal{W}_\mathbb{R}$ is a real form of $(X
\otimes \mathbb{C}^2) \oplus (X \otimes \mathbb{C}^2)^\ast$, the
dimension of $\mathcal{W}_\mathbb{R}$ must be a multiple of four
(for spinless particles it would only be a multiple of two).

\subsubsection{Class $D${\rm III}}

Let now time reversal be a symmetry of the quasiparticle system.
This means that magnetic fields and scattering by magnetic
impurities are absent.  On the other hand, spin-rotation
invariance is again required to be broken.

Known realizations of this situation exist in gapless superconductors,
say with spin-singlet pairing, but with a sufficient concentration of
spin-orbit impurities to cause strong spin-orbit scattering \cite{sf}.
In order for quasiparticle excitations to exist at low energy, the
spatial symmetry of the order parameter should be $d$-wave (more
precisely, a time-reversal invariant combination of the angular
momentum $l = + 2$ and $l = -2$ representations of $\mathrm{SO }_2$).
A noncharged realization occurs in the $B$-phase of ${}^3 \mathrm{He}$
\cite{vw}, where the order parameter is spin-triplet without breaking
time-reversal symmetry.  Another candidate are heavy-fermion
superconductors \cite{stewart}, where spin-orbit scattering often
happens to be strong owing to the presence of elements with large
atomic weights such as uranium and cerium.

Time-reversal invariance constrains the set of good Hamiltonians
$H$ by $H = T H T^{-1}$. Since $T^2 = - \mathrm{Id}$ for spin 1/2
particles, we are dealing with the case treated in \ref{sect:
4.3.2}.1. The space of time evolutions therefore is $\mathrm{SO}
(\mathcal{ W}_\mathbb{R}) / \mathrm{U}(\mathcal{V})$, which is a
symmetric space of the $D$III family. The standard form of the
Hamiltonians in this class is
\begin{equation}\label{HclassDIII}
    H = \begin{pmatrix} 0 &Z\\ Z^\ast &0 \end{pmatrix} \;,
\end{equation}
where $Z \in \mathrm{Hom}(\mathcal{V}^\ast, \mathcal{V})$ is skew.
(Note again that the dimension of $\mathcal{W}_\mathbb{R}$ is a
multiple of four, and would be a multiple of two for particles
with spin zero).

\subsubsection{Class $C$}

Next let the spin of the quasiparticles be conserved, and let
time-reversal symmetry be broken instead.  Thus magnetic fields
(or some equivalent $T$-breaking agent) are now present, while the
effect of spin-orbit scattering is absent. The symmetry group of
the physical system then is $G = G_0 = \mathrm{Spin}_3 = \mathrm
{SU}_2$.

This situation is realized in spin-singlet superconductors in the
vortex phase \cite{sfbn}.  Prominent examples are the cuprate (or
high-$T_c$) superconductors \cite{tsuei}, which are layered and
exhibit $d$-wave symmetry in their copper-oxide planes. It has been
speculated that some of these superconductors break time-reversal
symmetry spontaneously, by the generation of an order-parameter
component $\mathrm{i} d_{xy}$ or $\mathrm{i} s$ \cite{smf}. Other
realizations of this class include network models of the spin quantum
Hall effect \cite{glr}.

Following the general strategy of Sect.\ \ref{sect: reduction}, we
eliminate $G_0 = \mathrm{SU}_2$ from the picture by transferring
from $\mathcal{V} \oplus \mathcal{V}^\ast$ to the reduced space $X
\oplus X^\ast$.  In the process the bilinear form $b$ undergoes a
change of parity. To see this let $R = \mathbb{C}^2$ (a.k.a.\
spinor space) be the fundamental representation space of
$\mathrm{SU}_2$. $R$ is isomorphic to $R^\ast$ by $\psi : r
\mapsto \langle \mathrm{i} \sigma_2 \bar{r} , \cdot \rangle_R$
where $\sigma_2$ is the second Pauli matrix. This isomorphism
$\psi : R \to R^\ast$ is alternating. Therefore, by Prop.\
\ref{transfer b} the symmetric bilinear form of $\mathcal{V}
\oplus \mathcal{ V}^\ast$ gets transferred to the alternating form
$a$ of $X \oplus X^\ast$.

{}From Prop.\ \ref{g/k} we then infer that the space of time
evolutions is $\mathrm{USp}(X \oplus X^\ast)$ --- a symmetric
space of the $C$ family.  The standard form of the Hamiltonians
here is
\begin{displaymath}
    H=\begin{pmatrix}A &B\\ B^\ast &-A^\mathrm{t}\end{pmatrix}\;,
\end{displaymath}
with self-adjoint $A \in \mathrm{End}(X)$ and complex symmetric $B
\in \mathrm{Hom}(X^\ast,X)$.

\subsubsection{Class $C${\rm I}}\label{sect: classCI}

The next class is obtained by taking spin rotations as well as the
time reversal $T$ to be symmetries of the quasiparticle system.
Thus the symmetry group is $G = G_0 \cup T G_0$ with $G_0 =
\mathrm{Spin}_3 = \mathrm{SU}_2$.

Like in the previous symmetry class, physical realizations are
provided by the low-energy quasiparticles of unconventional
spin-singlet superconductors \cite{tsuei}.  The difference is that the
superconductor must now be in the Meissner phase where magnetic field
are expelled by screening currents.  In the case of superconductors
with several low-energy points in the first Brillouin zone, scattering
off hard impurities is needed to break additional conservation laws
that would otherwise emerge (see Sect.\ \ref{sect: 5.1.5}).

To identify the relevant symmetric space, we again transfer from
$\mathcal{V} \oplus \mathcal{V}^\ast$ to the reduced space $X
\oplus X^\ast$. As before, the bilinear form $b$ changes parity
from symmetric to alternating under this reduction. In addition
now, time reversal has to be transferred. As was explained in the
example following Lemma \ref{parity of beta}, the time-reversal
operator changes its involutory character from $T^2 = - \mathrm{
Id}_{\mathcal{V} \oplus \mathcal{V}^\ast}$ to $T^2 = + \mathrm{Id}
_{X \oplus X^\ast}$.

In the language of Sect.\ \ref{sect: classify} the block data are
$V = X$, $E = \mathrm{End}(V \oplus V^\ast)$, $b = a$, $T$
nonmixing, and $T^2 = \mathrm{Id}$.  This case was treated in
\ref{sect: 4.3.3}.3. From there, we know that the space of time
evolutions is $\mathrm{USp}(X \oplus X^\ast) / \mathrm{U}(X)$ -- a
symmetric space in the $C$I family.  The standard form of the
Hamiltonians in this class is the same as that given in
(\ref{HclassDIII}) but now with $Z \in \mathrm{Hom}(X^\ast,X)$
complex symmetric.

\subsubsection{Class $A${\rm III}}\label{sect: 5.1.5}

This class is commonly associated with random-matrix models for
the low-energy Dirac spectrum of quantum chromodynamics with
massless quarks (see Sect.\ \ref{sect: 5.2.1}). Here we review an
alternative realization, which has recently been identified
\cite{asz} in $d$-wave superconductors with soft impurity
scattering.

To construct this realization one starts from class $C$I, i.e.\
from quasiparticles in a superconductor with time-reversal
invariance and conserved spin, and enlarges the symmetry group by
imposing another $\mathrm{U}_1$ symmetry, generated by a
Hermitian operator $Q$ with $Q^2 = \mathrm{Id}$. The physical
reason for the extra conservation law is approximate momentum
conservation in a disordered quasiparticle system with a
dispersion law that has Dirac-type low-energy points at four
distinct places in the Brillouin zone.

Thus beyond the spin-rotation group $\mathrm{SU}_2$ there now
exists a one-parameter group of unitary symmetries $\mathrm {e}
^{\mathrm{i} \theta Q}$. The operators $\mathrm{e}^{\mathrm{i}
\theta Q}$ are defined on $\mathcal{V}$, and are diagonally
extended to $\mathcal{W} = \mathcal{V} \oplus \mathcal{V}^\ast$.
They are characterized by the property that they commute with
particle-hole conjugation $C$, time reversal $T$, and the spin
rotations $g \in \mathrm{SU}_2$.
%Such an operator exists only when $N$ is even.

The reduction to standard block data is done in two steps.  In the
first step, we eliminate the spin-rotation group $\mathrm{SU}_2$.
From the previous section, the transferred data are known to be $E =
\mathrm{End}(X \oplus X^\ast)$, $b = a$, $T$ nonmixing, and $T^2 =
\mathrm{Id}$.

The second step is to reduce by the $\mathrm{U}_1$ group generated by
$Q$.  For this consider the $\mathbb{C}$-linear operator $J :=
\mathrm{i}Q$ with $J^2 = - \mathrm{Id}$, and let the $J$-eigenspace
decomposition of $X$ be written $X = X_{+\mathrm{i}} \oplus
X_{-\mathrm{i}}$.  There is a corresponding decomposition $X^\ast =
{X^\ast}_{+\mathrm{i}} \oplus {X^\ast}_{-\mathrm{i}}$.  Since $J$
commutes with $T$, a complex structure is defined by it on the set of
$T$-real points of $X$. Therefore $\mathrm{dim}\, X_{+ \mathrm{i}} =
\mathrm{dim}\, X_{-\mathrm{i}}$.  Another consequence of $JT = TJ$ is
that the $\mathbb{C}$-antilinear operator $T$ exchanges $X_{+
  \mathrm{i}}$ with $X_{- \mathrm{i}}$.  Thus $T$ is mixing with
respect to the decompositions $X = X_{+ \mathrm{i}} \oplus X_{-
  \mathrm{i}}$ and $X^\ast = {X^\ast}_{+ \mathrm{i}} \oplus {X^\ast}_{-
  \mathrm{i}}$. The $\mathbb{C}$-antilinear operator $C$ maps $X_{\pm
  \mathrm{i}}$ to ${X^\ast}_{\mp \mathrm{i}}$.

The fully reduced block data now are $V := X_{+ \mathrm{i}} \oplus
{X^\ast}_{+ \mathrm{i}}$, $E = \mathrm{End}(V) \oplus \mathrm{End}
(V^\ast)$, $b = a$, $T$ mixing, and $T^2 = \mathrm{Id}$.  The
finite-dimensional version of this case was treated in \ref{sect:
4.3.1}.3.  Our answer for the space of time-evolution operators was
$\mathrm{SU}_{p+q} / \mathrm{S}( \mathrm{U}_p \times \mathrm{U}_q)$,
which is a symmetric space in the $A$III family.

Unlike the general case handled in \ref{sect: 4.3.1}.3, it here
follows from the fundamental physics definition of particle-hole
conjugation $C$ and time reversal $T$ that the operator $CT$
stabilizes a real subspace $V_\mathbb{R}$.  We also have $(CT)^2 = -
\mathrm{Id}$.  Therefore, the operator $CT$ defines a complex
structure of $V_\mathbb{R}$, and hence the integers $p$ and $q$, which
are the dimensions of the $CT$-eigenspaces in $V$, must be equal.

\subsubsection{Class $A$}

At this point a new symmetry requirement is brought into play:
conservation of the electric charge. Thus the global $\mathrm{U}_1$
gauge transformations of electrodynamics are now decreed to be
symmetries of the quasiparticle system.  This means that the system no
longer is a superconductor, where $\mathrm{U}_1$ gauge symmetry is
spontaneously broken, but is a metal or normal-conducting system.  If
all further symmetries are broken (time reversal by a magnetic field
or magnetic impurities, spin rotations by spin-orbit scattering,
etc.), the symmetry group is $G = G_0 = \mathrm{U}_1$.

All states (actually, field operators) in $\mathcal{V}$ have the
same electric charge.  Thus the irreducible $\mathrm{U}_1$
representations which they carry all have the same isomorphism
class, say $\lambda$. States in $\mathcal{V}^\ast$ carry the
opposite charge and belong to the dual class $\lambda^\ast$. Since
$\lambda \not= \lambda^\ast$, we are in the situation of Sect.\
\ref{sect: 4.3.1}, where $E = \mathrm{End}(\mathcal{V}) \oplus
\mathrm{End} (\mathcal{V}^\ast)$. With $T$ being absent, the space
of time evolutions is $\mathrm{U} (\mathcal{V})$ acting diagonally
on $\mathcal{V} \oplus \mathcal{ V}^\ast$.

In random-matrix theory, and in the finite-dimensional case where
$\mathrm{U}(\mathcal{V}) \cong \mathrm{U}_N$, one refers to these
matrix spaces as the circular Wigner-Dyson class of unitary
symmetry. The Hamiltonians in this class are represented by
complex Hermitian matrices.

If we make the restriction to traceless Hamiltonians, the space of
time evolutions becomes $\mathrm{SU}_N$, which is a type-II
irreducible symmetric space of the $A$ family.

\subsubsection{Class $A${\rm II}}

Beyond charge conservation or $\mathrm{U}_1$ gauge symmetry, time
reversal $T$ is now required to be a symmetry of the quasiparticle
system. Physical realizations of this case occur in metallic
systems with spin-orbit scattering.  The pioneering experimental
work (of the weak localization phenomenon in this class) was done
on disordered magnesium films with gold impurities
\cite{bergmann}.

The block data now is $E = \mathrm{End}(\mathcal{V}) \oplus
\mathrm{End}(\mathcal{ V}^\ast)$, $b = s$, $T$ nonmixing, $T^2 =
-\mathrm{Id}$.  This case was considered in \ref{sect: 4.3.1}.1.  The
main point there was that time reversal $T$ defines a $\mathbb
{C}$-linear symplectic structure $a$ on $\mathcal{V}$ by $a(v_1 , v_2)
= \langle T v_1 , v_2 \rangle$.  Conjugation by $T$ therefore fixes a
unitary symplectic group $\mathrm{USp}(\mathcal{V})$ inside of
$\mathrm{U}(\mathcal{V})$, and the space of good time evolutions is
$\mathsf{G} / \mathsf{K} = \mathrm{U}(\mathcal{V}) / \mathrm{USp}
(\mathcal{V})$.  In the finite-dimensional setting where $\mathsf{G} /
\mathsf{K} \cong \mathrm{U}_{2N} / \mathrm{USp}_{2N}$, this is called
the circular Wigner-Dyson class of symplectic symmetry in
random-matrix theory.  The Hamiltonians in this class are represented
by Hermitian matrices whose matrix entries are real quaternions. The
irreducible part $\mathrm{SU}_{2N} / \mathrm{USp}_{2N}$, obtained by
restricting to traceless Hamiltonians, is a type-I symmetric space in
the $A$II family.

\subsubsection{Class $A${\rm I}}\label{sect: classAI}

The next class is the Wigner-Dyson class of orthogonal symmetry.  In
the present quasiparticle setting it is obtained by imposing
spin-rotation symmetry, $\mathrm{U}_1$ gauge (or charge) symmetry and
time-reversal symmetry all at once.

Important physical realizations are by disordered metals in zero
magnetic field.  Families of quantum chaotic billiards also belong
to this class.

The group of unitary symmetries here is $G_0 = \mathrm{U}_1 \times
\mathrm{SU}_2$.  We eliminate the spin-rotation group $\mathrm{SU}_2$
from the picture by transferring from $\mathcal{V} = X \otimes
\mathbb{C}^2$ to the reduced space $X$. Again, the involutory
character of $T$ is reversed in the process: the transferred time
reversal satisfies $T^2 = + \mathrm{Id}$. The parity of the bilinear
form also changes, from symmetric to alternating; however, this turns
out to be irrelevant here, as there is still the $\mathrm{U}_1$ charge
symmetry and we are in the situation $\lambda \not= \lambda^\ast$.

The block data now is $E = \mathrm{End}(X) \oplus \mathrm{End}
(X^\ast)$, $b = a$, $T$ nonmixing, $T^2 = \mathrm{Id}$.  According to
\ref{sect: 4.3.1}.2 these yield (the Cartan embedding of)
$\mathrm{U}(X) / \mathrm{O}(X)$ as the space of good time
evolutions. The irreducible part $\mathrm{SU}(X) / \mathrm{SO}(X)$, or
$\mathrm{SU}_N / \mathrm{SO}_N$ in the finite-dimensional setting, is
a symmetric space in the $A$I family. The Hamiltonian matrices in this
class can be arranged to be real symmetric.

\subsection{The Euclidean Dirac operator for chiral fermions}
\label{sect: 5.2}

We now explore the physical examples afforded by Dirac fermions in
a random gauge field background.  These examples include the Dirac
operator of quantum chromodynamics, i.e., the theory of strong
$\mathrm{SU}_3$ gauge interactions between elementary particles
called quarks.

The mathematical setting for this has already been described in
Sect.\ \ref{sect: Dirac}. Recall that one is given a twisted
spinor bundle $S \otimes R$ over Euclidean space-time, and that
$\mathcal{V}$ is taken to be the Hilbert space of $L^2$-sections
of that bundle. One is interested in the Dirac operator $D_A$ in a
gauge field background $A$ and in the limit of zero mass:
\begin{displaymath}
    D_A = \mathrm{i} \gamma^\mu (\partial_\mu - A_\mu) \;.
\end{displaymath}
We extend the self-adjoint operator $D_A$ diagonally from
$\mathcal{V}$ to the fermionic Nambu space $\mathcal{W} = \mathcal{V}
\oplus \mathcal{V}^\ast$ by the condition $D_A = - C D_A C^{-1}$. The
chiral `symmetry' $\Gamma D_A + D_A \Gamma = 0$, where $\Gamma =
\gamma_5$ is the chirality operator, then becomes a true symmetry $D_A
= T D_A T^{-1}$ with an antiunitary operator $T = C \Gamma = \Gamma
C$, which mixes $\mathcal{V}$ and $\mathcal{ V^\ast}$.

\subsubsection{Class $A${\rm III}}\label{sect: 5.2.1}

Let now the complex vector space $R = \mathbb{C}^N$ be the
fundamental representation space for the gauge group $\mathrm
{SU}_N$ with $N \ge 3$. ($N$ is called the number of colors in
this context.) Quantum chromodynamics is the special case $N = 3$.

The fact that the extended Dirac operator $D_A$ acts diagonally on
$\mathcal{W} = \mathcal{V} \oplus \mathcal{V^\ast}$ is attributed
to a symmetry group $G_0 = \mathrm{U}_1$ which has $\mathcal{V}$
and $\mathcal{V}^\ast$ as inequivalent representation spaces. For
a generic gauge-field configuration there exist no further
symmetries; thus the total symmetry group is $G = G_0 \cup T G_0$.

The block data here is $V = \mathcal{V}$, $E = \mathrm {End}(V)
\oplus \mathrm{End}(V^\ast)$, $b = s$, $T$ mixing, $T^2 = \mathrm
{Id}$, which is the case considered in \ref{sect: 4.3.1}.3. If $n
= \mathrm{dim}\, V$, we have
\begin{displaymath}
    \mathfrak{p} \cong \mathfrak{su}_n / \mathfrak{s}
    (\mathfrak{u}_p \oplus \mathfrak{u}_q) \;.
\end{displaymath}
The difference of integers $p - q$ is to be identified with the
difference between the number of right and left zero modes of
$D_A^2$. (`Right' and `left' in this context pertain to the
$(+1)$- and $(-1)$-eigenspaces of the chirality $\Gamma =
\gamma_5$.) The latter number is a topological invariant called
the index of the Dirac operator.

\subsubsection{Class $BD${\rm I}}\label{sect: BDI}

We retain the framework from before, but now consider the gauge
group $\mathrm{SU}_2$, where the number of colors $N = 2$. In this
case the massless Dirac operator $D_A$ has an additional
antiunitary symmetry \cite{v1}, which emerges as follows.

Recall that the unitary $\mathrm{SU}_2$-representation space $R =
\mathbb{C}^2$ is isomorphic to the dual representation space
$R^\ast$ by a $\mathbb{C }$-linear mapping $\psi : R \to R^\ast$.
Combining the inverse of this with $\iota : R \to R^\ast$ defined
by $\iota(r) = \langle r , \cdot \rangle_R$, we obtain a
$\mathbb{C }$-antilinear mapping $\beta := \psi^{-1} \circ \iota :
R \to R$. The map $\beta$ thus defined commutes with the
$\mathrm{SU }_2$-action on $R$.  By Lemma \ref{parity of beta} it
satisfies $\beta^2 = - \mathrm{Id}_R$ since $\psi$ is alternating.

Now, on the (untwisted) spinor bundle $S$ over Euclidean
space-time $M$ there exists a $\mathbb{C}$-antilinear operator
$\alpha$, called {\it charge conjugation} in physics, which
anticommutes with the Clifford action $\gamma : T^\ast M \to
\mathrm{End}(S)$; thus $\alpha \mathrm{i} \gamma = \mathrm{i}
\gamma\, \alpha$. Since $\gamma_5 = \gamma^0 \gamma^1 \gamma^2
\gamma^3$, this implies that $\alpha$ commutes with $\gamma_5 =
\Gamma$ and stabilizes the $\Gamma$-eigenspace decomposition $S =
S_+ \oplus S_-$ into half-spinor components $S_\pm$.  The charge
conjugation operator has square $\alpha^2 = - \mathrm{Id}_S$.

For the case of three or more colors, the existence of $\alpha$ is
of no consequence from a symmetry perspective, as the fundamental
and antifundamental representations of $\mathrm{SU}_N$ are
inequivalent for $N \ge 3$. For $N = 2$, however, we also have
$\beta$, and $\alpha$ combines with it to give an antiunitary
symmetry $T_1 = \alpha \otimes \beta$. Indeed,
\begin{displaymath}
    T_1^{\vphantom{-1}} D_A^{\vphantom{-1}} T_1^{-1} =
    (\alpha \otimes \beta) D_A (\alpha \otimes \beta) = \alpha
    (\mathrm{i} \gamma^\mu) \alpha^{-1} \otimes \beta
    (\partial_\mu - A_\mu) \beta^{-1} \;.
\end{displaymath}
Since gauge transformations $g(x) \in \mathrm{SU}_2$ commute with
$\beta$, so do the components $A_\mu(x) \in \mathfrak{su}_2$ of
the gauge field.  Thus $\beta A_\mu \beta^{-1} = A_\mu\,$, and
since $\alpha (\mathrm{i}\gamma) \alpha^{-1} = \mathrm{i} \gamma$,
we have
\begin{displaymath}
    T_1^{\vphantom{-1}} D_A^{\vphantom{-1}} T_1^{-1} = D_A \;.
\end{displaymath}
Note that the antiunitary symmetry $T_1 : \mathcal{V} \to \mathcal
{V}$ is nonmixing, and $T_1^2 = \mathrm{Id}$.  As usual, the
extension to an operator $T_1 : \mathcal{W} \to \mathcal{W}$ is
made by requiring $C T_1 = T_1 C$.

Thus we now have two antiunitary symmetries, $T$ and $T_1$.  Because
$T$ is mixing and $T_1$ nonmixing, the unitary operator $P = T T_1 =
T_1 T$ mixes $\mathcal{V}$ with $\mathcal{V}^\ast$.  Since $T^2 =
T_1^2 = \mathrm{Id}$, and $(CP)^2 = \mathrm{Id}$, this is the case
treated in \ref{sect: 5.1.1}.1, where we found
\begin{displaymath}
    \mathfrak{p} \cong \mathfrak{so}( \mathcal{V}_\mathbb{R} ) /
    (\mathfrak{so} (\mathcal{V}_\mathbb{R}^+ ) \oplus \mathfrak{so}
    (\mathcal{V}_\mathbb{R}^-)) \;.
\end{displaymath}
After truncation to finite dimension this is $\mathfrak{so}_{p+q}
/ (\mathfrak{so}_p \oplus \mathfrak{so}_q)$. The difference $p -
q$ still has a topological interpretation as the index of the
Dirac operator.

Although our considerations explicitly referred to the case of the
gauge group being $\mathrm{SU}_2$, the only specific feature we
used was the existence of an alternating isomorphism $\psi : R \to
R^\ast$. The same result therefore holds for any gauge group
representation $R$ where such an isomorphism exists.  In
particular it holds for the fundamental representation of the
whole series of symplectic groups $\mathrm{USp}_{2N}$ (which
includes $\mathrm{SU}_2 \cong \mathrm{USp}_2$).

\subsubsection{Class $C${\rm II}}\label{sect: CII}

Now take $R$ to be the adjoint representation of any compact Lie
(gauge) group $K$ with semisimple Lie algebra.  This case is called
`adjoint fermions' in physics. A detailed symmetry analysis of it was
presented in \cite{halasz}.

The Cartan-Killing form on $\mathrm{Lie}(K)$,
\begin{displaymath}
    B(X,Y) = \mathrm{Tr}\, \mathrm{ad}(X) \mathrm{ad}(Y) \;,
\end{displaymath}
is nondegenerate, invariant, complex bilinear, and symmetric. $B$
therefore defines an isomorphism $\psi : R \to R^\ast$ by $\psi(X)
= B(X, \cdot)$. Since $B$ is symmetric, so is $\psi$.

The change in parity of $\psi$ reverses the parity of the
antiunitary operator $\beta = \psi^{-1} \circ \iota$, which now
satisfies $\beta^2 = +\mathrm{Id}_R$.  By $\alpha^2 = -\mathrm
{Id}$ this translates to $T_1^2 = (\alpha \otimes \beta)^2 = -
\mathrm {Id}$.

Thus we now have two antiunitary symmetries $T$ and $T_1$ with $T^2 =
\mathrm{Id} = - T_1^2$, and $(CP)^2 = (CTT_1)^2 = - \mathrm{Id}$.
This case was handled in \ref{sect: 5.1.2}.1 where we found
\begin{displaymath}
    \mathfrak{p} \cong \mathfrak{usp}( \mathcal{V} ) /
    (\mathfrak{usp} (\mathcal{V}^+ ) \oplus \mathfrak{usp}
    (\mathcal{V}^-)) \;.
\end{displaymath}
In a finite-dimensional setting this would be $\mathfrak{usp}_{2p+
2q} / (\mathfrak{usp}_{2p} \oplus \mathfrak{usp}_{2q})$.

In summary, the physical situation is ruled by a mathematical
trichotomy: the isomorphism $\psi : R \to R^\ast$ is either symmetric,
or alternating, or does not exist.  The corresponding symmetry class
of the massless Dirac operator is $C$II, $BD$I, or $A$III,
respectively.  As was first observed by Verbaarschot \cite{jjmv}, this
is the same trichotomy that ruled Dyson's threefold way.

\medskip\noindent {\bf Acknowledgment}. This work was carried out
under the auspices of the Deutsche Forschungsgemeinschaft, SFB/TR12.
Major portions of the article were prepared while M.R.Z.\ was visiting
the Institute for Advanced Study (Princeton, USA) and the Newton
Institute for Mathematical Sciences (Cambridge, UK).  The support of
these institutions is gratefully acknowledged.

\end{document}